\documentclass[]{JHEP3}

\usepackage{epsfig} 

\newcommand{\Hh}{\lower1.2ex\hbox{$\stackrel{\textstyle 
H}{\footnotesize\sim}$}} 
\newcommand{\Hho}{\lower1.2ex\hbox{$\stackrel{\textstyle 
H_1}{\footnotesize\sim}$}} 
\newcommand{\Hhw}{\lower1.2ex\hbox{$\stackrel{\textstyle 
H_2}{\footnotesize\sim}$}} 
\newcommand{\h}{\lower1.2ex\hbox{$\stackrel{\textstyle 
h}{\footnotesize\sim}$}}

\newcommand{\gsim}{\lower.7ex\hbox{$\;\stackrel{\textstyle>}{\sim}\;$}} 
\newcommand{\lsim}{\lower.7ex\hbox{$\;\stackrel{\textstyle<}{\sim}\;$}} 
\newcommand{\be}{\begin{equation}} \newcommand{\ee}{\end{equation}} 
\newcommand{\beq}{\begin{equation}} \newcommand{\eeq}{\end{equation}} 
\newcommand{\bea}{\begin{eqnarray}} \newcommand{\eea}{\end{eqnarray}} 
 
\def\simlt{\stackrel{<}{{}_\sim}} \def\simgt{\stackrel{>}{{}_\sim}}

\title{Implications for New Physics from Fine-Tuning Arguments: II. Little 
Higgs Models} 
\author{J. Alberto Casas, Jos\'e Ram\'on Espinosa and Irene Hidalgo\\ 
IFT-UAM/CSIC, Facultad de Ciencias, C-XVI,\\ 
UAM, Cantoblanco, 28049 Madrid, Spain 
\\ E-mail: \email{alberto.casas@uam.es, 
jose.espinosa@uam.es, irene.hidalgo@uam.es}} 
\abstract{We examine the 
fine-tuning associated to electroweak breaking in Little Higgs 
scenarios and find it to be always substantial and, generically, much 
higher 
than suggested by the rough estimates usually made. This is due to 
implicit tunings between parameters that can be overlooked at first
glance but show up in a more systematic analysis. Focusing on four 
popular and 
representative Little Higgs scenarios, we find that the fine-tuning is 
essentially comparable to that of the Little Hierarchy problem of the 
Standard Model 
(that these scenarios attempt to solve) and higher than in supersymmetric 
models.  This does not demonstrate that all Little Higgs models are 
fine-tuned, but stresses the need of a careful analysis of this issue in 
model-building before claiming that a particular model is not 
fine-tuned.
In this respect we identify the main sources of 
potential fine-tuning that should be watched out for, in order to 
construct a successful Little Higgs model, which seems to be  a 
non-trivial goal. } 
\preprint{IFT-UAM/CSIC-05-07}

\begin{document}

\section{Introduction}

In this paper we continue the exam of the implications for new physics 
from fine-tuning arguments. In a previous paper \cite{CEHI} we revisited 
the use of the Big Hierarchy problem of the Standard Model (SM) to 
estimate the scale of new physics, $\Lambda_{\rm SM}$, illustrating our 
results with two physically relevant examples: right handed (see-saw) 
neutrinos and supersymmetry (SUSY).  Here we study Little Higgs (LH) 
scenarios as the new physics beyond the SM.

LH models were introduced as an alternative to SUSY in order to 
solve the Little Hierarchy problem. Very briefly, the latter consists in 
the following: in the SM (treated as an effective theory valid below 
$\Lambda_{\rm SM}$) the mass parameter $m^2$ in the Higgs potential 
\be 
V={1\over 2} m^2 h^2 +{1\over 4}\lambda h^4\ , \ee 
receives important 
quadratically-divergent contributions \cite{veltman}. At one-loop, 
\be 
\label{quadrdiv0} \delta_{q} m^2 = {3\over 64\pi^2}(3g^2 + g'^2 + 
8\lambda - 8\lambda_t^2)\Lambda_{\rm SM}^2\ , \ee 
where $g, g', \lambda$ 
and $\lambda_t$ are the $SU(2)\times U(1)_Y$ gauge couplings, the quartic 
Higgs coupling and the top Yukawa coupling, respectively.  The requirement 
of no fine-tuning between the above contribution and the tree-level value 
of $m^2$ sets an upper bound on $\Lambda_{\rm SM}$. E.g. for a Higgs mass 
$m_h = 115-200$ GeV, 
\be \label{quadrft} \left|{\delta_{ q} m^2 \over 
m^2}\right| \leq 10\ \Rightarrow \ \Lambda_{\rm SM} \simlt 2-3\ {\rm TeV} 
\ . \ee 
This upper bound on $\Lambda_{\rm SM}$ is in a certain tension 
with the experimental lower bounds on the suppression scale $\Lambda$ of 
higher order operators, derived from fits to precision electroweak data 
\cite{LEParadox}, which typically require $\Lambda\simgt$ 10 TeV; and this 
is known as the Little Hierarchy problem.

Let us briefly outline the general structure of LH models.  Their two 
basic ingredients are, first, that the SM Higgs is a Goldstone boson of a 
spontaneously broken global symmetry; and second, the explicit breaking of 
this symmetry by gauge and Yukawa couplings in a collective way (a 
coupling alone is not able to produce enough breaking to give a mass to 
the SM Higgs). In consequence, the SM Higgs is a pseudo-Goldstone boson, 
with mass protected at 1-loop from quadratically divergent contributions. 
In principle, this is enough to avoid the Little Hierarchy problem: if the 
quadratic corrections to $m^2$ appear at the 2-loop level, the extra $(4 
\pi)^{-2}$ suppression factor in $\delta_{ q}m^2$ allows for a 10 TeV 
cut-off with no fine-tuning price.

It should be noticed that the above argument does not imply that in LH 
models there are no extra states at all below 10 TeV.  As discussed in the 
previous paper \cite{CEHI}, even if the quadratic divergences cancel 
exactly, the new physics states do contribute with logarithmic and finite 
corrections to $m^2$, so their masses should still not be larger than 2--3 
TeV (as happens in the supersymmetric case). This is also the case for LH 
models: the lightest extra states have masses in the TeV range (the scale 
at which the global symmetry is spontaneously broken), but their 
contributions to the electroweak observables are calculable and 
(hopefully) under control.  Besides, this effective description is valid 
up to a 
cut-off scale, $\Lambda\simeq 10$ TeV, beyond which some unspecified UV 
completion takes over \cite{UV,UVmore}.

Despite the good prospects, the absence of fine-tuning in particular LH 
scenarios should be checked in practice. More precisely, the fine-tuning 
must be computed for the different LH models with the same level of rigor 
employed for the supersymmetric models in the past. A systematic attempt 
of this kind has not been done up to now, and it is the main goal of 
this paper. We will focus only on the naturalness of the electroweak 
breaking, although LH models may have other (model-dependent) problems.

\FIGURE[t]{
\psfig{file=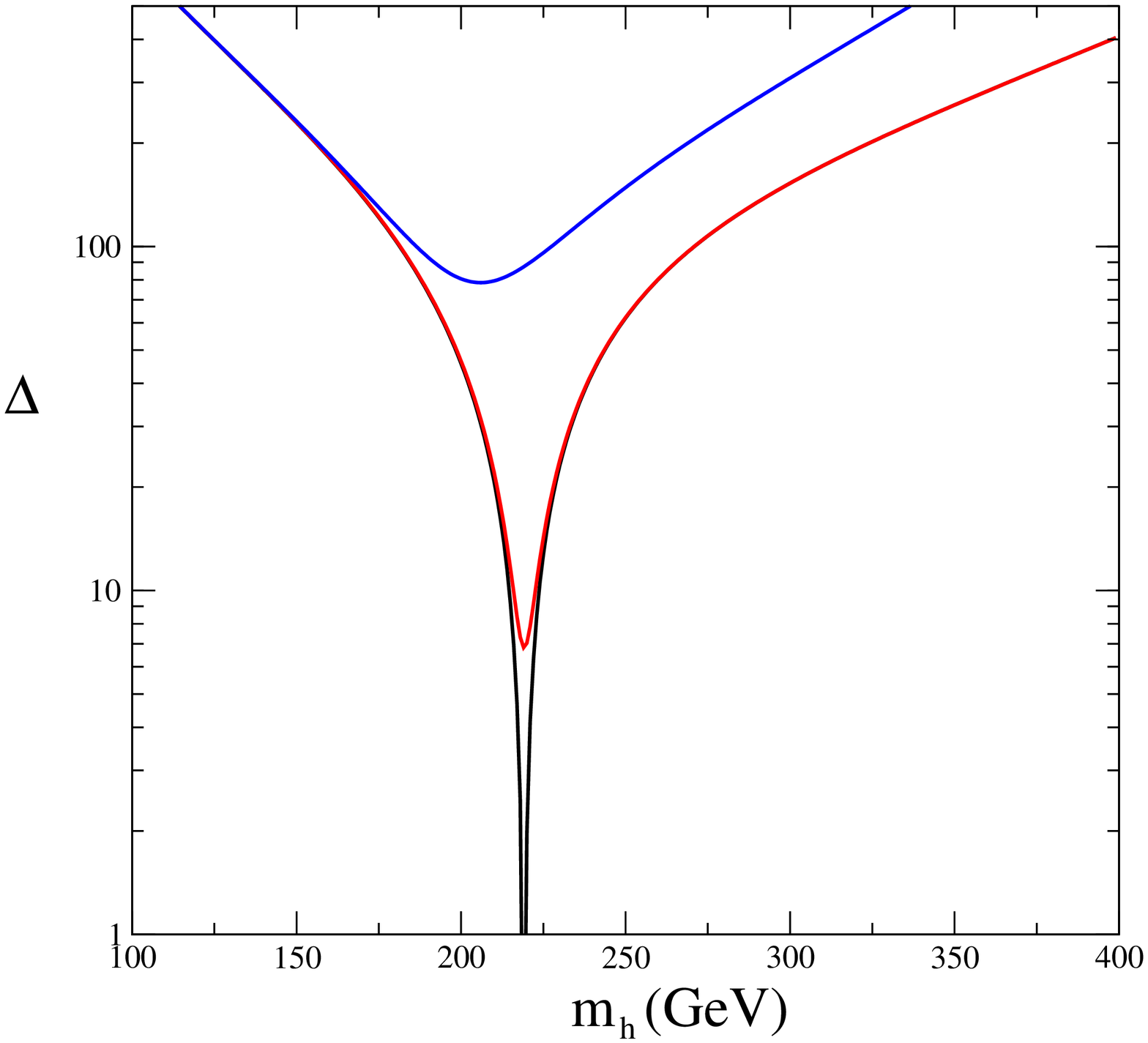,width=7cm,bbllx=2.cm,bblly=0.cm,bburx=20.cm,bbury=23.cm,angle=-90}
\caption{\footnotesize Fine-tuning contours as a function of the Higgs 
mass in the SM with a cut-off $\Lambda=10$ TeV. This can be considered as 
the fine-tuning of the Little Hierarchy problem in the SM. Different 
curves correspond to progressively more sophisticated definitions of 
$\Delta$ (from black [bottom line] to red to blue [top line], see text for 
details).} 
\label{fig:velt-a}} 

To quantify the fine tuning we follow Barbieri and Giudice \cite{BG,BC}: 
we write the Higgs VEV as $v^2=v^2(p_1, p_2, \cdots)$, where $p_i$ are input 
parameters of the model under study, and define $\Delta_{p_i}$, the 
amount of fine tuning associated to $p_i$, by 
\be \label{ftBG} {\delta M_Z^2\over M_Z^2}= 
{\delta v^2\over v^2} = \Delta_{p_i}{\delta p_i\over p_i}\ , 
\ee 
where $\delta M_Z^2$ (or $\delta v^2$) is the change induced 
in $M_Z^2$ (or $v^2$) by a change $\delta p_i$ in $p_i$. Roughly speaking, 
$|\Delta^{-1}_{p_i}|$ measures the probability of a cancellation among 
terms of a given size to obtain a result which is $|\Delta_{p_i}|$ times 
smaller.  Due to the statistical meaning of $\Delta_{p_i}$, we define the 
total fine-tuning as 
\be \label{Deltatot} \Delta 
\equiv\left[\sum_i\Delta_{p_i}^2\right]^{1/2}\ .  
\ee 
It is important to recall that the Little Hierarchy problem of the SM, 
which the LH 
models attempt to solve, is itself a fine-tuning problem: one could simply 
assume $\Lambda_{\rm SM}\simgt 10$ TeV with the 'only' price of tuning 
$\delta_{ q} m^2$, as given by eq.~(\ref{quadrdiv0}), at the 0.4--1 \% 
level (or, equivalently, $\Delta = 100-250$).  Therefore, to be of 
interest, the LH models should at least improve this degree of 
fine-tuning. In order to perform a fair comparison, this estimate can be 
refined following the lines explained in refs.~\cite{CEHI,KM,EJ}. First of 
all, eq.~(\ref{quadrdiv0})  should be renormalization-group improved.  
Then, the value of $\Delta$ $(=\Delta_\Lambda)$ vs. $m_h$ is given by the 
(bottom) black line 
of fig.~\ref{fig:velt-a}. The deep throat at $m_h\sim 220$ GeV results 
from an accidental cancellation between the various terms in 
eq.~(\ref{quadrdiv0}). This throat is cut when the fine-tuning parameter 
associated to the top mass ($\Delta_{\lambda_t}$) is added in quadrature 
as explained above (for details see ref.~\cite{CEHI}), giving the (middle) 
red line.  Finally, once the fine-tuning parameter 
associated to the Higgs mass itself ($\Delta_\lambda$) is included as 
well, the value of $\Delta$ is given by the (top) blue line, which thus 
represents the fine-tuning associated to the Little Hierarchy problem.  
This has to be compared with the tuning of LH models.  On the other hand, 
for the minimal supersymmetric standard model (MSSM) the degree of 
fine-tuning is presently at the few percent level ($\Delta \simeq 20-40$ 
for $m_h\simlt 125$ GeV), while for other supersymmetric models the 
situation is much better \cite{CEH0,RETAHILA}. Hence, in order to be 
competitive with supersymmetry, LH models should not worsen the MSSM 
performance. We will use these criteria in order to analyze the success 
of several representative LH models.

Due to the great variety of LH models we do not attempt to perform here 
an exhaustive analysis of them. Rather, we have focused on four LH 
scenarios \cite{Littlest,Peskin,ChengLow,Sch} which are probably the most 
popular ones, and tried to extract 
general lessons for other models.  The prototype LH scenario is the 
so-called Littlest Higgs model \cite{Littlest}. This model is a very good 
example to start with, due to its simplicity and because it shares many 
features with more elaborate LH constructions. Actually, many of those 
models are simply modifications of the Littlest Higgs model. The Littlest 
Higgs has some phenomenological problems with the 
constraints from precision electroweak observables. (Incidentally, this 
illustrates the fact that the impact of the TeV--mass states of LH models 
on electroweak observables is not always under control \cite{pew}.) Since 
our focus is the naturalness of electroweak breaking, we will ignore those 
constraints, although the strongest results would come from combining both 
analyses. On the other hand, there exist modifications of the Littlest 
Higgs (also studied in this paper)  able to overcome those difficulties.

The paper is organized as follows. In sect.~2 we analyze the structure, 
and evaluate the fine-tuning, of the Littlest Higgs model. Sections 3 and 
4 are devoted respectively to the computation of the fine-tuning in two 
popular modifications of the Littlest Higgs proposed in 
refs.~\cite{Peskin} and \cite{ChengLow}. The latter corresponds to the 
so-called Littlest Higgs model with $T$-parity. In sect.~5 we study a 
recent proposal (the so-called ``Simplest Little Higgs'' \cite{Sch}), 
whose structure differs substantially from the Littlest Higgs.  In all 
these cases the fine-tuning turns out to be essentially comparable with 
that of the Little Hierarchy problem of the SM (that LH models attempt to 
solve) and higher than in supersymmetric models, and we discuss the 
reasons for this fact.  Finally, in sect.~6 we summarize our results and 
present some conclusions.  In addition we present in Appendix A a simple 
recipe to evaluate the fine-tuning when the various parameters of a model 
are subject to constraints. Appendix~B contains details on the structure 
of the different Little Higgs models studied.

\section{The Littlest Higgs \cite{Littlest}}

\subsection{Structure of the model}

The Littlest Higgs model is a non-linear sigma model based on a global 
$SU(5)$ symmetry, spontaneously broken to $SO(5)$ at a scale 
$f\sim 1$ TeV, and explicitly broken by the gauging of an $[SU(2)\times 
U(1)]^2$ subgroup.  After the spontaneous breaking, the latter gets broken 
to its diagonal subgroup, identified with the SM electroweak gauge group, 
$SU(2)_L\times U(1)_Y$.  From the 14 (pseudo)-Goldstone bosons of the 
$SU(5)\rightarrow SO(5)$ breaking, 4 degrees of freedom (d.o.f.) are true 
Goldstones [eaten by the gauge bosons of the spontaneously broken 
(``axial") $SU(2)\times U(1)$, which thus acquire masses $\sim f\sim 1$ 
TeV through the Higgs mechanism] and the remaining 10 d.o.f.  correspond 
to the SM Higgs doublet, $H=(h^0,h^+)$, (4 d.o.f.) and a complex $SU(2)_L$ 
scalar triplet, $\phi$ (6 d.o.f.) with $Y=1$;  in vectorial notation, 
$\phi=(\phi^{++},\phi^+,\phi^0)$. All these fields can be treated 
simultaneously in a nonlinear matrix field $\Sigma$ (see Appendix~B for 
details).

The $[SU(2)\times U(1)]^2$ gauge interactions give a radiative mass to the 
SM Higgs, but only when the couplings of both groups are simultaneously 
present, as explained in more detail below. Hence, the quadratically 
divergent contributions only appear at two-loop order, and the high-energy 
cut-off can be pushed up to a scale $\Lambda\sim 4\pi f \sim 10$ TeV, as 
explained in the introduction.  For the potentially dangerous top-Yukawa 
interactions things work in a similar way: the spectrum is enlarged with 
two extra fermions of opposite chiralities, and the conventional 
top-Yukawa coupling, $\lambda_t$, is not an input parameter but results 
from two independent couplings $\lambda_1, \lambda_2$. Both must be 
present in order to generate a radiative correction to the Higgs mass, and 
this again forbids quadratically divergent corrections to $m_h^2$ at 
one--loop.

For our purposes, the relevant states besides those of the SM are:  the 
pseudo-Goldstone bosons $H,\phi$; the heavy gauge bosons, $W', B'$, of the 
axial $SU(2)\times U(1)$; and the two extra (left and right) fermionic 
d.o.f. that combine in a vector-like ``heavy Top", $T$. The relevant part 
of the Lagrangian can be found in Appendix B, eqs.~(\ref{Lkin}) and 
(\ref{Lf}). It consists of two pieces 
\be \label{L1} 
{\cal L} = {\cal L}_{kin}(g_1, g_2, g_1', g_2')\ +\ {\cal L}_{f}(\lambda_1, 
\lambda_2)  \ , \ee 
where $g_1, g'_1$ ($g_2, g'_2$) are the gauge 
couplings of the first (second) $SU(2)\times U(1)$ factor, and $\lambda_1, 
\lambda_2$ are the two independent fermionic couplings. These couplings 
are constrained by the relations with the SM couplings, 
\bea 
\label{relations} {1\over g^2} = {1\over g_1^2}+{1\over g_2^2}\ , 
\;\;\;\;\;\ {1\over g'^2} = {1\over g_1'^2}+{1\over g_2'^2}\ ,\;\;\;\;\; 
{2\over\lambda_t^2} = {1\over\lambda_1^2}+{1\over\lambda_2^2} \ , \eea 
where $g$ and $g'$ are the $SU(2)$ and $U(1)_Y$ gauge couplings, 
respectively, and $\lambda_t$ is the top Yukawa coupling.  The Lagrangian 
(\ref{L1}) gives ${\cal O}(f)$ masses to $W', B'$ and $T$. These heavy 
masses have a non-trivial dependence on the full non-linear field 
$\Sigma$, which contains the $H$ and $\phi$ fields.  In particular, 
retaining only the dependence on $h\simeq {\rm Re}(h^0)\sqrt{2}$ we get 
\bea \label{masas1} m_{W'}^2(h)&=&M_{W'}^2+{\cal O}(h^2)={1\over 
4}(g_1^2+g_2^2)f^2 - {1\over 4}g^2h^2+ {\cal O}(h^4/f^2)\ ,\nonumber\\ 
m_{B'}^2(h)&=&M_{B'}^2+{\cal O}(h^2)={1\over 20}(g_1'^2+g_2'^2)f^2-{1\over 
4}g'^2h^2+ {\cal O}(h^4/f^2)\ , \nonumber\\ m_{T}^2(h)&=&M_{T}^2+{\cal 
O}(h^2)=(\lambda_1^2+\lambda_2^2)f^2- {1\over 2} \lambda_t^2 h^2 + {\cal 
O}(h^4/f^2)\ .  \eea 

At this level, $H$ and $\phi$ are massless, but they get massive 
radiatively. The simplest way to see this is by using the effective 
potential. Let us consider first the quadratically divergent contribution 
to the one-loop scalar potential, given by \be \label{V1gen} V_1^{\rm 
quad} = {1\over 32 \pi^2}\Lambda^2\ {\rm Str} {\cal M}^2 \ , \ee where the 
supertrace ${\rm Str}$ counts degrees of freedom with a minus sign for 
fermions, and ${\cal M}^2$ is the (tree-level, field-dependent) 
mass-squared matrix. In our case, the previous formula gives 
\be 
\label{V1} V_1^{\rm quad} = {1\over 32 \pi^2}\Lambda^2\ \left[ 6 
m_W^2+9m_{W'}^2 + 3 m_Z^2 + 3 m_{B'}^2 - 12 (m_t^2+m_{T}^2)\right]\ .  
\ee 
By looking at the $h$-dependence of the masses above it is easy to 
check that $V_1^{\rm quad}$ does not contain a mass term for $h$ (this 
will be generated by the logarithmic and finite contributions to the 
potential, to be discussed shortly). The reason for this result is the 
following.  If $\lambda_1=g_2=g_2'=0$, the Lagrangian (\ref{L1}) recovers 
a global $SU(3)$ [$SU(3)_1$, living in the upper corner of $SU(5)$] that 
protects the mass of the Higgs (which transforms by a shift under that 
symmetry). On the other hand, if $\lambda_2=g_1=g_1'=0$, then a different 
$SU(3)$ symmetry [$SU(3)_2$, living in the lower corner of $SU(5)$] is 
recovered that also protects the Higgs mass. A non-zero value for the 
Higgs mass can only be generated by breaking both $SU(3)$'s and therefore 
both type-1 and type-2 couplings should be present.  Quadratically 
divergent diagrams involve only one type of coupling and therefore cannot 
contribute to the Higgs mass. This is the so-called collective breaking of 
the original $SU(5)$ symmetry and is one of the main ingredients of Little 
Higgs models.

These symmetries do not protect the mass of the triplet. In fact, if we 
include the full dependence of the bosonic ($W', B'$) and fermionic ($T$) 
masses on the $\Sigma$ field, $V_1^{\rm quad}$ contains operators, ${\cal 
O}_V(\Sigma)$ and ${\cal O}_F(\Sigma)$ respectively, that produce a mass 
term for the triplet $\phi$ of order $\Lambda^2/(16 \pi^2)\sim f^2$. 
Explicit expressions for these operators are given in Appendix~B.  Then, 
following \cite{Littlest}, it is reasonable to assume that ${\cal 
O}_V(\Sigma)$ and ${\cal O}_F(\Sigma)$ are already present at tree-level, 
as a remnant of the heavy physics integrated out at $\Lambda$ (a threshold 
effect). These effects can be accounted for by adding an extra 
piece to the Lagrangian, 
\be 
\label{DeltaL1} -\Delta{\cal L} = c\ {\cal 
O}_V(\Sigma) \ +\ c'\ {\cal O}_F(\Sigma)\ , \ee 
were $c$ and $c'$ are 
unknown coefficients [see eq.~(\ref{potential}) in Appendix B for an 
explicit expression of $\Delta{\cal L}$].  For future use, it is 
convenient to discuss here what is the natural size of $c$ and $c'$. Naive 
dimensional analysis \cite{NDA} has been used to estimate $c, c'\sim {\cal 
O}(1)$. We can make a more precise evaluation by computing the one-loop 
contributions to $c$ and $c'$ coming from (\ref{V1}), keeping the full 
dependence of the masses on $\Sigma$. Then we get 
\bea \label{ccp} c &=& 
c_0 \ +\ c_1 = c_0 \ + 3/4\ , \nonumber\\ c' &=& c_0' \ +\ c_1' = c_0' \ - 
{24}\ .  \eea 
where the subindex 0 labels the unknown threshold contributions from 
physics beyond $\Lambda$.

Besides giving a mass to $\phi$, the operators in eq.~(\ref{DeltaL1}) 
produce a coupling $\sim h^2\phi$\footnote{This coupling induces a 
tadpole for $\phi$ after electroweak symmetry breaking. Keeping the VEV of 
$\phi$ small enough is a necessary requirement to obtain an acceptable 
model and we ensure that this is the case in our numerical analysis. Then, 
it is a good approximation to neglect the effect of that small VEV in most 
places.} and a quartic coupling for $h$. This quartic coupling is modified 
by the presence of the $h^2\phi$ term once the heavy triplet is integrated 
out.  After that is done, the Higgs quartic coupling $\lambda$ can be 
written in the simplest manner as 
\be \label{lambda} {1\over 
\lambda}={1\over \lambda_a}+{1\over \lambda_b}\ , \ee 
with \be 
\lambda_a\equiv c(g_2^2+g_2'^2)-c'\lambda_1^2\ , \;\;\;\;\; 
\lambda_b\equiv c(g_1^2+g_1'^2)\ . \ee We see that the structure of 
(\ref{lambda}) is similar to that of (\ref{relations}) for the fermion and 
gauge boson couplings, with $\lambda_a$ ($\lambda_b$) being a type-1 
(type-2) coupling.

In order to write the one-loop Higgs potential, we need explicit 
expressions for the $h$-dependent masses of the spectrum. In the scalar 
sector, we decompose $h^0\equiv (h^{0r}+ih^{0i})/\sqrt{2}$ and 
$\phi^{0}\equiv 
i(\phi^{0r}+i\phi^{0i})/\sqrt{2}$. In the $CP$-even sector we write the 
relevant part of the mass matrix in the basis $\{h^{0r},\phi^{0r}\}$; in 
the $CP$-odd sector we use the basis $\{h^{0i},\phi^{0i}\}$ and finally, 
in the charged sector the basis $\{h^+,\phi^+\}$. The three mass matrices 
are very similar in structure and can be written simultaneously 
as\footnote{At this point there is no tree-level mass term for the Higgs 
field but the presence of a quartic coupling gives it a nonzero mass in a 
background $h$.} \be M^2_\kappa(h)=\left[ \begin{array}{cc} {1\over 
4}a_\kappa\lambda_+h^2+{1\over \sqrt{2}}s_\kappa\lambda_- f t+{\cal 
O}(h^4/f^2) & b_\kappa\lambda_-fh+{\cal O}(h^2)\\ & \\ 
b^*_\kappa\lambda_-fh+{\cal O}(h^2)& \lambda_+(f^2-c_\kappa h^2)+{\cal 
O}(h^4/f^2) \end{array} \right]\ , \label{massmat} \ee where the index 
$\kappa=\{0r,0i,+\}$ labels the different sectors, $a_\kappa=\{3,1,1\}$, 
$s_\kappa=\{1,-1,0\}$, $b_\kappa=\{1/\sqrt{2},1/\sqrt{2},i/2\}$, 
$c_\kappa=|b_\kappa|^2$, and we have defined $\lambda_+\equiv 
\lambda_a+\lambda_b$, $\lambda_-\equiv \lambda_a-\lambda_b$. We have also 
included in these mass matrices the contribution of the triplet VEV, 
$t\equiv\langle\phi^{0r}\rangle$, with \be t\simeq -{1\over 
2\sqrt{2}}{\lambda_-h^2\over\lambda_+f} \ .  \ee The off-diagonal entries 
in (\ref{massmat}) are due to the $h^2\phi$ coupling and they cause mixing 
between $h$ and $\phi$.  Concerning the masses, the effect of this mixing 
is negligible for the triplet [at order $h^2$, the masses of  
$\phi^{0r}$ and $\phi^{0i}$ are the same, and these fields can still be 
combined in a complex  field $\phi^{0}$]. Explicitly, these masses are 
\be \left[ \begin{array}{c} m_{\phi^0}^2(h)\\ m_{\phi^+}^2(h)\\ 
m_{\phi^{++}}^2(h) \end{array}\right] =M_{\phi}^2+{\cal 
O}(h^2)=(\lambda_a+\lambda_b)f^2 -\left[
  \begin{array}{c} 2\\ 1\\ 0 \end{array}\right] \lambda h^2+{\cal 
O}(h^4/f^2)\ . 
\label{masatrip} 
\ee 
We will call ${h'}^{0r}$, 
${h'}^{0i}$ and 
${h'}^+$ the light mass eigenstates of (\ref{massmat}) in the different 
sectors, for which we get  
\be 
\left[ \begin{array}{c} m^2_{{h'}^{0r}}(h)\\ m^2_{{h'}^{0i}}(h)\\ 
m^2_{{h'}^+}(h) \end{array}\right] =\left[ \begin{array}{c} 3\\ 1\\ 1 
\end{array}\right] \lambda h^2+{\cal O}(h^4/f^2)\ .  
\ee 
From the previous 
expressions it is straightforward to check that, in the contribution of 
scalars to $V_1^{\rm quad}$, 
\be 
{\Lambda^2\over 
32\pi^2}\left(m_{{h'}^{0r}}^2+m_{{h'}^{0i}}^2+2m_{{h'}^+}^2 
+2m_{\phi^0}^2+2m_{\phi^+}^2+2m_{\phi^{++}}^2\right)\ , 
\ee 
there is also 
a cancellation of $h^2$ terms. This is due to the fact that the operators 
of (\ref{DeltaL1}) still respect the same $SU(3)_i$ symmetries of the 
original Lagrangian as they originate from quadratically divergent 
one--loop corrections.

Finally, a non-vanishing mass parameter for $h$ arises from the 
logarithmic and finite contributions to the effective potential. In the 
$\overline{\rm MS}$ scheme, in Landau gauge, and setting the 
renormalization scale $Q=\Lambda$, 
\bea 
\label{m2} m^2 &=& {3\over 
64\pi^2}\left\{3g^2M_{W'}^2\left[\log{\Lambda^2\over M_{W'}^2} + {1\over 
3}\right] + g'^2M_{B'}^2\left[\log{\Lambda^2\over M_{B'}^2}+ {1\over 
3}\right] \right\} \nonumber\\ &+&{3\lambda\over 8\pi^2}M_\phi^2 
\left[\log{\Lambda^2\over M_\phi^2}+ 1\right] -{3\lambda_t^2\over 
8\pi^2}M_T^2 \left[\log{\Lambda^2\over M_{T}^2}+ 1\right]\ , 
\eea 
where 
we have included the contribution from the $\phi$ masses.

In summary, the effective potential of the Higgs field can be written in 
the SM-like form \be V={1\over 2} m^2 h^2 +{1\over 4}\lambda h^4\ , 
\label{V} \ee where $\lambda$ and $m^2$ are given by eqs.~(\ref{lambda}) 
and (\ref{m2}).  The Higgs VEV is simply \be v^2=-{m^2\over \lambda}\ . 
\label{v2} \ee

\subsection{Fine-tuning analysis}

A rough estimate of the fine-tuning associated to electroweak breaking in 
the Littlest Higgs model can be obtained from eq.~(\ref{m2}). The 
contribution of the heavy Top, $T$, to the Higgs mass parameter is 
\be 
\label{deltatmh2} \delta_T m^2 = -{3\lambda_t^2\over 8\pi^2}M_T^2 
\left[\log{\Lambda^2\over M_{T}^2}+ 1\right]\ .  \ee 
 Using 
eqs.~(\ref{relations}) and (\ref{masas1}), it follows\footnote{Similar 
bounds, based on the same type of coupling structure, hold for the rest of 
heavy states: $M_{W'}^2\geq g^2 f^2$, $M_{B'}^2\geq g'^2f^2/5$ and 
$M^2_\phi\geq 4\lambda f^2$.} that $M_T^2\geq 2 \lambda_t^2 f^2$, and thus 
$\delta_T m^2 \geq 0.37 f^2$ (the minimum corresponds to 
$\lambda_1=\lambda_2=\lambda_t$). Thus the ratio $\delta_T m^2 / m^2$, 
tends to be quite large: {\it e.g.} for $f= 1$ TeV and $m_h=115, 150, 250$ 
GeV, one gets $|\delta_T m^2 / m^2|\geq 56,\ 33,\ 12$ respectively. Since 
there are other potential sources of fine-tuning, this should be 
considered as a lower bound on the total fine-tuning. Actually, the 
overall fine tuning is usually much larger than this estimate, as we show 
below.  (Eventually we will go back to this rough argument to 
improve it in a simple way.)

In order to perform a complete fine-tuning analysis we determine first the 
input parameters, $p_i$, and then calculate the associated fine-tunings, 
$\Delta_{p_i}$, according to eq.~(\ref{ftBG}), i.e.  
$\Delta_{p_i}=(p_i/v^2)(\partial v^2/ \partial p_i)$. For the Littlest 
Higgs model the input parameters of the Lagrangian [eqs.~(\ref{L1}) and 
(\ref{DeltaL1})] are 
\be \label{indepparLH} p_i \ =\ \{g_1,\ g_2,\ 
g_1',\ g_2',\ \lambda_1,\ \lambda_2,\ c,\ c',\ f \}\ .  \ee 
We have not 
included $\Lambda$ among these parameters since we are assuming $\Lambda 
\simeq 4\pi f$. On the other hand, the parameter 
$f$ basically appears as a multiplicative factor in the mass parameter, 
$m^2$, so $\Delta_f$ is always ${\cal O}(1)$, and can be 
ignored\footnote{Now it is clear that the assumption $\Lambda
\simeq 4\pi f$ reduces the 
amount of fine-tuning. Had we kept $\{\Lambda,f\}$ as input parameters, 
variations of $\Lambda$ or $f$ would have produced large changes in $m^2$, 
and thus in $v^2$. Therefore, this assumption is a conservative one.}.  
Finally, the above parameters are constrained by the measured values of 
the top mass and the gauge couplings $g, g'$, according to 
eq.~(\ref{relations}). The procedure to estimate the fine-tuning in the 
presence of constraints is discussed in Appendix~A. The net effect is a 
reduction of the ``unconstrained" total fine-tuning, $\Delta = (\sum_i 
\Delta_{p_i}^2)^{1/2}$, according to eq.~(\ref{constrDelta}). In this 
particular case, that equation gives 
\bea \label{DeltaLH} \Delta = 
\left[\sum_i \Delta_{p_i}^2- \sum_{\alpha=1}^3 {1\over N_\alpha^2} 
\left(\sum_i p_i {\partial G^{(0)}_\alpha\over \partial p_i} \Delta_{p_i} 
\right)^2 \right]^{1/2}\ , \eea 
where $G^{(0)}_\alpha =\{ g^2, g'^2, 
\lambda_t^2\}$ are functions of the $p_i$ as given in 
eq.~(\ref{relations}), and \be N_\alpha^2\equiv \sum_i 
p_i^2\left({\partial G^{(0)}_\alpha\over \partial p_i}\right)^2\ , \ee are 
normalization constants.

As announced before, $\Delta$ is in general much larger than the initial 
rough estimate, although the precise magnitude depends strongly on the 
region of parameter space considered and decreases significantly as $m_h$ 
increases.  Let us discuss how this comes about.
The negative contribution from $M_T^2$ to $m^2$ in eq.~(\ref{m2}) must be 
compensated by other positive contributions. Typically, this requires a 
large value of the triplet mass, $M_\phi^2=(\lambda_a+\lambda_b)f^2$, 
which requires a large value of $(\lambda_a+\lambda_b)$, but keeping 
$1/\lambda=1/\lambda_a+1/\lambda_b$ fixed for a given $m_h$. There are two 
ways of achieving this\footnote{The existence of two separate regions of 
solutions can be also understood from the fact that the minimization 
condition (\ref{v2}) becomes quadratic in $c$, for given values of 
$\lambda, \lambda_1, g_1, g_1'$, and in the approximation $\log 
(\Lambda^2/M_\phi^2)\simeq h$-independent.}: 
\bea \label{ab} {\rm 
a)}&&\;\; \lambda\simeq \lambda_b\ll \lambda_a\simeq M_\phi^2/f^2\ 
,\nonumber\\ {\rm b)}&&\;\; \lambda\simeq \lambda_a\ll \lambda_b\simeq 
M_\phi^2/f^2\ .  \eea 
Notice that the one-loop $m^2$ is a symmetric 
function of $\lambda_a$ and $\lambda_b$, so cases ${\rm a)}$ and ${\rm 
b)}$ are simply related by $\lambda_a\leftrightarrow\lambda_b$.  This 
means that the triplet and Higgs masses are exactly the same in both cases 
although the fine-tuning may be different (since the dependence of 
$\lambda_{a,b}$ on $p_i$ is not the same), and indeed it is, as we discuss 
next.

\FIGURE[t]{\mbox{ 
\psfig{file=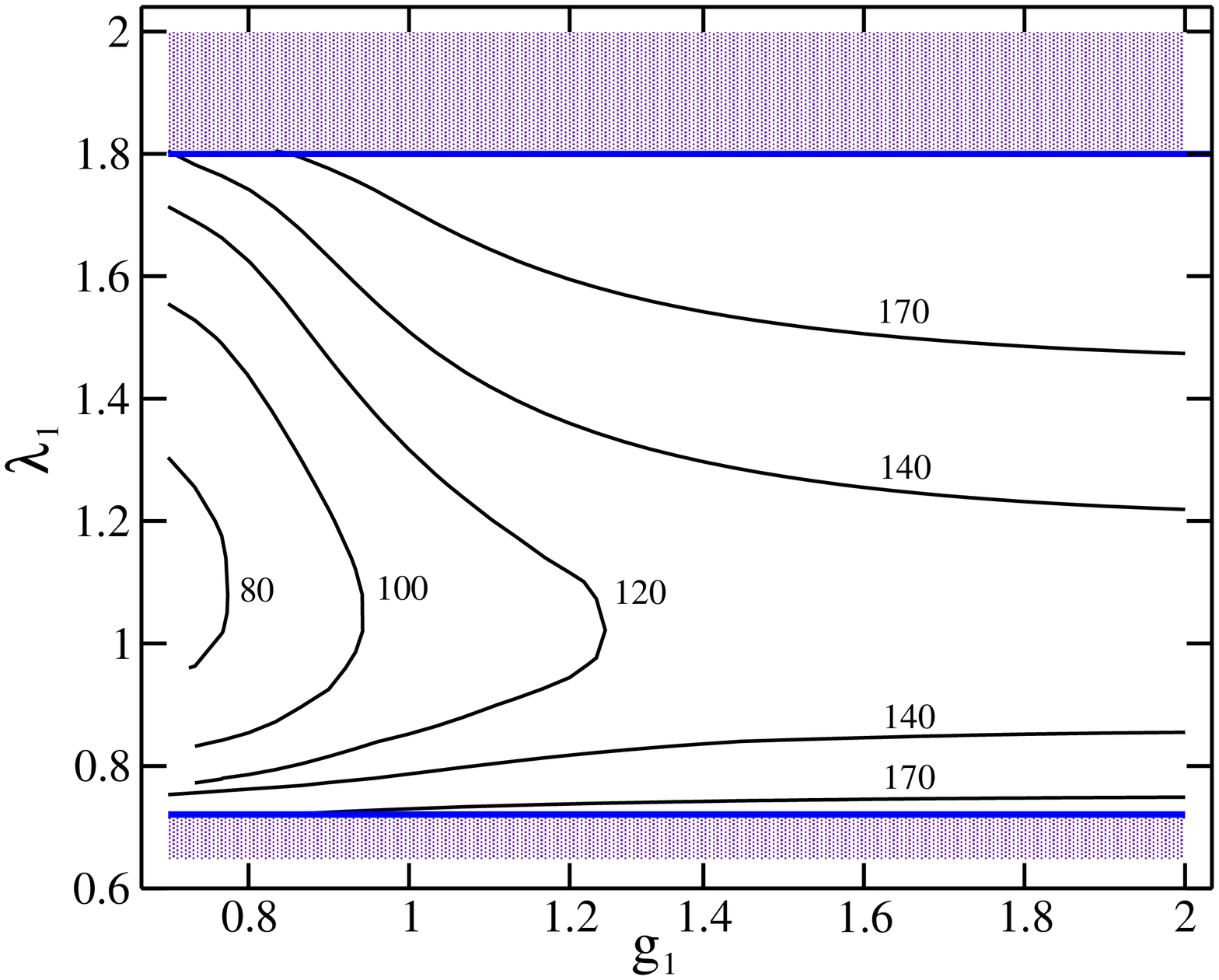,width=5cm,bbllx=4.cm,bblly=0.cm,bburx=20.cm,bbury=23.cm,angle=-90} 
\psfig{file=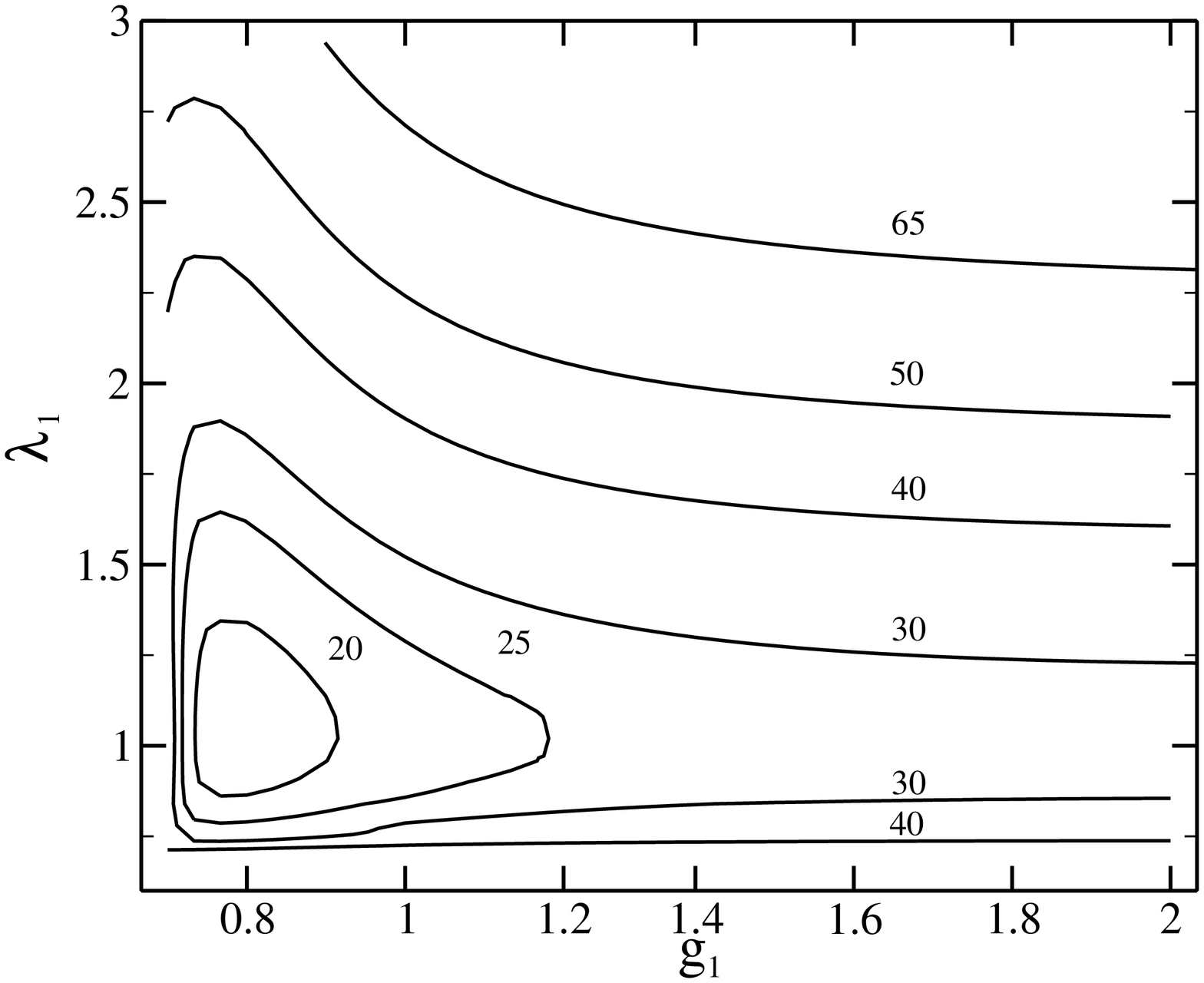,width=5cm,bbllx=4.cm,bblly=0.cm,bburx=20.cm,bbury=23.cm,angle=-90}} 
\caption{\footnotesize Preliminary fine-tuning contours for the Littlest 
Higgs model, case a) of eq.~(\ref{ab}), for two different values of the 
Higgs mass: $m_h=115$ GeV (left) and $m_h=250$ GeV (right).} 
\label{fig:littlest-a}} 

For case a), the value of $\Delta$ is shown by the contour plots of 
fig.~\ref{fig:littlest-a} which correspond to two different values of the 
Higgs mass. We present our results in the plane $\{g_1,\lambda_1\}$. In 
each point of this plane, $g_2$ and $\lambda_2$ are then fixed by 
eq.~(\ref{relations}); the values of $c$ and $c'$ are fixed by the 
minimization condition for electroweak breaking and the choice of Higgs 
mass. The value of $g_1'$ has been taken at $g_1'^2=g_2'^2 = g'^2/2$, 
which nearly minimizes the fine-tuning. (Note also that $g_1\geq g$ and 
thus smaller values of $\Delta$ cannot be reached by lowering $g_1$ in 
fig.~\ref{fig:littlest-a}.) The shaded areas correspond to regions that do 
not give a correct electroweak symmetry breaking (in these regions, 
$M_\phi\geq \Lambda$, which besides being beyond the range of validity of 
the effective theory, makes negative the triplet contribution to $m^2$).
 These plots illustrate the large size of $\Delta$, which is significantly 
larger than the previous rough estimate. This is not surprising since, as 
stated before, besides the heavy top contribution to $m^2$ (on which the 
estimate was based), there are other contributions that depend in various 
ways on the different input parameters. This gives additional 
contributions to the total fine-tuning, increasing its value. The plots 
also show how $\Delta$ decreases for increasing $m_h$. This is due to the 
fact that the larger $m_h$, and thus $\lambda$, the larger the required 
value of $m^2$ in (\ref{v2}), which reduces the level of cancellation 
needed between the various contributions to $m^2$ in (\ref{m2}) 
\cite{CEHI}. Although 
the fine-tuning is substantial, it could be considered as tolerable [i.e. 
${\cal O}(10)$], for some (small) regions of parameter space, at least for 
large $m_h$.  However, on closer examination the fine-tuning turns out to 
be larger than shown by fig.~\ref{fig:littlest-a}. From the condition a) 
in (\ref{ab}) \be \lambda\simeq c(g_1^2+g_1'^2)=\lambda_b\ll 
\lambda_a=c(g_2^2+g_2'^2)-c'\lambda_1^2\ , \ee it is clear that in this 
case $c'$ is large (and negative), while $c$ is small. But then, 
eq.~(\ref{ccp}) shows that there is an implicit tuning between $c_0$ and 
$c_1$ to get the small value of $c$.  In fact, it makes more sense to 
include $c_0$ and $c_0'$, rather than $c$ and $c'$, among the unknown input 
parameters 
appearing in (\ref{indepparLH}). Then, since $\Delta_{c_0} = |(c_0/ 
c)\Delta_{c}|$ (and similarly for $\Delta_{c_0'}$), the global fine-tuning 
becomes much larger. This is illustrated in fig.~\ref{fig:littlest-a2} 
(upper plots), where $\Delta$ is systematically above ${\cal O}(10)$, even 
for large $m_h$.

\FIGURE[t]{\mbox{ 
\epsfig{file=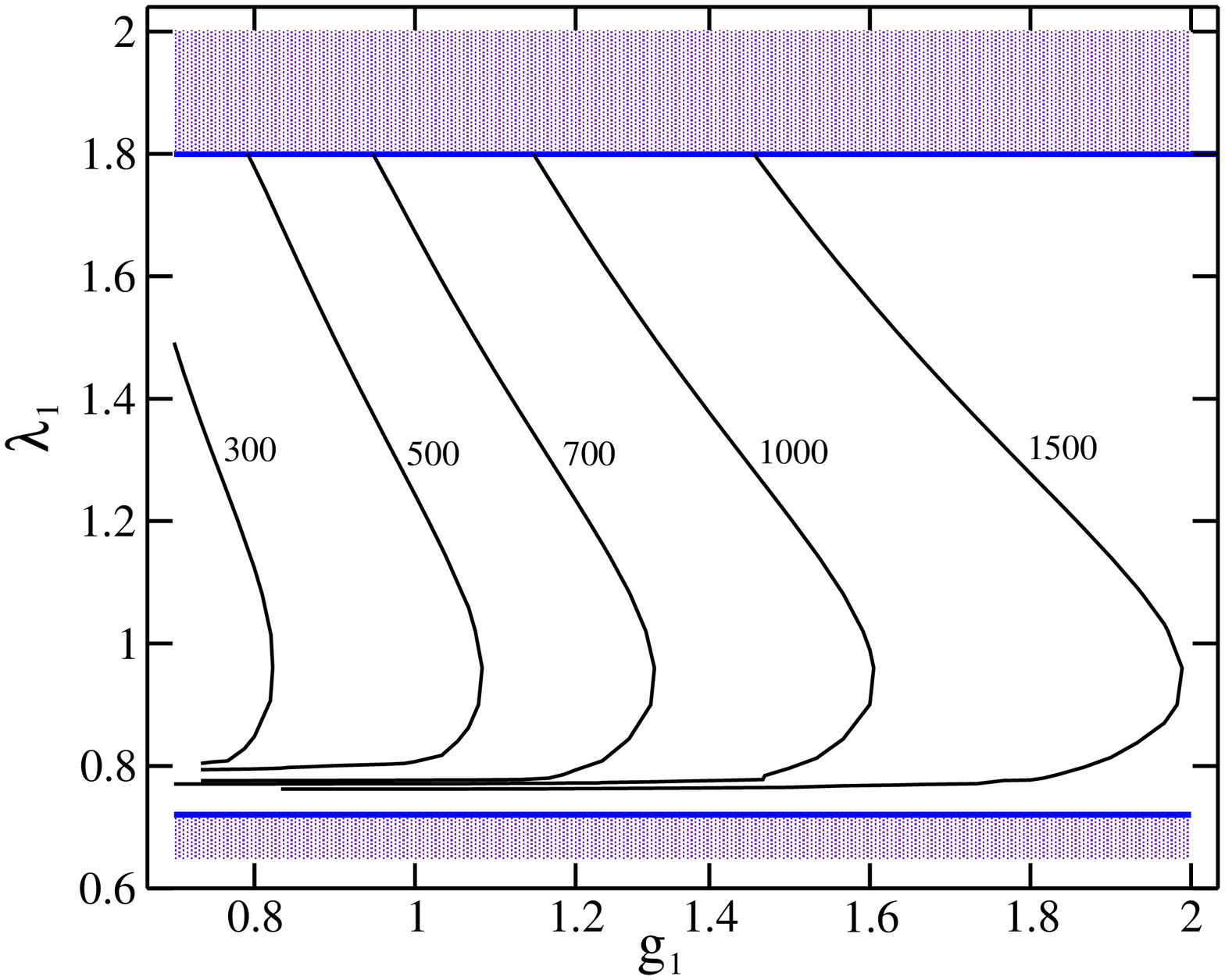,width=5cm,bbllx=3.cm,bblly=1.cm,bburx=19.cm,bbury=24.cm,angle=-90} 
\epsfig{file=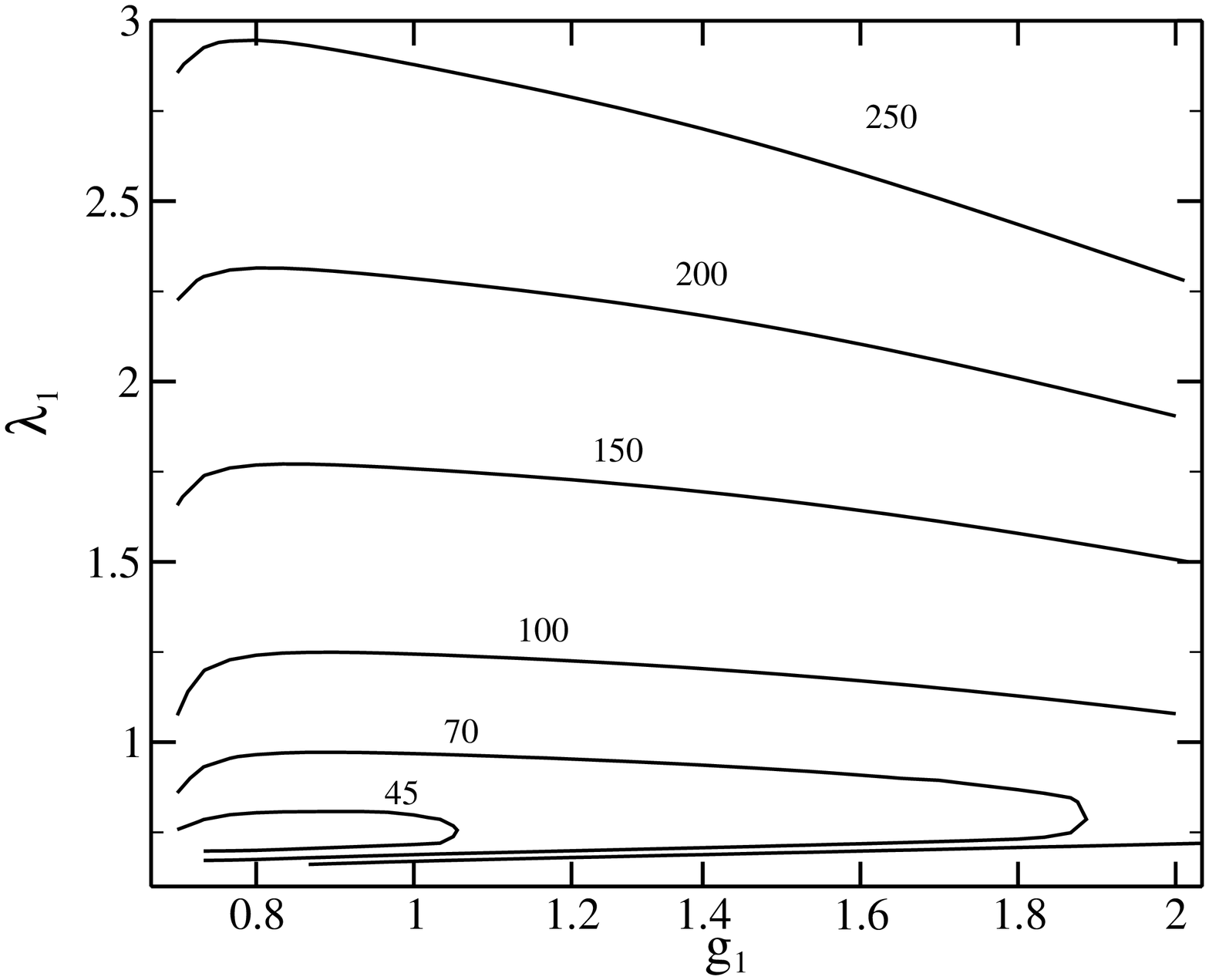,width=5cm,bbllx=3.cm,bblly=1.cm,bburx=19.cm,bbury=24.cm,angle=-90} 
} \vspace{2.cm} \mbox{ 
\epsfig{file=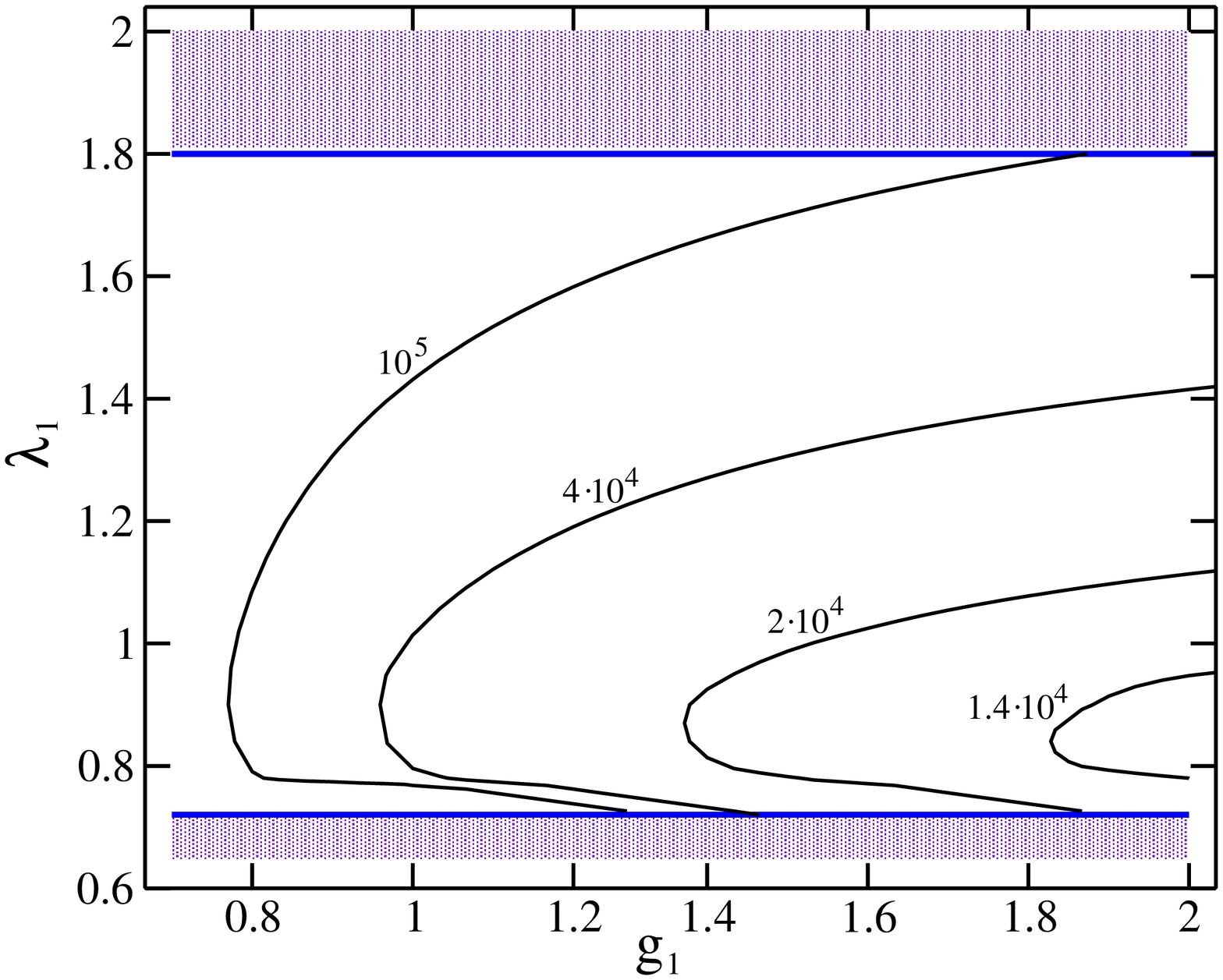,width=5cm,bbllx=-1.cm,bblly=1.cm,bburx=15.cm,bbury=24.cm,angle=-90} 
\epsfig{file=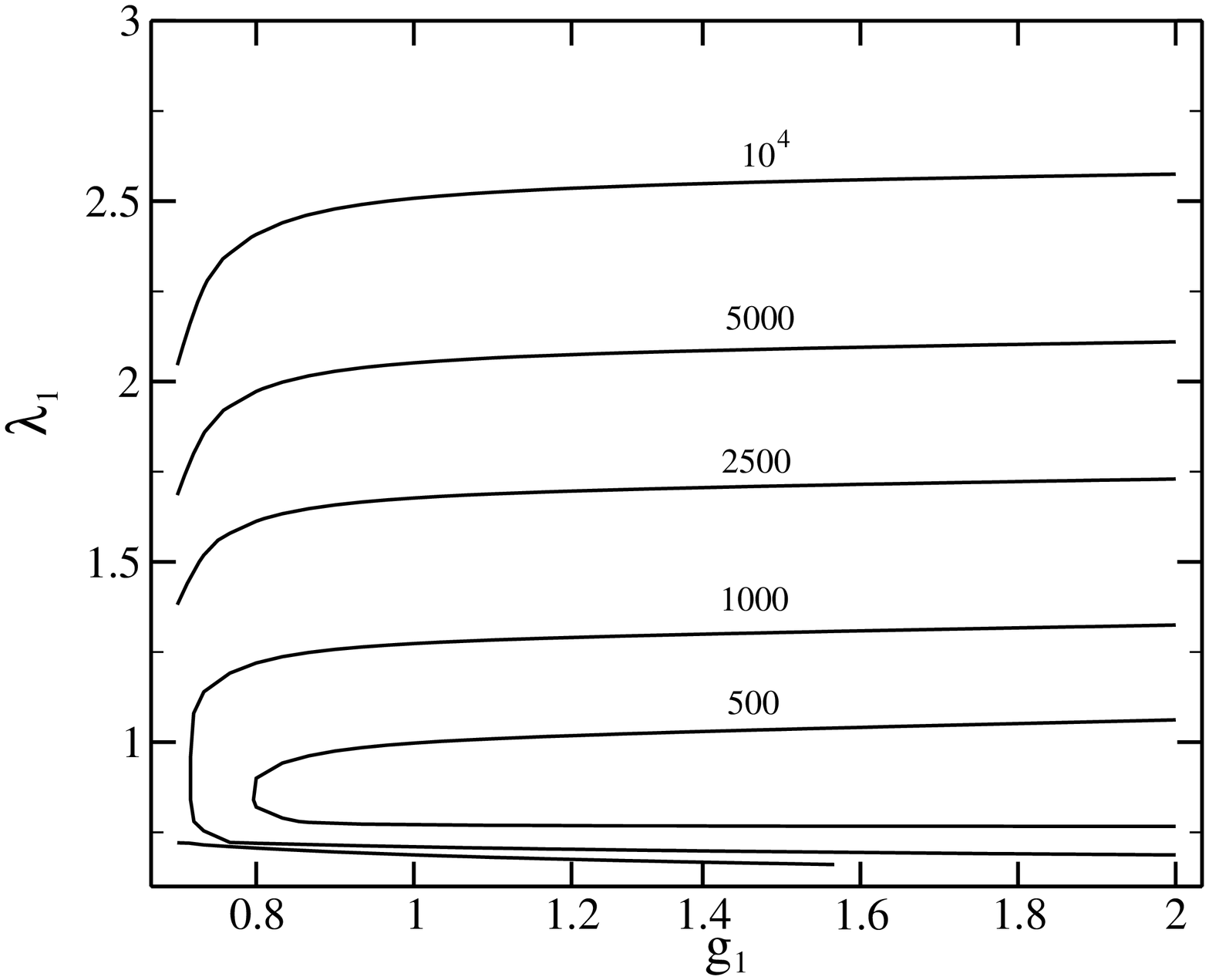,width=5cm,bbllx=-1.cm,bblly=1.cm,bburx=15.cm,bbury=24.cm,angle=-90} 
} \caption{\footnotesize Final fine-tuning contours for the Littlest Higgs 
model, using $c_0$ and $c'_0$ of eq.~(\ref{ccp}) as unknown parameters to 
improve the analysis.  TOP: Case a) of eq.~(\ref{ab}) for $m_h=115$ GeV 
(left) and $m_h=250$ GeV (right).  BOTTOM: The same but for case b) of 
eq.~(\ref{ab}).} \label{fig:littlest-a2} } 

There is a simple way of understanding the order of magnitude of $\Delta$.  
We can repeat the rough argument at the beginning of this subsection, but 
considering now the contribution of the triplet to the Higgs mass 
parameter in (\ref{m2}). More precisely, since 
$M_\phi^2=(\lambda_a+\lambda_b)f^2 
=[c(g_1^2+g_1'^2+g_2^2+g_2'^2)-c'\lambda_1^2]f^2$, we can focus on the 
contribution proportional to $c'$: 
\be \label{deltacpmh2} \delta_{c'} 
m^2 = -{3\lambda\over 8\pi^2}c'\lambda_1^2 \left[\log{\Lambda^2\over 
M_{\phi}^2}+ 1\right]\ .  \ee 
Now, $c'$ itself contains a radiative 
piece $c'_{1}= -24$ [see eq.~(\ref{ccp})], whose relative contribution to 
$m^2$ is then given by 
\be \label{deltacpmh22} \left|{\delta_{c'_{1}} 
m^2\over m^2} \right|\ge {9 \over 2\pi^2}\lambda_t^2 {f^2\over v^2} 
\left[\log{\Lambda^2\over M_{\phi}^2}+ 1\right]\ \simeq 45\ , \ee 
where 
we have first used $\lambda_1^2\ge \lambda_t^2/ 2$ and then $M_\phi\sim 
f$. Hence we easily expect ${\cal O}(100)$ contributions to $\Delta$, as 
reflected in fig.~\ref{fig:littlest-a2}.

It is interesting to note that this rough argument holds even if there are 
additional contributions to $m^2$, since it is based on the size of 
contributions that are present anyway. In particular, two-loop corrections 
or `tree-level' (i.e. threshold) corrections to $m^2$ are not likely to  
help in improving the fine-tuning.  Of course, it might happen that they 
have just 
the right size to cancel the known large contributions, such as those of 
eqs.~(\ref{deltatmh2}) and (\ref{deltacpmh22}). However, in the absence of 
a theoretical argument for that cancellation, this possibility can only be 
understood a priori as a fortunate accident. The chances for the latter 
are precisely what the fine-tuning analysis evaluates.

For case b) in eq.~(\ref{ab}) things are much worse, as illustrated in 
fig.~\ref{fig:littlest-a2} (lower plots), which shows huge values of 
$\Delta$. The reason is the following. In case ${\rm b)}$, both $c$ and 
$c'$ are sizeable, so there is no implicit tuning between $c_0$ ($c_0'$) 
and $c_1$ ($c_1'$), but this implies a cancellation to get 
$\lambda_a=c(g_2^2+g_2'^2)-c'\lambda_1^2\simeq 
\lambda$, which requires a delicate tuning. This ``hidden fine-tuning" is 
responsible for the unexpectedly large values of $\Delta$. In other words, 
small changes in the input parameters of the model produce large 
changes in the value of $\lambda$, and thus in the value of $v^2$.

\FIGURE[t]{\centerline{ 
\epsfig{file=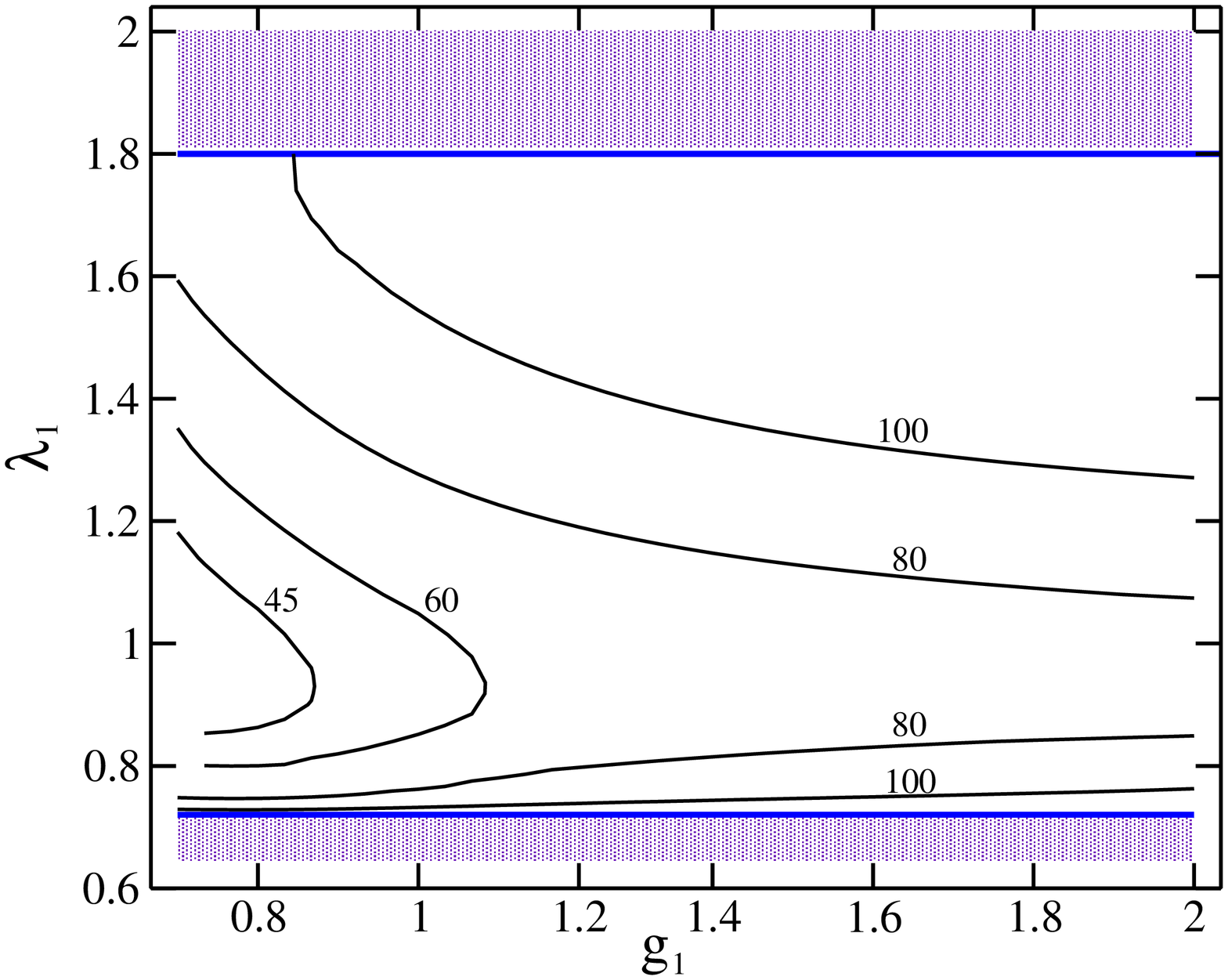,width=5cm,bbllx=4.cm,bblly=0.cm,bburx=20.cm,bbury=23.cm,angle=-90} 
\epsfig{file=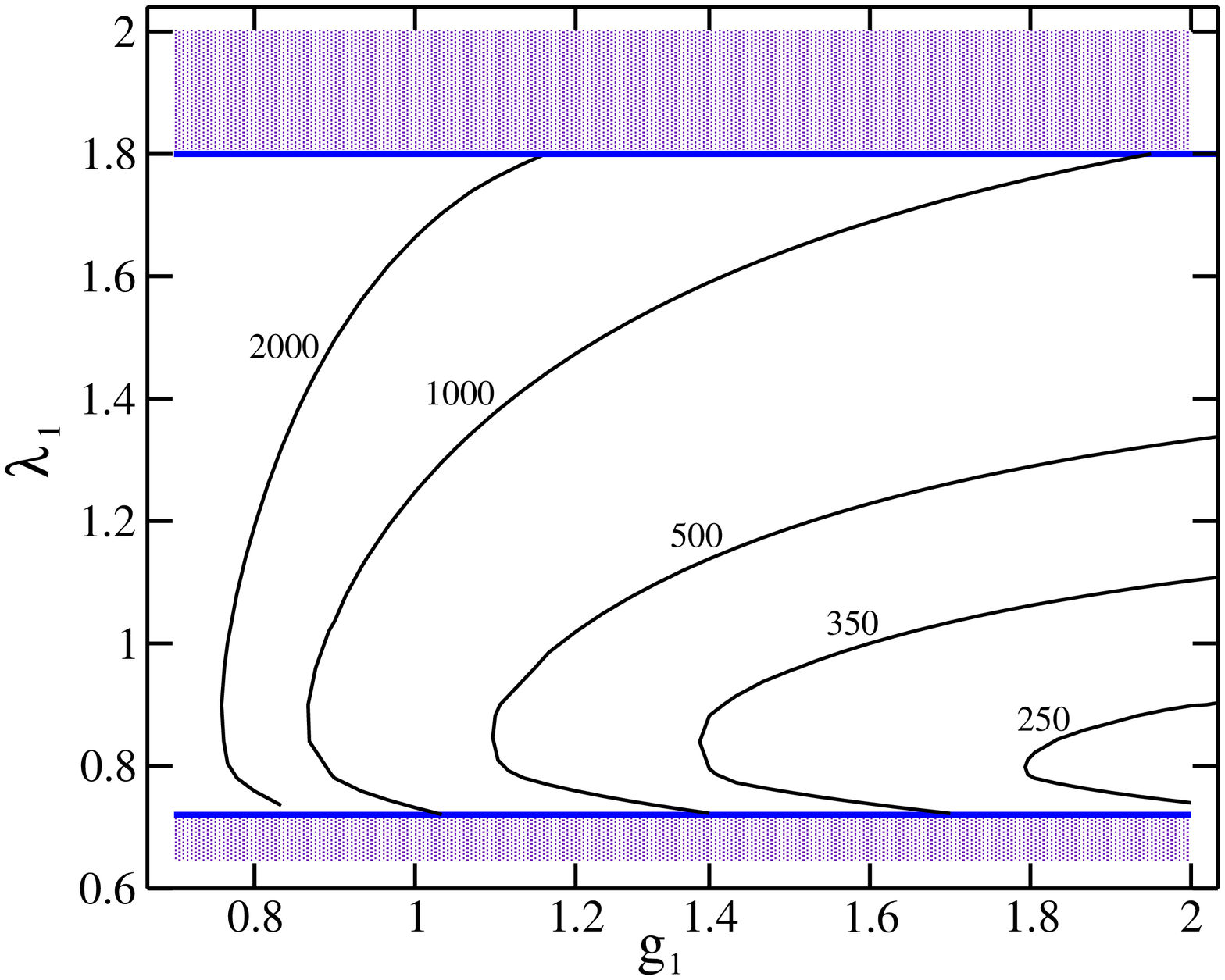,width=5cm,bbllx=4.cm,bblly=0.cm,bburx=20.cm,bbury=23.cm,angle=-90} 
} \caption{\footnotesize LEFT: The same as the bottom-left plot of 
fig.~\ref{fig:littlest-a2}, but keeping $\lambda$ fixed.  RIGHT: 
Fine-tuning associated to $\lambda$ itself in the same case.} 
\label{fig:littlest-b} } 

Now, imagine some future time after the Higgs mass has already been 
measured so that the parameter $\lambda$ takes a particular value and the 
other parameters of the model can only be varied in such a way that 
$\lambda$ 
remains constant. Then, according to the above discussion, the fine-tuning 
for case b) should be dramatically reduced and, apparently, this is 
exactly what happens. The condition of constant $\lambda$ can be 
incorporated in the computation of $\Delta$ using eq.~(\ref{constrDelta})  
with an additional constraint $G_4^{(0)}=\lambda$.\footnote{The constraint 
$G_4^{(0)}=\lambda$ is not independent of the others (for $g^2$, $g'^2$ 
and $\lambda_t$). A Gramm-Schmidt orthonormalization of the different 
constraints is enough to deal with this complication (see Appendix~A). } 
The new ``constrained" fine tuning in case b) (for $m_h=115$ GeV), is 
shown in the left plot of fig.~\ref{fig:littlest-b}, to be compared with 
the bottom-left plot of fig.~\ref{fig:littlest-a2}. Although still 
sizeable, the fine-tuning is now much smaller.

However, this behaviour does not alleviate the fine-tuning problems. If 
the Higgs mass is measured, one can also consider what is the fine-tuning 
between the input parameters of the model to produce such value of 
$m_h$, in the same way that one examines the fine-tuning to produce the 
measured value of $v^2$. Let us denote the fine-tuning in $m_h^2$ (or 
equivalently in $\lambda$) associated to a parameter $p_i$ by 
$\Delta_{p_i}^{(\lambda)}$. It is given by 
\be \label{BG2} 
{\delta \lambda\over \lambda} = 
\Delta_{p_i}^{(\lambda)}{\delta p_i\over p_i}\ .  \ee 
The right plot in 
fig.~\ref{fig:littlest-b} shows that the values of $\Delta^{(\lambda)}$ 
are quite large, as expected.
If $\Delta^{(\lambda)} > {\cal O}(1)$, this fine-tuning must be taken into 
account and, since $\Delta$ and $\Delta^{(\lambda)}$ represent 
independent inverse probabilities, they should be multiplied to estimate 
the total fine-tuning $\Delta \cdot \Delta^{(\lambda)}$ in the model.
This fine-tuning turns out to be very large, comparable to the values of 
$\Delta$ before the measurement of $m_h$.

The final conclusion is that the ``standard" Littlest Higgs model has 
built-in a significant fine-tuning problem, especially for $m_h<250$ GeV, 
even if other problems with electroweak observables are ignored. In this 
range the fine-tuning is typically $\Delta\simgt {\cal O}(100)$, i.e. 
essentially of 
the same order (or higher) than that of the Little Hierarchy problem of 
the SM [see fig.~\ref{fig:velt-a}] and  more severe than the 
MSSM one. For larger values of $m_h$, which is not so attractive 
from the point of view of fits to electroweak observables 
\cite{langacker}, the situation is better, although still $\Delta > 10$. 
The final results of this section are summarized by 
fig.~\ref{fig:littlest-a2}.

Let us finish this subsection with two additional comments.  First, notice 
that the plots presented correspond to $f=1$ TeV, which is a desirable and 
standard value in Little Higgs models. For other values of $f$, the 
parametric dependence of the fine-tuning is $\Delta \propto f^2$. In 
fact, precision electroweak observables in the Littlest Higgs model 
require larger values of the masses of the new 
particles and therefore of $f$ \cite{pew}, which makes the fine-tuning 
even more 
severe. The second comment concerns perturbativity. We have just seen that
 a large value of $c'$ [and also $c$ for region b) in eq.~(\ref{ab})] is 
generically required for a correct electroweak breaking. Actually, from 
eq.~(\ref{ccp}), it seems indeed natural to expect large values of $c'$, 
which might be a problem for perturbativity.  One way of obtaining a 
smaller value of $c'$ would be to lower $\Lambda$, making it smaller than 
$4\pi f$, which reduces the low-energy radiative contribution to $c'$. In 
fact it is well known \cite{SS} that chiral perturbation theory as a low 
energy description of technicolor theories with a large number of 
technifermions, $N$, breaks down at the scale $4\pi f/\sqrt{N}$. In the 
Littlest Higgs model we do have a large number of degrees of freedom ({\it 
e.g.} 12 only from $T$) so, the low-energy effective theory would not be 
reliable all the way up to $4\pi f$.  Conversely, if one insists in 
keeping $\Lambda\simeq 4\pi f/\sqrt{N}\simeq 10$ TeV to solve the Little 
Hierarchy problem, one would need $f$ larger than 1 TeV. This would help 
with the fits to precision electroweak measurements but would worsen 
significantly the fine-tuning.

\section{A Modified Version of the Littlest Higgs Model \cite{Peskin}}

This model \cite{Peskin} is very similar to the 
Littlest Higgs, except for the fact that the gauged subgroup of $SU(5)$ is 
$[SU(2)\times SU(2)\times U(1)_Y]$, rather than $[SU(2)\times U(1)]^2$.  
The absence of the heavy $B'$ gauge boson helps with precision electroweak 
fits \cite{Peskin}, which is the main motivation for this model. The price 
to pay for not doubling the gauged $U(1)$ is that the Higgs mass is not 
protected from quadratically divergent radiative corrections involving 
$U(1)_Y$ interactions even at one-loop level.  However, those corrections 
are not especially dangerous, due to the smallness of the $g'$ coupling. 
Otherwise, the structure of the model is very similar to the Littlest 
Higgs [in particular, the Lagrangian contains pieces similar to (\ref{L1}) 
and 
(\ref{DeltaL1}), see Appendix~B.2 for details].  The input parameters of 
the model 
are now 
\be 
\label{parpesk} 
p_i \ =\ \{g_1,\ g_2,\ \lambda_1,\ \lambda_2,\ c,\ c',\ f \}\ ,  
\ee 
to be compared with (\ref{indepparLH}) for the Littlest Higgs model. As 
in that model, $f$ can be ignored for the fine-tuning analysis.

For the fine-tuning analysis we need the $h$-dependent masses, which enter 
the one-loop effective potential. These are collected in Appendix~B.2. 
Besides the absence of $g_1'$ and $g_2'$, the main difference with the 
original Littlest Higgs model is that the Higgs mass parameter $m^2$ gets 
an additional positive 
contribution from the operator $c\ {\cal O}_V(\Sigma)$ (the form of this 
operator is dictated by the quadratically divergent contribution from 
gauge boson loops, see Appendix~B.2), 
\be \label{deltam2} \delta m^2= c 
g'^2{\Lambda^2\over 16\pi^2}=c g'^2f^2\ . \ee This contribution involves 
$g'$ as anticipated. Adding the one-loop logarithmic corrections we get 
\bea \label{m2p} m^2 &=& cg'^2f^2+{9g^2\over 
64\pi^2}M_{W'}^2\left[\log{\Lambda^2\over M_{W'}^2} + {1\over 3}\right] 
-{3\lambda_t^2\over 8\pi^2}M_T^2 \left[\log{\Lambda^2\over M_{T}^2}+ 
1\right] \\ &+&{3\over 8\pi^2}\left\{\left(\lambda+{17\over 
12}cg'^2\right) M_\phi^2 \left[\log{\Lambda^2\over M_\phi^2}+ 1\right] - 
\left(\lambda+{1\over 12}cg'^2\right) M_s^2 \left[\log{\Lambda^2\over 
M_s^2}+ 1\right] \right\} \nonumber \ , \eea 
where the Higgs quartic 
coupling is now \be \lambda={1\over 4}\left[\lambda'_a+\lambda'_b-{4\over 
3}c g'^2- {(\lambda'_a-\lambda'_b)^2\over(\lambda'_a+\lambda'_b+4cg'^2)} 
\right]\ , \label{lambdap} \ee with $\lambda_a'\equiv c 
g_2^2-c'\lambda_1^2$ and $\lambda_b'\equiv c g_1^2$.  The expression for 
$M_T$ is as for the Littlest Higgs, the triplet mass is $M_\phi^2=( 
\lambda'_a+ \lambda'_b+4cg'^2)f^2$ and $M_s^2=c g'^2f^2$ is the squared 
mass associated to the light Higgses (see Appendix B.2). 
Eqs.~(\ref{m2p}) and (\ref{lambdap}) have to be compared with (\ref{m2}) 
and (\ref{lambda}) for the Littlest Higgs.

\FIGURE[t]{\centerline{ 
\epsfig{file=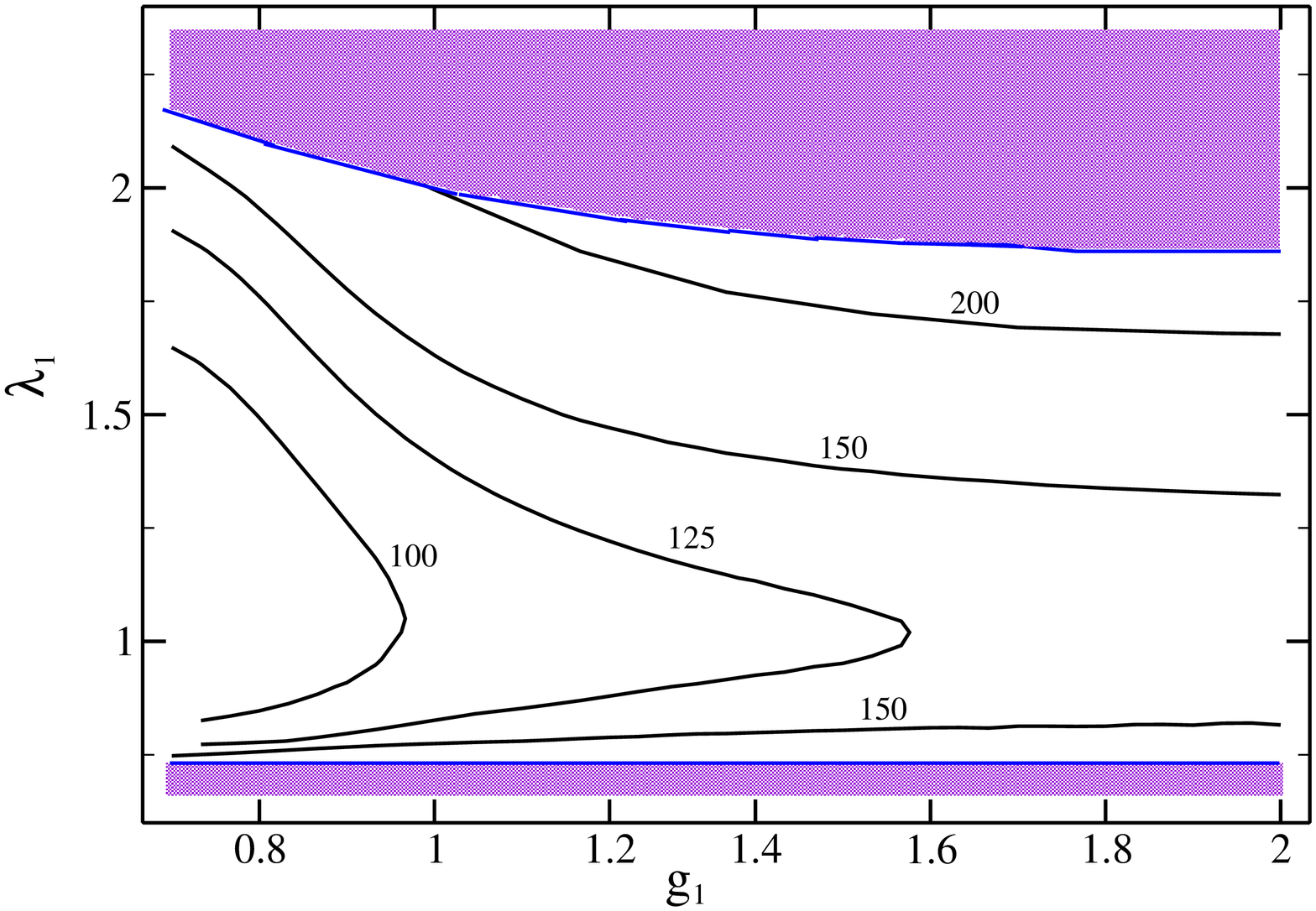,width=5cm,bbllx=4.cm,bblly=2.cm,bburx=20.cm,bbury=27.cm,angle=-90} 
\epsfig{file=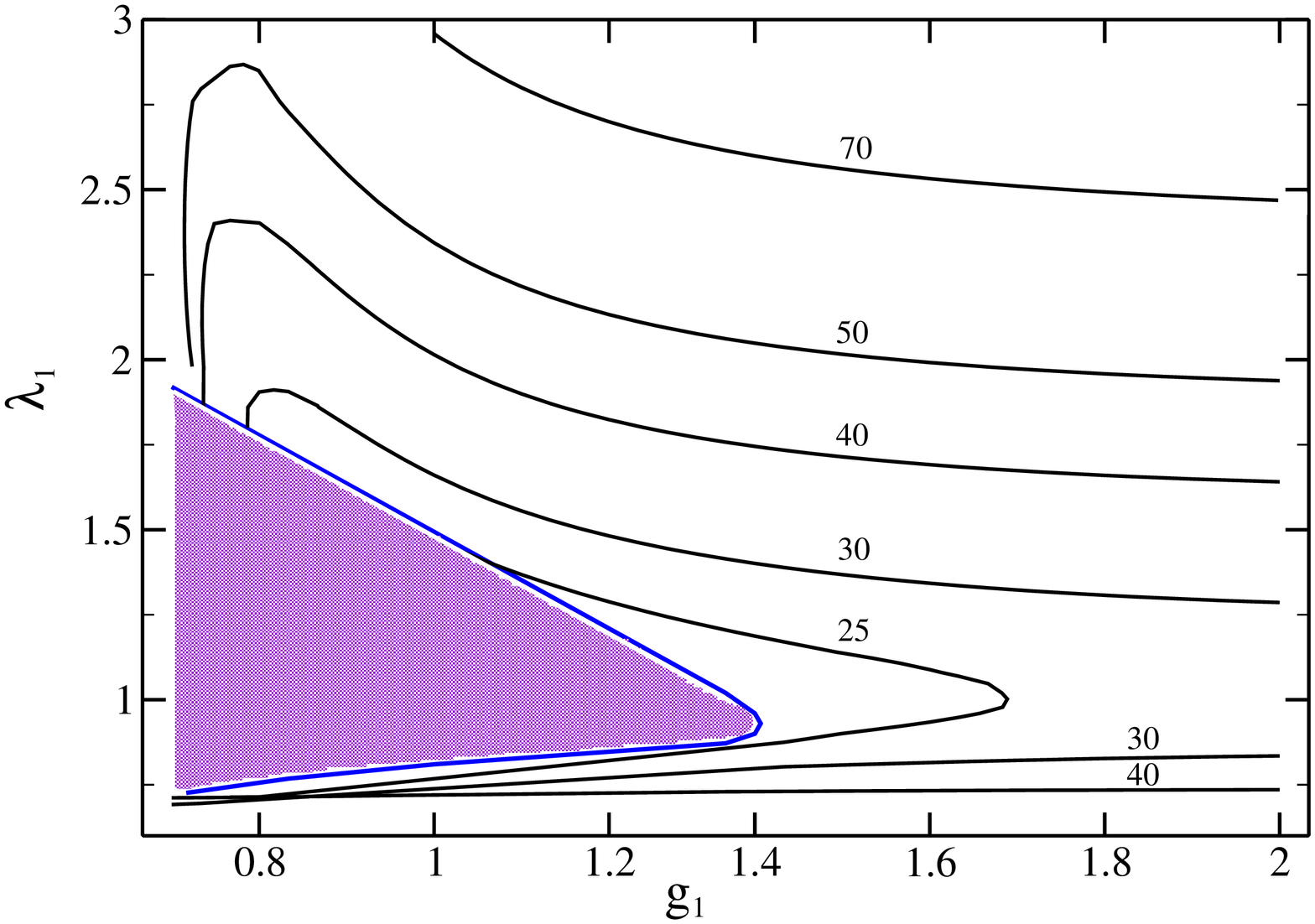,width=5cm,bbllx=4.cm,bblly=2.cm,bburx=20.cm,bbury=27.cm,angle=-90} 
} \caption{\footnotesize Preliminary fine-tuning contours, using $c$ and 
$c'$ as unknown parameters, for the Little Higgs model of \cite{Peskin}, 
case a), with $m_h=115$ GeV (left plot) and $m_h=250$ GeV (right plot).} 
\label{fig:peskin} } 

\FIGURE[t]{\mbox{ 
\epsfig{file=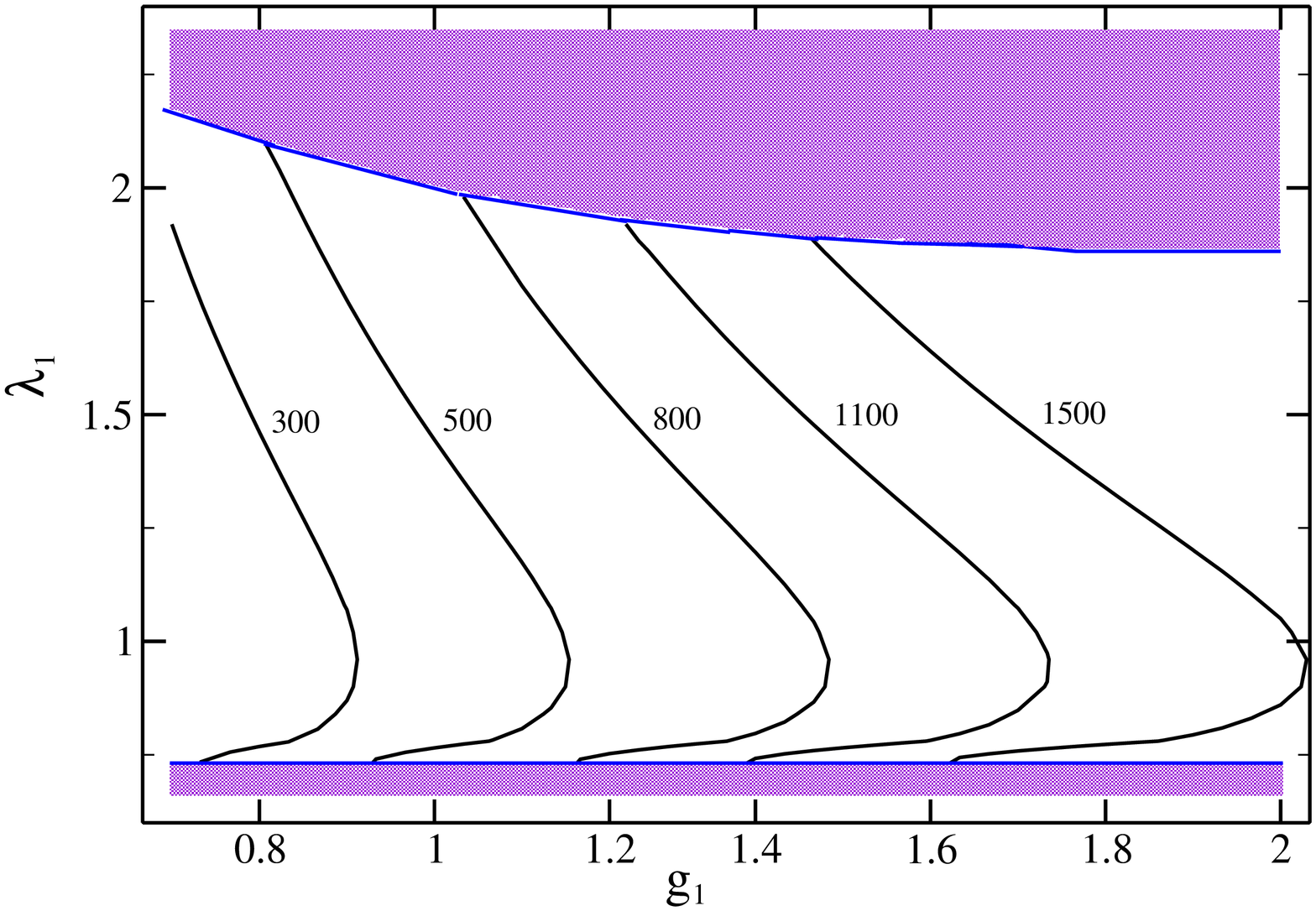,width=5cm,bbllx=3.cm,bblly=2.cm,bburx=19.cm,bbury=26.cm,angle=-90} 
\epsfig{file=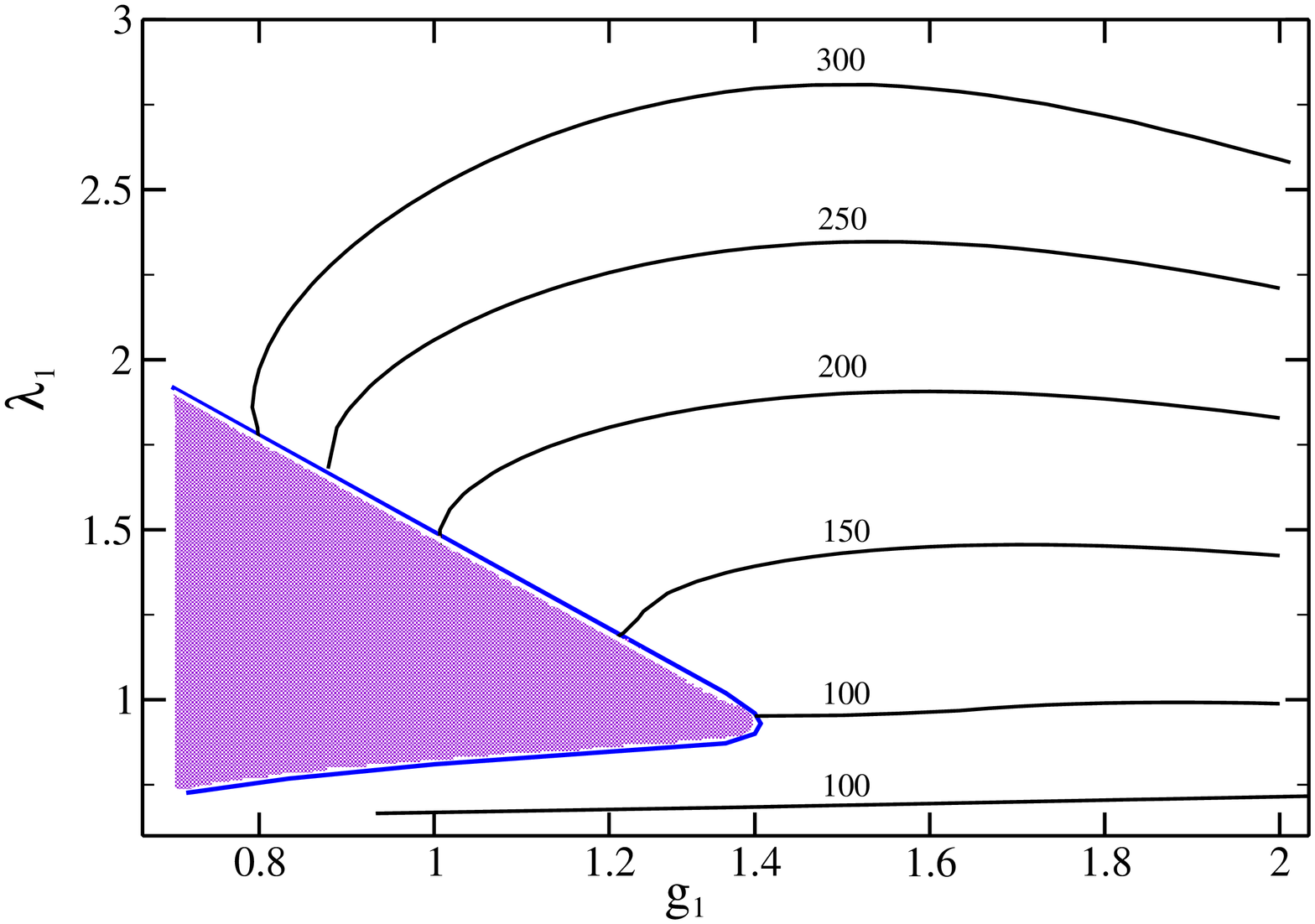,width=5cm,bbllx=3.cm,bblly=1.cm,bburx=19.cm,bbury=24.cm,angle=-90} 
} \vspace{2.cm} \mbox{ 
\epsfig{file=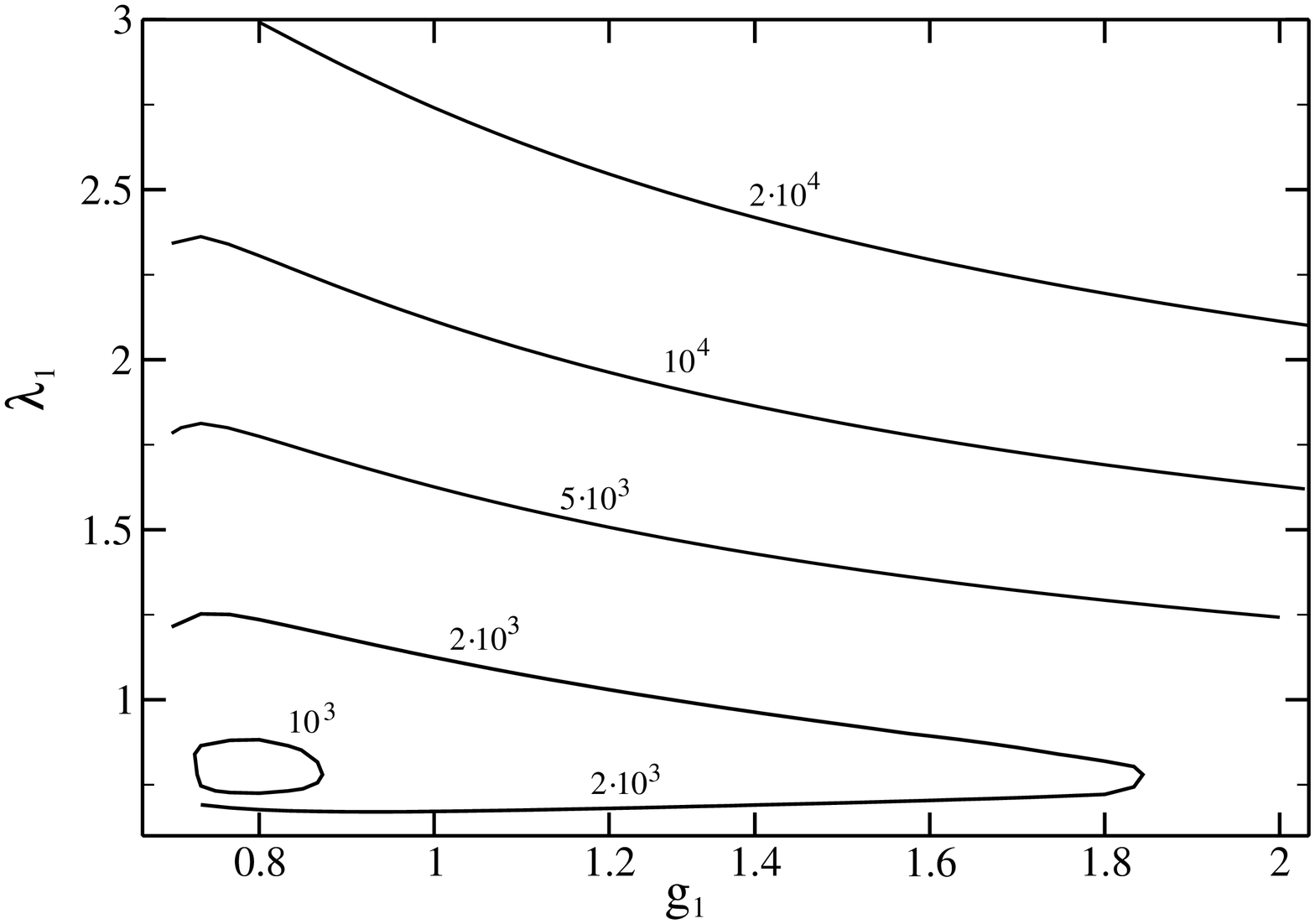,width=5cm,bbllx=-1.cm,bblly=2.cm,bburx=15.cm,bbury=26.cm,angle=-90} 
\epsfig{file=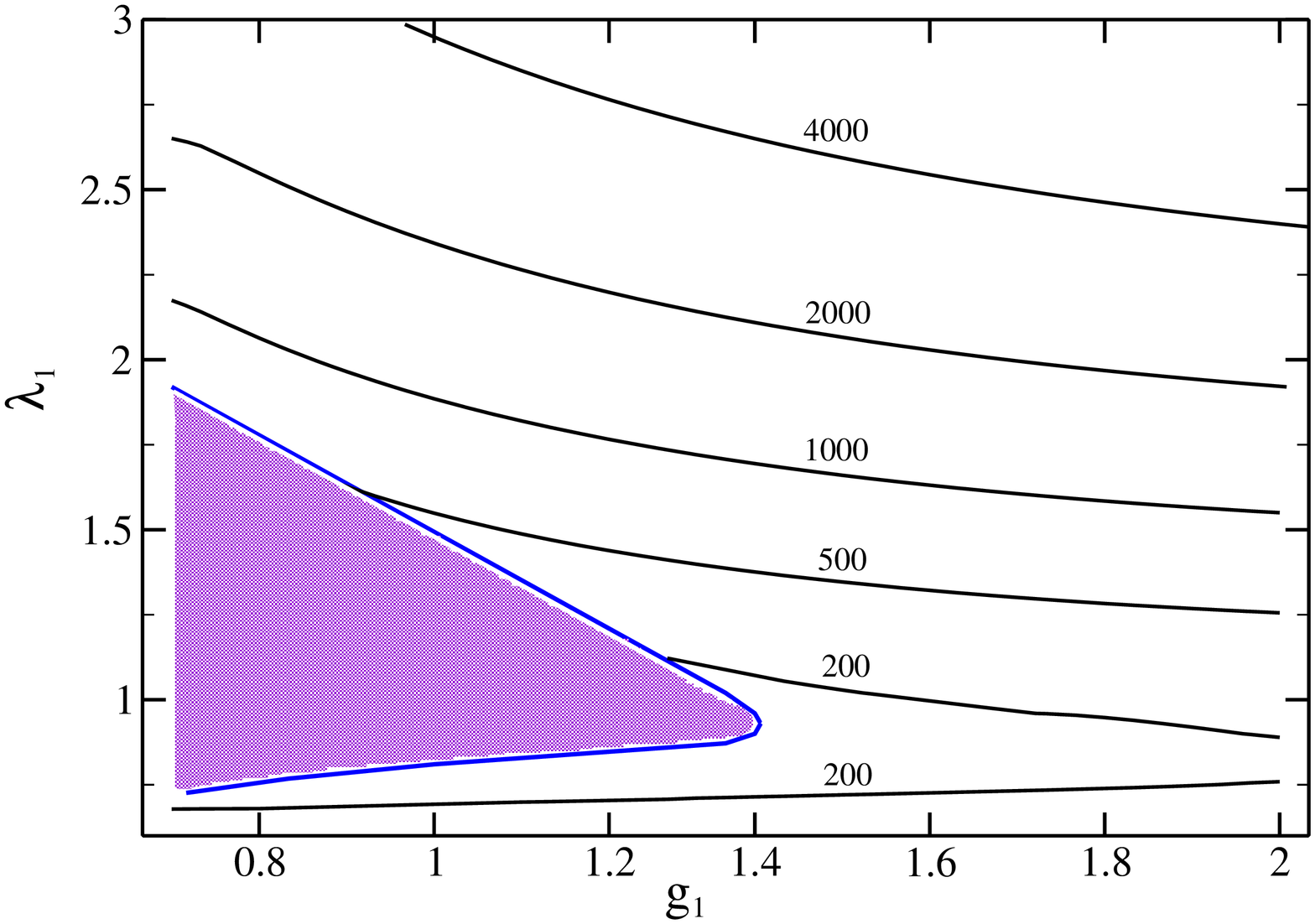,width=5cm,bbllx=-1.cm,bblly=1.cm,bburx=15.cm,bbury=24.cm,angle=-90} 
} \caption{\footnotesize Final fine-tuning contours for the Little Higgs 
model of \cite{Peskin}, using $c_0$ and $c_0'$ as unknown parameters, for 
the two regions of solutions: a) (top) and b) (bottom), and two different 
values of the Higgs mass: $m_h=115$ GeV (left) and $m_h=250$ GeV (right).} 
\label{fig:peskin2} } 

The presence of the $g'$ terms in $m^2$ complicates the parameter 
dependence of the minimization condition for electroweak breaking: $c$ and 
$c'$ do no longer enter 
in $m^2$ just through $\lambda'_a$ and $\lambda'_b$. Nevertheless, there 
are 
still two separate regions of solutions, which are the respective heirs of 
the two regions named a) and b) for the Littlest Higgs model 
[eq.~(\ref{ab})]\footnote{Again, the existence of these two regions can be 
understood here using the approximation explained in footnote 5.}; thus we 
keep the same notation.

The fine-tuning $\Delta$ for the region a), using $c$ and $c'$ as input
 parameters, is shown in fig.~\ref{fig:peskin}. The 
magnitude of $\Delta$ is similar to that in the Littlest Higgs 
model, fig.~\ref{fig:littlest-a}.  In the present case the tree-level 
contribution $cg'^2f^2$ in (\ref{m2p}), which is positive\footnote{For 
$c<0$ one breaks the electroweak symmetry at tree-level. However, this 
possibility leads to a large VEV for the triplet and therefore we focus on 
$c>0$.}, helps in compensating the negative correction from the heavy Top, 
so that the contribution from the triplet, and thus the triplet mass 
$M_\phi^2$, is not required to be as large as before. Consequently, the 
values of $c$ and $c'$ will be smaller, as happened (for $c$) in the 
region a) of the Littlest Higgs model. However, as discussed in the 
previous section, small $c$ and $c'$ cause additional 
fine-tuning\footnote{Note that eq.~(\ref{ccp}) holds also in this model.}, 
which 
can be taken into account by using $c_0$ and  $c_0'$, rather than $c$ 
and $c'$ as the input parameters appearing in 
(\ref{parpesk}). 
This enhancement of the fine-tuning can be 
appreciated in the corresponding plots [both for a) and b) regions]
in fig.~\ref{fig:peskin2}.

Fig.~\ref{fig:peskin2} represents our final results for the model analyzed 
in this section. The fine-tuning is quite similar to that for the Littlest 
Higgs model, as summarized in fig.~\ref{fig:littlest-a2}. Therefore, the 
same 
comments apply here: the fine-tuning is always substantial ($\Delta > 10$) 
and for $m_h < 250$ GeV is essentially of the same order as (or higher 
than) that of the 
Little Hierarchy problem [$\Delta \simgt {\cal O}(100)$] and worse than 
in the MSSM. As in the Littlest Higgs, two-loop or `tree-level' 
contributions to $m^2$ are not likely to improve the situation [note in 
particular that eqs.~(\ref{deltatmh2}) and (\ref{deltacpmh22}) remain the 
same in this scenario].

\section{A Little Higgs model with $T$-parity \cite{ChengLow}}

This model \cite{ChengLow} is still based on the same $SU(5)/SO(5)$ 
structure of the Littlest Higgs model (with a gauged $[SU(2)\times 
U(1)]^2$ subgroup) and the gauge and scalar field content is the same, as 
described in Appendix~B.1 (although extended versions are possible 
\cite{ChengLow}).  However, the Lagrangian is different: a $T$-parity is 
imposed such that the triplet and the heavy gauge bosons are $T$-odd while 
the Higgs doublet is $T$-even. This $T$-parity plays a role similar to 
$R$-parity in SUSY: it has the welcome effect of forbidding a number of 
dangerous couplings (like the $h^2\phi$ one responsible for the triplet 
VEV, as discussed in previous sections; or direct couplings of the SM 
fields to the new gauge bosons) improving dramatically the fit to 
electroweak data.

The gauge kinetic part of the Lagrangian is as in eq.~(\ref{Lkin}) but 
$T$-parity imposes the equalities \be g_1=g_2=\sqrt{2}g\ ,\;\;\;\;\;\; 
g'_1=g'_2=\sqrt{2}g'\ , \label{gT} \ee where $g$ and $g'$ are the gauge 
coupling constants of the SM. Imposing $T$-invariance on the fermionic 
sector requires the introduction of several new degrees of freedom, and 
the scalar operators of (\ref{potential}) are replaced by a $T$-symmetric 
expression given by (\ref{potentialT}).

The squared masses to ${\cal O}(h^2)$ in this model are similar to those 
in the Littlest Higgs model. In the gauge boson sector they are exactly 
the same as in (\ref{masas1}), with gauge couplings related by 
eq.~(\ref{gT}). In the fermion sector, despite the inclusion of extra 
degrees of freedom, the only mass relevant for our purposes is that of the 
heavy Top which, to order $h^2$, remains the same as in the Littlest 
Higgs model [see eq.~(\ref{masas1})]. The squared masses of the other 
fermions do not have an $h$-dependence [they can be relatively heavy (in 
the multi-TeV range) and are irrelevant for low-energy phenomenology].

In the scalar sector, an important difference with respect to the Littlest 
Higgs model is that now there is no $\phi h^2$-coupling. As a result, the 
Higgs quartic coupling does not get modified after decoupling the triplet 
field and is simply given by:  \be \lambda={1\over 
4}(\lambda_a+\lambda_b)\ , \label{lambdaT} \ee [now 
$\lambda_a=2c(g^2+g'^2)-c'\lambda_1^2$ and $\lambda_b=2c(g^2+g'^2)$] to be 
compared with eq.~(\ref{lambda}) for the Littlest Higgs. Another direct 
consequence of not having a $\phi h^2$-coupling is the absence of the 
off-diagonal entries in the scalar mass matrices in the $CP$-even, 
$CP$-odd and charged sectors (see Appendix~B.3 for details).

The one-loop-generated Higgs mass parameter, $m^2$, is given by the same 
expression as that of the Littlest Higgs model [eq.~(\ref{m2})] but, as we 
have seen, $T$-parity imposes strong relations between the parameters of 
the model. In particular, we have now \be M_{W'}^2=g^2f^2\ , \;\;\; 
M_{B'}^2={1\over 5}g'^2f^2\ , \;\;\; M_{\phi}^2=4 \lambda f^2\ .  \ee The 
model is therefore much more constrained than the Littlest Higgs.

\FIGURE[t]{\centerline{ 
\epsfig{file=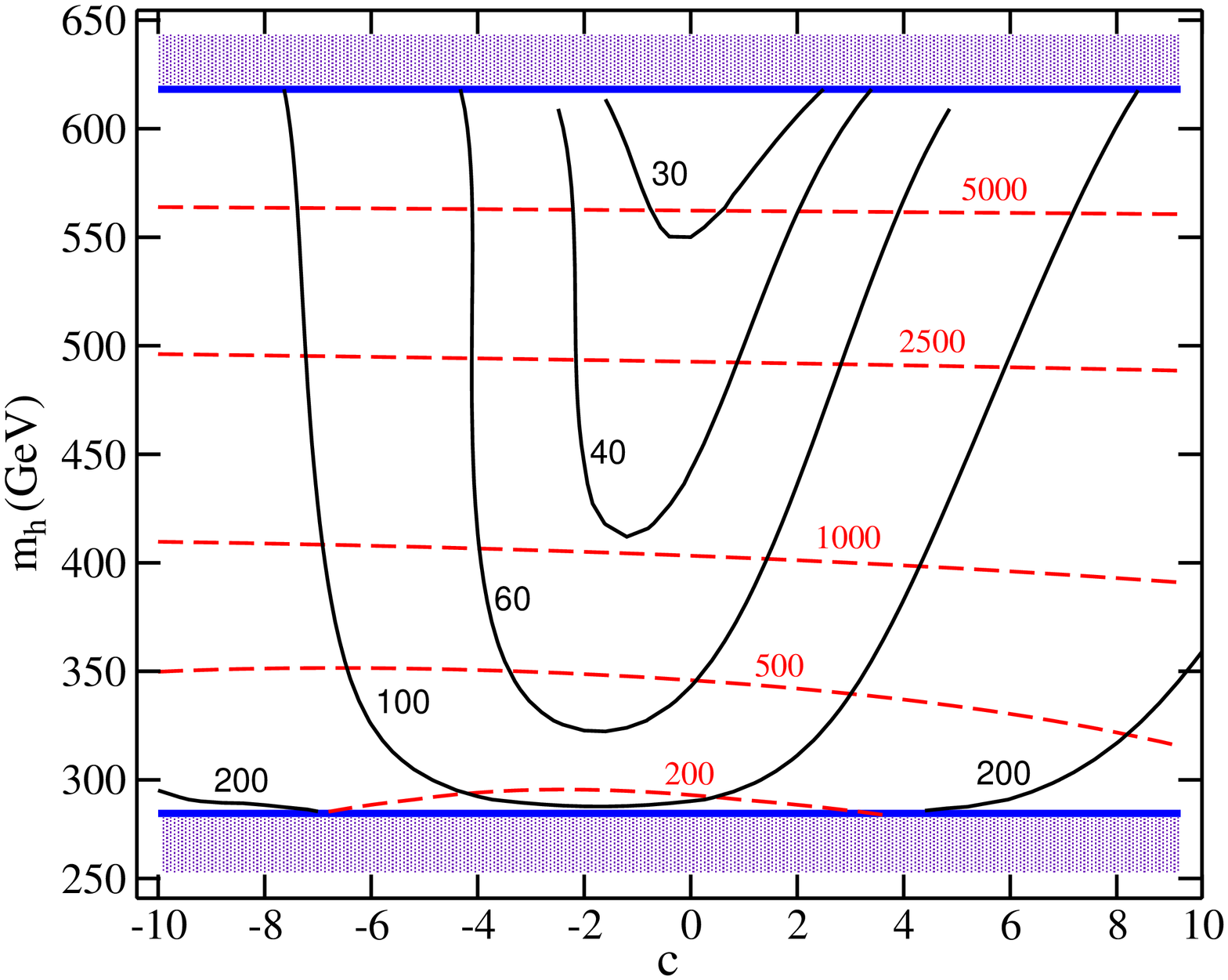,width=5cm,bbllx=3.cm,bblly=0.cm,bburx=19.cm,bbury=23.cm,angle=-90} 
} \caption{\footnotesize Fine-tuning contours in a Little Higgs model with 
$T$-parity, using $c_0$ and $c'_0$ of eq.~(\ref{ccp}) as unknown 
parameters. Solid (dashed) lines correspond to case 1 (2) of 
eq.~(\ref{abT}).} \label{fig:littlesTa} } 

For the fine-tuning analysis, we start by identifying the input 
parameters, which are now 
\be 
\label{indepparLHT} 
p_i \ =\ \{\lambda_1,\ \lambda_2,\ c,\ c',\ f \}\ ,  
\ee 
to be compared with (\ref{indepparLH}) and (\ref{parpesk}). Again, we can 
leave $f$ aside as explained after 
(\ref{indepparLH}). The couplings $\lambda_{1,2}$ are related by the usual 
top-Yukawa constraint in eq.~(\ref{relations}) while $c$ and $c'$ are 
related to $\lambda$ through eq.~(\ref{lambdaT}). For a given value of the 
Higgs mass (and therefore of the coupling $\lambda$) the minimization 
condition for electroweak breaking can be solved for $M_T^2$, which fixes 
$\lambda_1^2+\lambda_2^2$, but not $\lambda_1$ or $\lambda_2$ separately.  
From this continuum of solutions, the top mass constraint 
[eq.~(\ref{relations})] leaves only two of them, simply related by 
$\lambda_1\leftrightarrow\lambda_2$. We will refer to these two solutions  
as 
\be 
\label{abT} 
{\rm 1)}\; \lambda_1\leq \lambda_2 \ , 
\;\; \;\; \;\;  
{\rm 2)}\; \lambda_2\leq \lambda_1 \ .  
\ee 
If $\lambda$ is small, $M_\phi$ is not large enough to compensate the 
negative heavy Top contribution 
to the one-loop Higgs mass and the minimization condition is not 
satisfied. If, on the other hand, $\lambda$ is too large then the Top 
contribution, which cannot be arbitrarily large (it grows with $M_T$, 
but 
only up to $M_T=\Lambda$), is also unable to satisfy the minimization 
condition. Thus, we obtain a limited range for $m_h$: 280~GeV $\simlt m_h 
\simlt$ 625~GeV, for $f=1$ TeV. This result has interest of itself for the 
phenomenology of the Littlest Higgs model with $T$-parity, with the caveat 
that possible two-loop (or `tree-level') contributions to the Higgs mass 
parameter can change the limits of that interval for $m_h$, as we discuss 
in more detail below.

The resulting constrained fine-tuning [using $c_0$ and $c'_0$ of 
eq.~(\ref{ccp}) as unknown parameters] is shown in 
figure~\ref{fig:littlesTa}. As $g_1$ is not a free-parameter anymore, we 
present our results in the plane $\{c, m_h\}$.  The black solid lines 
correspond to case 1) and the red dashed ones to case 2). At the lower 
bound for $m_h$, which is determined by the minimal possible value of 
$M_T^2=(\lambda_1^2+\lambda^2_2)f^2$, one has 
$\lambda_1=\lambda_2=\lambda_t$ and therefore cases 1) and 2) give the 
same results for the fine-tuning, as can be seen in the figure. At the 
upper bound on $m_h$ one has $M_T^2=\Lambda^2$, which implies 
$\lambda_i\simeq 4\pi$ for $i=1$ or 2, at the limit of perturbativity. We 
see that the fine-tuning is sizeable throughout all parameter space in 
spite of the large values of the Higgs mass. It is always larger for case 
2)  because a larger value of $\lambda_1$ affects directly the parameter 
$\lambda_a$ and therefore the value of $\lambda$. In fact, as will be 
clearer shortly, the largest contribution to the fine-tuning comes, in 
most cases, through the dependence of $\lambda$ on $c,c'$ and $\lambda_1$.

\FIGURE[t]{\centerline{ 
\epsfig{file=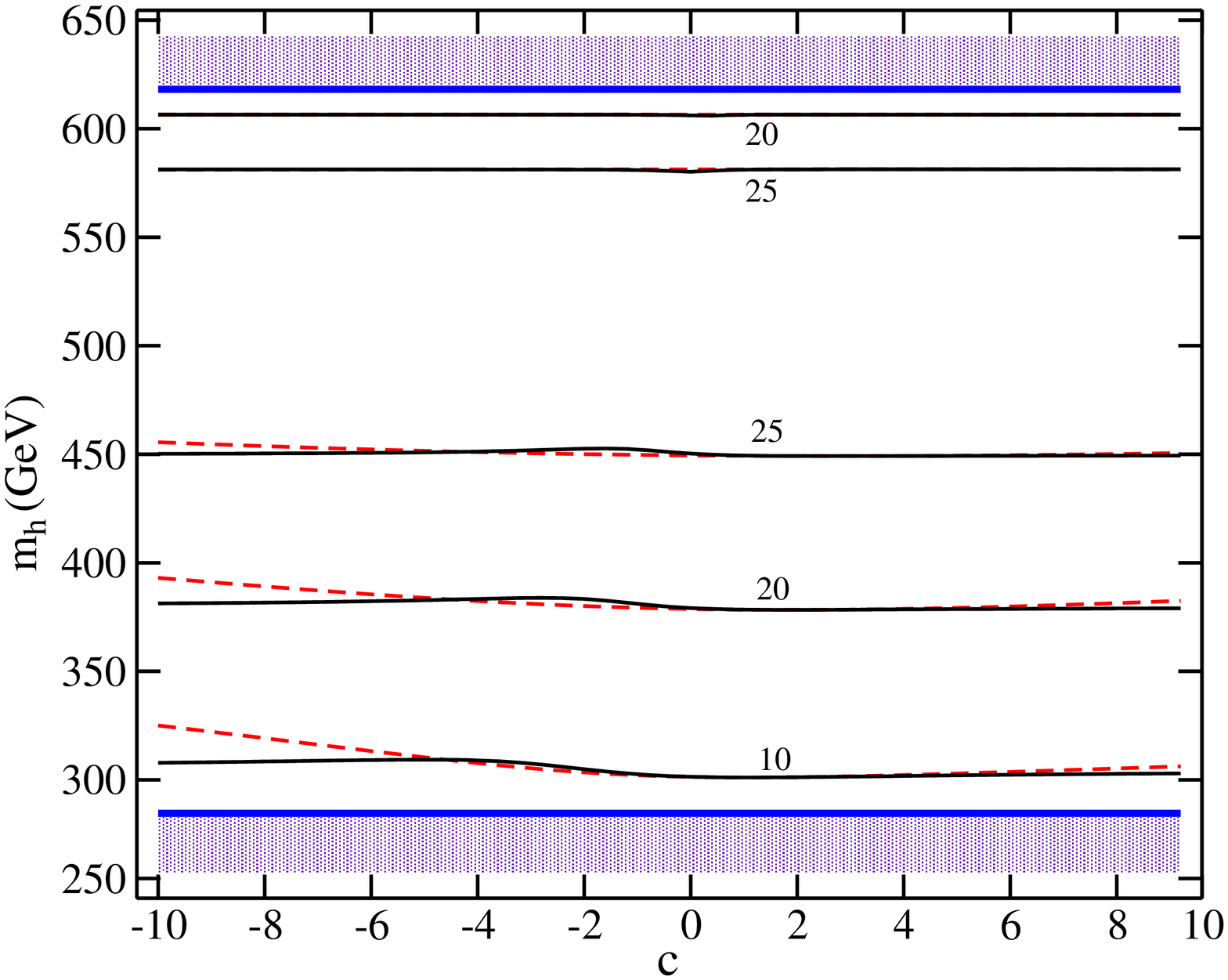,width=5cm,bbllx=3.cm,bblly=0.cm,bburx=19.cm,bbury=24.cm,angle=-90} 
\epsfig{file=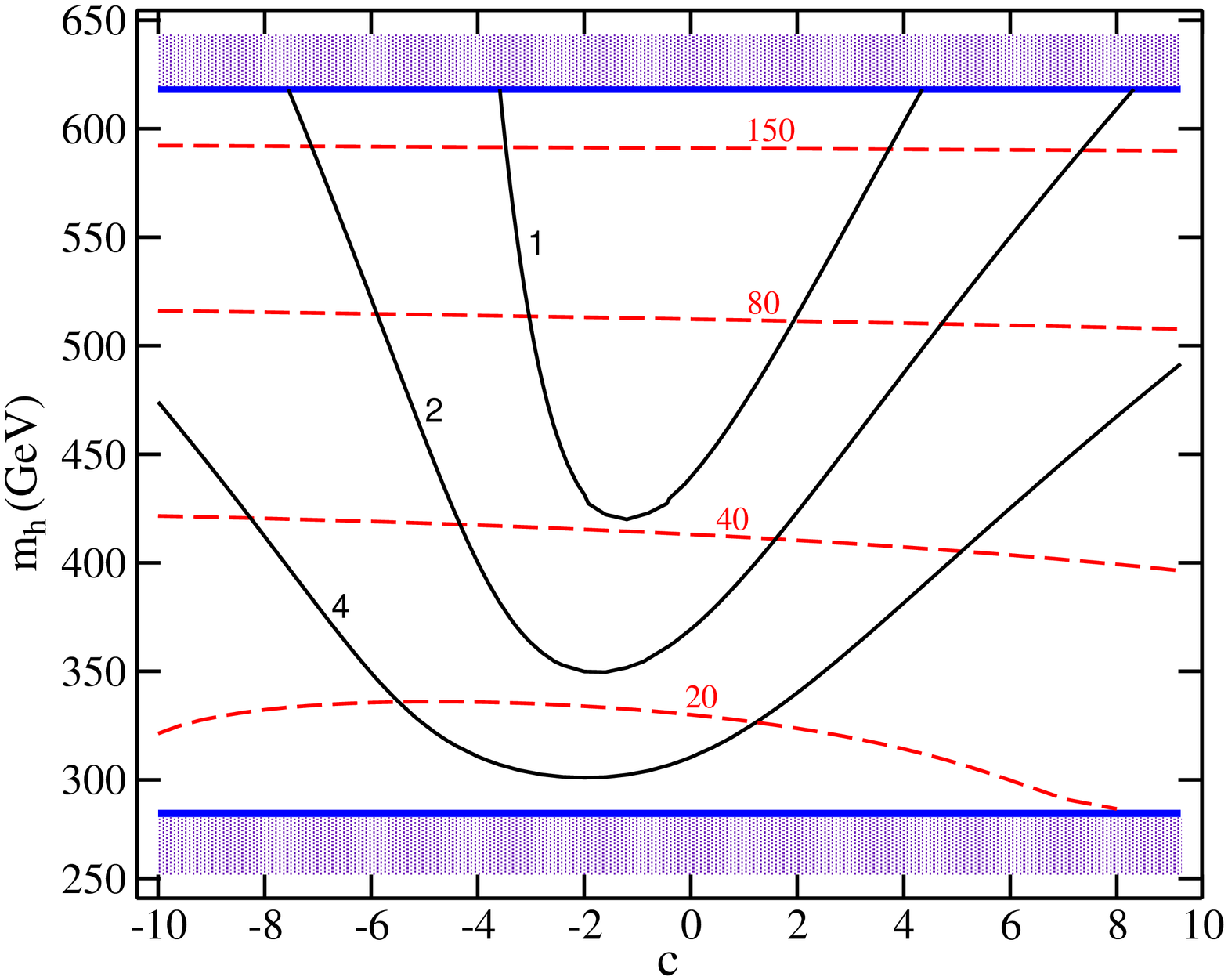,width=5cm,bbllx=3.cm,bblly=0.cm,bburx=19.cm,bbury=24.cm,angle=-90} 
} \caption{\footnotesize Left: Same as in fig.~\ref{fig:littlesTa} but 
keeping fixed $\lambda$.  Right: Fine-tuning associated to $\lambda$ 
itself.  [Solid (dashed) lines correspond to case 1 (2) of 
eq.~(\ref{abT})].  } \label{fig:littlesTb} } 

From the previous discussion, it follows that at some future time, after 
the Higgs mass has already been measured (and thus $\lambda$ gets fixed), 
the fine-tuning would get dramatically reduced, especially in case 2). 
This is shown by fig.~\ref{fig:littlesTb}, left plot, which presents the 
fine-tuning when the constraint of fixed $\lambda$ is enforced. The 
fine-tuning is nearly independent of $c$, and varies only through the 
values of $\lambda_{1,2}$, getting the smallest values at the boundaries 
of parameter space. This can be understood from the simple analytical 
approximation 
\be 
\Delta \simeq {M_T^2\over 2\lambda v^2} 
{|\lambda_1^2-\lambda_2^2|\over \sqrt{\lambda_1^4+ 
\lambda_2^4}}{3\lambda_t^2\over 2\pi^2} \log{\Lambda^2\over M_T^2} \ , 
\label{ap} 
\ee 
which is easy to derive and explains why cases 1) and 2) 
give very similar values for the fine-tuning\footnote{The small 
sensitivity to $c$ and the small difference between scenarios 1) and 2) 
which can be appreciated in fig.~\ref{fig:littlesTb} is a subtle effect 
[not captured by the approximation (\ref{ap})] due to the dependence of 
$\lambda$ on $c, c'$ and $\lambda_1$ (even though we are fixing 
$\lambda$). Such effects are discussed in Appendix A.}. Although the 
fine-tuning is moderate, we still have to worry 
about the tuning in $\lambda$ itself, as we did in section~3 for the model 
of ref.~\cite{Peskin}. We show that tuning in the right plot of 
fig.~\ref{fig:littlesTb}.  Analytically we find 
\be 
\Delta^{(\lambda)}\simeq {\lambda_1^2\over 4\lambda} 
\left[{4 c'^2\lambda_1^4\over \lambda_1^4+ \lambda_2^4} + (c'-c'_0)^2+ 
16(c-c_0)^2{(g^2+g'^2)^2\over \lambda_1^4} \right]^{1/2}\ .  
\ee 
We see that there is a big difference between cases 1) and 2). In case 1), the 
coupling $\lambda_1$ varies between $\lambda_t$ at the lower limit of 
$m_h$ and $\lambda_t/\sqrt{2}$ at the upper limit, and it does not cost 
much to get $\lambda$ right. Therefore the associated tuning is always 
small.  In case 2), $\lambda_1$ is of moderate size ($\sim\lambda_t$) near 
the lower limit on $m_h$ but grows significantly when $m_h$ increases 
(reaching $\lambda_1\sim 4\pi$ near the upper limit). Then, getting 
$\lambda$ right requires small values of $c'$ and, being unnatural, this 
causes a sizeable tuning. Coming back to fig.~\ref{fig:littlesTa}, one can 
easily check that the dependence of the fine-tuning in that plot on $c$ 
and $m_h$ can be understood as a particular combination of the two effects 
shown in fig.~\ref{fig:littlesTb}.

\FIGURE[t]{\centerline{ 
\epsfig{file=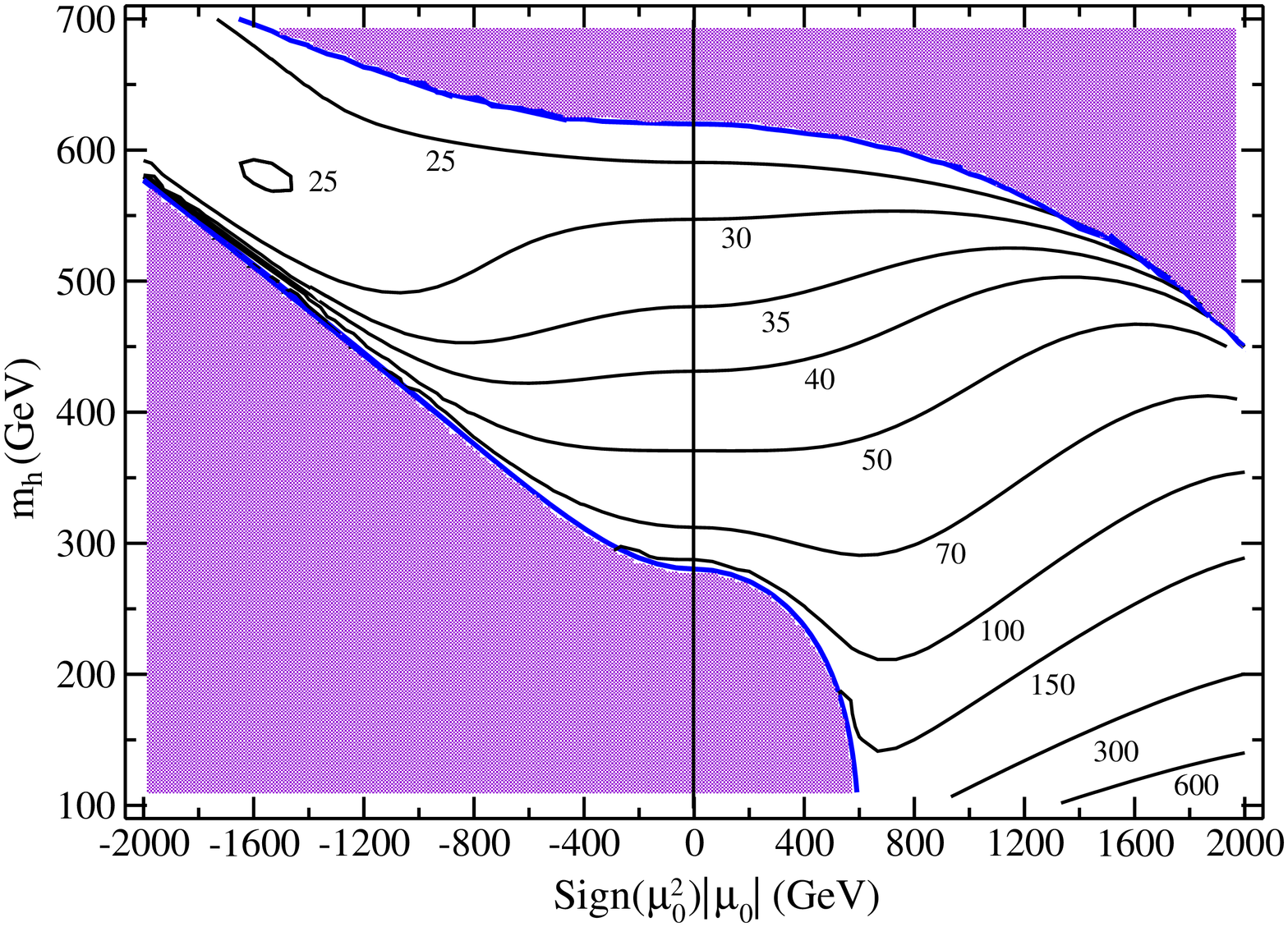,width=4.5cm,bbllx=3cm,bblly=0.cm,bburx=19.cm,bbury=28.cm,angle=-90} 
\epsfig{file=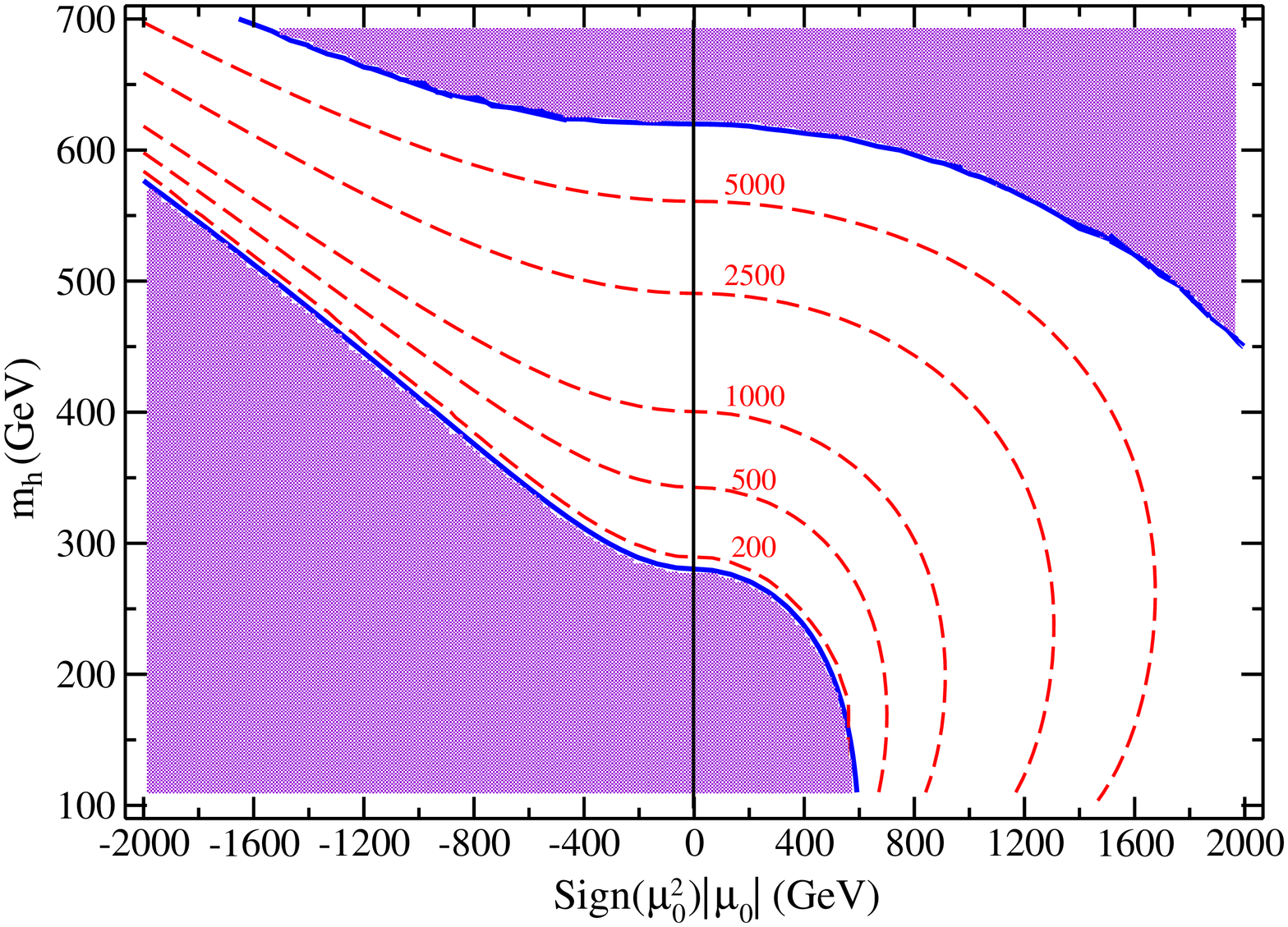,width=4.5cm,bbllx=3cm,bblly=0.cm,bburx=19.cm,bbury=28.cm,angle=-90} 
} \caption{\footnotesize Fine-tuning contours in a Little Higgs model with 
$T$-parity, with a `tree-level' $\mu_0^2$ mass parameter, using $c_0$ and 
$c'_0$ of eq.~(\ref{ccp}) as unknown parameters and setting $c=0$.  The 
left (right) plot corresponds to case 1 (2) of eq.~(\ref{abT}).  } 
\label{fig:littlesTmu} } 

Finally, let us consider the effect of two-loop (or `tree-level') 
contributions to the Higgs mass parameter which, as mentioned, can allow 
Higgs masses below the (quite high) lower limit $m_h\geq 280$ GeV of 
fig.~\ref{fig:littlesTa}. We mimic this effect by adding a constant mass 
term $1/2 \mu_0^2 h^2$ to the Higgs potential (allowing both signs of 
$\mu_0^2$).  From the arguments given in previous sections, we do not 
expect big changes in the fine-tuning but it is interesting to consider 
this possibility as a way of accessing regions of lower Higgs mass, which 
are more attractive phenomenologically.  Notice that 
eq.~(\ref{indepparLHT}) is now enlarged by one more parameter, namely 
$\mu_0^2$.  The resulting fine-tuning for cases 1) and 2) of 
eq.~(\ref{abT}) is shown in fig.~\ref{fig:littlesTmu}, (left and right 
plots, respectively), setting $c=0$ (which nearly minimizes the 
fine-tuning).  For Higgs masses accessible already with $\mu_0=0$, the 
fine-tuning does not change much, as expected, while for lower Higgs 
masses the fine-tuning increases [case 1)] or remains large [case 2)]. We 
see that case 1) continues to be the best option.

Figs.~\ref{fig:littlesTa} and \ref{fig:littlesTmu} summarize our results 
for the model analyzed in this section. As for the models of sections 2 
and 3, the fine-tuning is always substantial ($\Delta > 10$) and usually 
comparable to (or higher than) that of the Little Hierarchy problem 
[$\Delta \simgt {\cal O}(100)$] and worse than in the MSSM. Notice also 
that 
the lowest fine-tuning, $\Delta \sim 25$, is obtained for large values of 
the Higgs mass, $m_h\simgt 500$ GeV, which is generically disfavoured from 
fits to precision electroweak observables \cite{langacker}. In addition, 
such large values of $m_h$ are less satisfactory from the point of view of 
the Little Higgs philosophy: the Little Higgs mechanism is interesting 
because it might explain the lightness of the Higgs compared to the TeV 
scale.

\section{The Simplest Little Higgs Model \cite{Sch}}

We now depart from the group structure of the Littlest Higgs and 
consider a model, proposed in \cite{Sch}, that is based on a global 
$[SU(3)\times U(1)]^2/[SU(2)\times U(1)]^2$. The initial gauged subgroup 
is $[SU(3)\times U(1)_X]$ which gets broken to the electroweak subgroup, 
with 
\be {1\over g'^2}={1\over 3g^2}+{1\over g_x^2}\ . \label{gpgx} 
\ee 
This symmetry breaking is triggered by the VEVs $f_1$ and $f_2$ of 
two $SU(3)$ triplets, $\Phi_1$ and $\Phi_2$. For later use we define 
\be 
f^2\equiv f_1^2+f_2^2\ , 
\ee 
which measures the total amount of breaking. 
This spontaneous breaking produces 10 Goldstone bosons, 5 of which are 
eaten by the Higgs mechanism to make massive a complex $SU(2)$ doublet of 
extra $W'$s, $(W'^{\pm}, W'^0)$, and an extra $Z'$. The remaining 5 
degrees of freedom are: $H$ [an $SU(2)$ doublet to be identified with the 
SM Higgs] and $\eta$ (a singlet). Details about this breaking are left for 
Appendix~B.4. The initial tree-level Lagrangian has a structure similar to 
eq.~(\ref{L1}). In particular, $m^2$ and $\lambda$ are zero at this level.

As in previous models, in order to study the electroweak breaking, we need 
to consider the one-loop Higgs potential, for which we have to
to compute the $h$-dependent masses of the model. We collect here these 
masses leaving again details for Appendix~B.4. In the gauge sector, 
besides the massless photon, the rest 
of gauge bosons have the following masses. For the charged 
$(W^\pm,W'^\pm)$ pair, one has, expanding in powers of $h$, 
\bea 
m_{W'^\pm}^2(h)&=&M_{W'}^2-{1\over 4} g^2 h^2 + {\cal O}(h^4/f^2)\ , 
\nonumber\\ m_{W^\pm}^2(h)&=&{1\over 4} g^2 h^2 + {\cal O}(h^4/f^2)\ ,
\label{mWWp} 
\eea 
with 
\be M_{W'}^2\equiv {1\over 2} g^2 f^2\ . 
\ee 
For 
the $(Z'^0,Z^0)$ pair, 
\bea m_{Z'^0}^2(h)&=&M_{Z'}^2-{1\over 4} 
(g^2+g'^2)h^2 + {\cal O}(h^4/f^2)\ , \nonumber\\ m_{Z^0}^2(h)&=&{1\over 4} 
(g^2+g'^2) h^2 + {\cal O}(h^4/f^2)\ , \label{mZZp} 
\eea 
with 
\be 
M_{Z'}^2\equiv {2 g^2\over 3-t_w^2} f^2\ , 
\ee 
where $t_w\equiv g'/g$. Finally, 
the complex $W'^0$ has mass 
\be m_{W'^0}^2(h)=M_{W'}^2\ . 
\ee

The fermion sector is enlarged as usual. The states relevant for 
electroweak breaking are the SM top quark and a heavy Top, with 
masses squared 
\bea 
m_{T}^2(h)&=&M_T^2-{1\over 2} \lambda_t^2 h^2 + {\cal O}(h^4/f^2)\ , 
\nonumber\\ 
m_{t}^2(h)&=&{1\over 2} \lambda_t^2 h^2 + {\cal O}(h^4/f^2)\ ,
\label{mtT} 
\eea 
where 
\be 
M_T^2\equiv \lambda_1^2f_1^2+\lambda_2^2f_2^2\ ,
\ee 
where $f_{1,2}$ are the triplet VEVs. Here $\lambda_{1,2}$ are new 
Yukawa couplings of the Little Higgs model, and $\lambda_t$ is the SM top 
Yukawa coupling, given by the relation 
\be 
{f^2\over \lambda_t^2} = {f_1^2\over \lambda_2^2}+ {f_2^2\over 
\lambda_1^2}=f_1^2f_2^2 \left({1\over \lambda_1^2f_1^2}+ 
{1\over \lambda_2^2f_2^2} \right)\ .
\ee 

One can trivially check the cancellation of $h^2$ terms in ${\rm Str} M^2$ 
from the explicit expressions of the masses given above. In fact, the 
cancellation holds to all orders in $h$ (and $\eta$), as is clear from the 
more general formula for the masses presented in Appendix~B.4 [see 
eq.~(\ref{genmass})]. Therefore, and in contrast with previous models, 
one-loop quadratically divergent corrections from gauge or fermion loops 
do not induce scalar operators to be added to the Lagrangian. Then, no 
Higgs quartic coupling is present at this level.

Less divergent one-loop corrections do induce both a mass term and a 
quartic coupling for the Higgs. Using again the $\overline{\rm MS}$ 
scheme in Landau gauge\footnote{Our scheme differs from that used in 
\cite{Sch}, but the 
difference is numerically small.} and setting the renormalization 
scale $Q=\Lambda$, it is straightforward to compute the one-loop potential 
including fermion and gauge boson loops once the masses are known as a 
function of $h$. Performing an expansion of this potential in powers of 
$h$, one gets \cite{Sch} 
\be 
\label{Vrad} 
V(h)={1\over 2} \delta m^2 h^2 +{1\over 
4}\left[\delta_1 \lambda(h)-{\delta m^2\over 3} {f^2\over f_1^2 
f_2^2}\right]h^4 +... 
\ee 
with 
\bea 
\delta m^2 & = & {3\over 32 \pi^2} 
\left[g^2 M_{W'}^2\left(\log{\Lambda^2\over M_{W'}^2}+{1\over 3}\right) 
+{1\over 2}(g^2+g'^2)M_{Z'}^2\left(\log{\Lambda^2\over M_{Z'}^2}+{1\over 
3}\right)\right] \nonumber\\ &-&{3\over 8 \pi^2}\lambda_t^2 
M_{T}^2\left(\log{\Lambda^2\over M_{T}^2}+1\right) +... \ , \label{dm2} 
\eea and \bea \delta_1\lambda(h)&=& - {3\over 128 \pi^2} \left[g^4 
\left(\log{M_{W'}^2\over m_{W}^2(h)}-{1\over 2}\right) +{1\over 
2}(g^2+g'^2)^2\left(\log{M_{Z'}^2\over m_{Z}^2(h)}-{1\over 
2}\right)\right] \nonumber\\ &+&{3\over 16 \pi^2}\lambda_t^4 
\left(\log{M_{T}^2\over m_{t}^2(h)}-{1\over 2}\right)+... \ , 
\label{deltal} 
\eea 
where the dots in (\ref{dm2}) and (\ref{deltal}) stand 
for subdominant contributions (in particular those from the $\eta$ and the 
Higgs field itself, which was also subdominant in previous models).

The radiatively induced Higgs mass, $\delta m^2$, is dominated as usual by 
the negative heavy Top contribution, which is again too large (being 
$M_T^2\geq 4\lambda_t^2 f_1^2 f_2^2/f^2$) and now there is no bosonic 
contribution that can be used to compensate it. This problem is solved 
\cite{Sch} by adding to the tree-level potential a mass $\mu^2$ for the 
triplets $\Phi_{1,2}$ (see Appendix~B.4). Such operator contributes to the 
Higgs potential the piece \be \delta_0 V= {1\over 2}\mu_0^2 h^2-{1\over 
48} {\mu_0^2f^2\over f_1^2f_2^2}h^4+... \label{delta0V} \ee where 
$\mu_0^2$ is given in terms of the fundamental mass parameter $\mu^2$ by 
\be \mu_0^2=\mu^2{f^2\over f_1f_2}\ . \ee By choosing $\mu^2_0>0$ we get a 
positive contribution to the Higgs mass parameter that can compensate the 
heavy Top contribution in $\delta m^2$. The tree-level value of the Higgs 
quartic coupling from (\ref{delta0V}) is then negative but the large (and 
positive) radiative corrections in (\ref{deltal}) can easily overcome that 
effect.

\FIGURE[t]{\centerline{ 
\epsfig{file=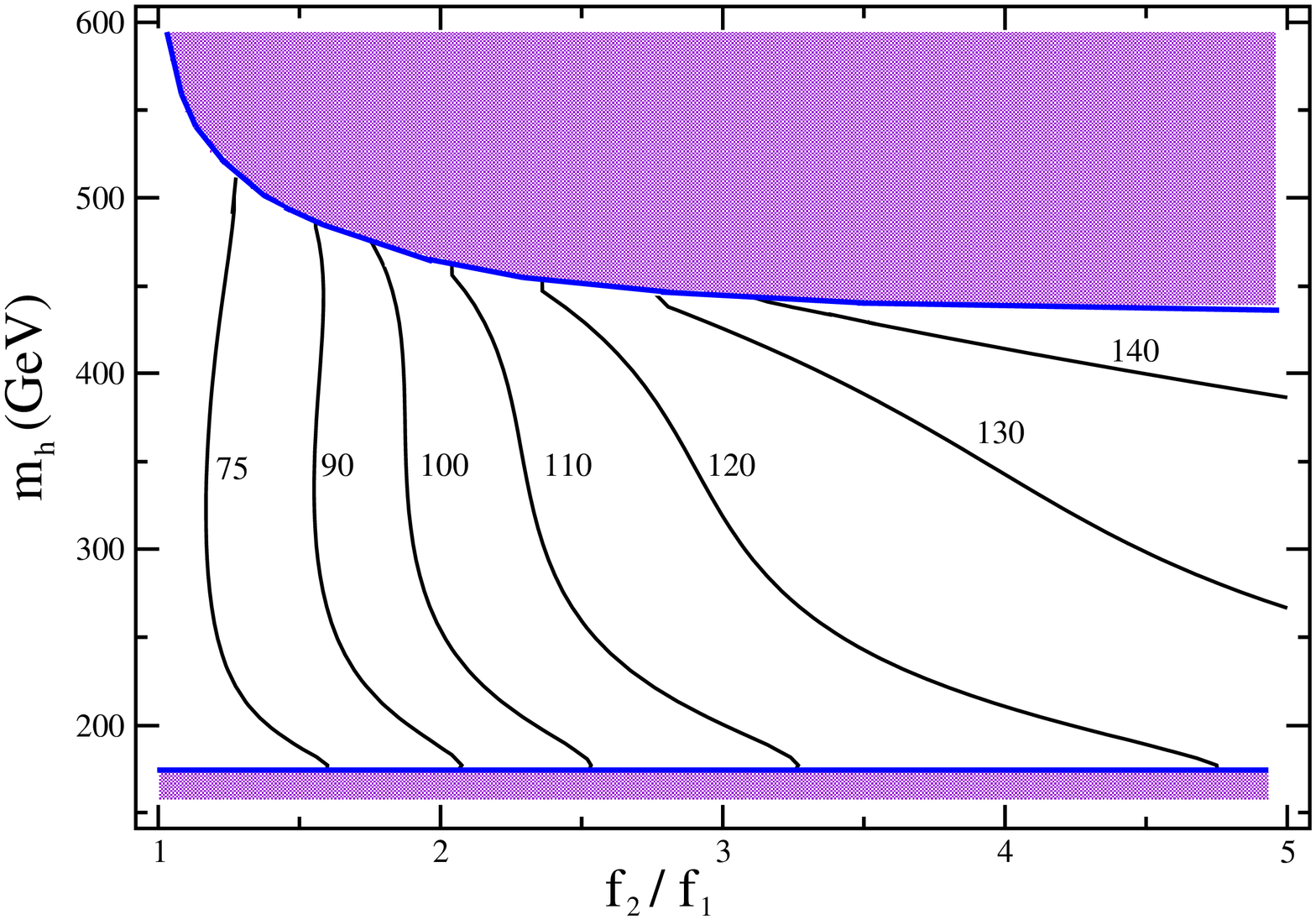,width=4.75cm,bbllx=4.cm,bblly=0.cm,bburx=19.cm,bbury=27.cm,angle=-90} 
\epsfig{file=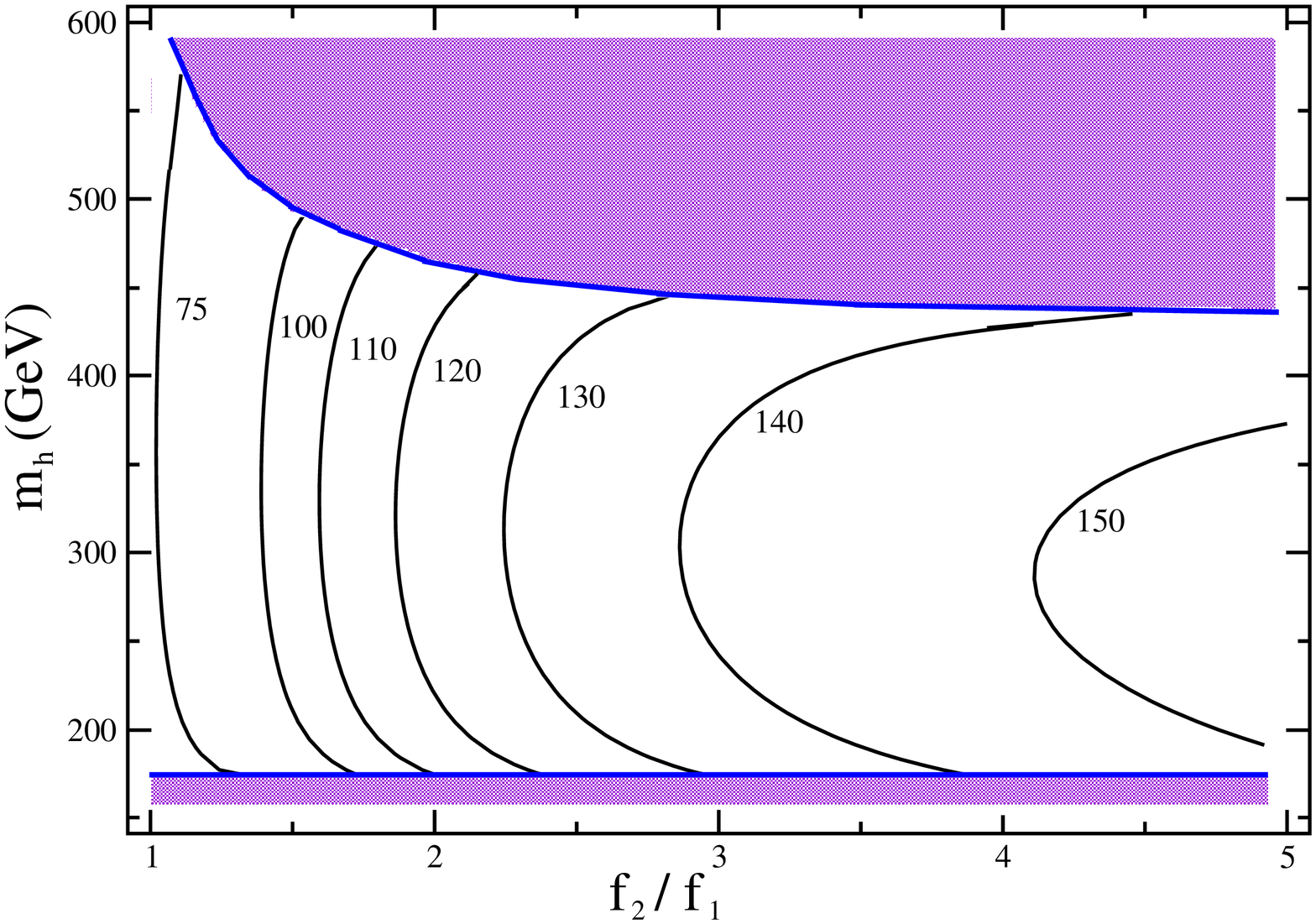,width=4.75cm,bbllx=4.cm,bblly=0.cm,bburx=19.cm,bbury=27.cm,angle=-90} 
} \caption{\footnotesize Fine-tuning contours for the Simplest Little 
Higgs model for cases 1) (left plot) and 2) (right plot) of 
eq.~(\ref{simplestcases}).} \label{fig:simplestf2f1} } 

In order to compute the fine-tuning in this model we use the previous 
potential, (\ref{Vrad}) plus (\ref{delta0V}): 
\be 
V(h)={1\over 2} 
(\mu_0^2+\delta m^2) h^2 +{1\over 4}\left[\delta_1 \lambda(h)-{f^2\over 
3f_1^2 f_2^2}\left(\delta m^2+{\mu_0^2\over 4}\right)\right]h^4+... 
\label{Vsch} 
\ee 
As mentioned, it does not contain the subdominant 
contributions from $\eta$ and the Higgs field. The input parameters are 
now: 
\be 
\{\lambda_1,\lambda_2,\mu^2,f_1,f_2\}\ . 
\ee 
Without loss of generality we can choose $f_1\leq f_2$, in which case the 
UV cut-off is $\Lambda=4\pi f_1$. Since we want $\Lambda=10$ TeV (the 
scale of the Little Hierarchy problem) we also set $f_1=1$ TeV. As $f_1$ 
and $f_2$ are 
not the only mass scales in the problem (there is $\mu^2$ as well) it is 
important to include the fine-tuning associated to them, which might be 
large now.

The Higgs mass that results from the potential (\ref{Vsch}), after trading 
$\mu_0^2$ by $v$ using the minimization condition, can be computed as a 
function of $M^2_T$ for fixed $f_2/f_1$. For any pair 
$\{\lambda_1,\lambda_2\}$ that gives a particular value of $M_T^2$, there 
is 
another pair $\{\lambda_1,\lambda_2\}\rightarrow \{\lambda_2 
f_2/f_1,\lambda_1 f_1/f_2\}$ that gives the same $M_T^2$. Therefore each 
choice of $M_T^2$ (to get a particular value of $m_h$) corresponds to two 
different solutions in terms of $\lambda_{1,2}$. We will refer to them as 
\be 
{\rm 1)}\; \lambda_1 f_1 \leq Â\lambda_2 f_2\ , \;\;\; \;\;\; 
{\rm 2)}\; \lambda_1 f_1 \geq \lambda_2 f_2\ . 
\label{simplestcases} 
\ee 
As mentioned above, these two solutions 
are related by the interchange $\lambda_1 f_1\leftrightarrow \lambda_2 
f_2$.
Fig.~\ref{fig:simplestf2f1} gives the fine-tuning in the plane 
$\{m_h,f_2/f_1\}Ã$ for these two cases 

\FIGURE[t]{\centerline{ 
\epsfig{file=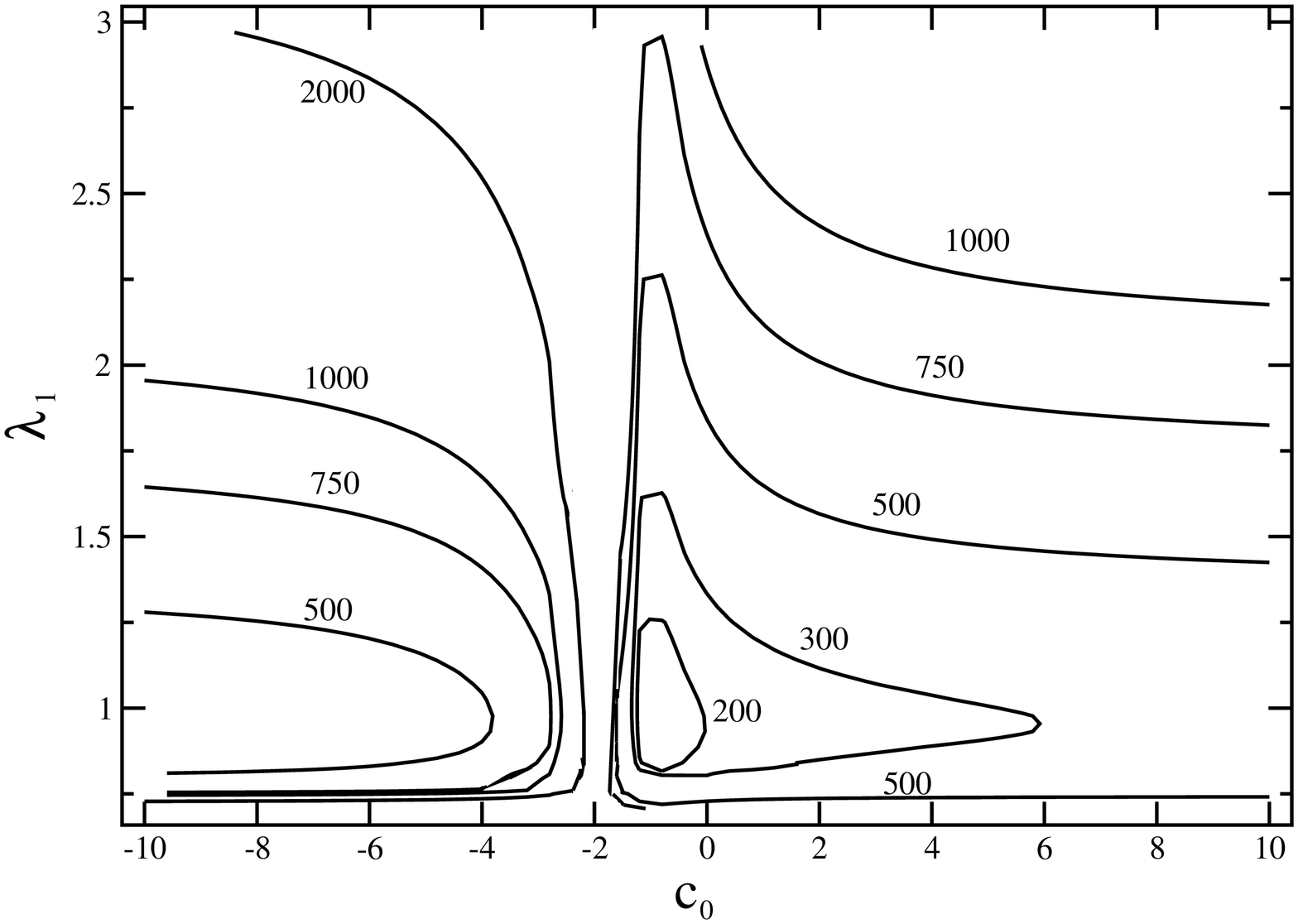,width=4.75cm,bbllx=4.cm,bblly=0.cm,bburx=20.cm,bbury=27.cm,angle=-90} 
\epsfig{file=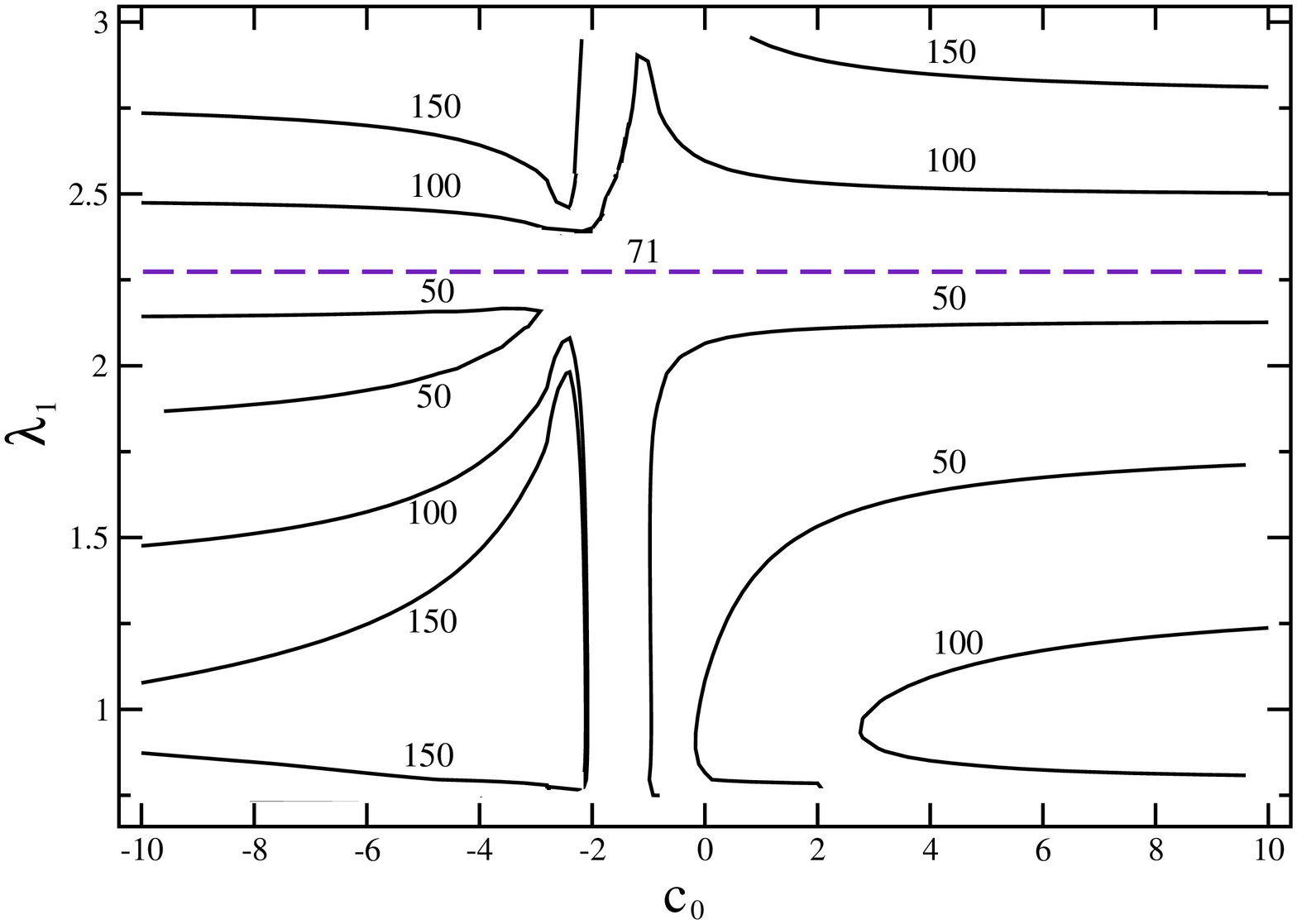,width=4.75cm,bbllx=4.cm,bblly=0.cm,bburx=20.cm,bbury=27.cm,angle=-90} 
} \caption{\footnotesize Fine-tuning contours for the Simplest Little 
Higgs model augmented by a `tree-level' quartic coupling $\lambda_0$, with 
$m_h=115$ GeV (left plot) and $m_h=250$ GeV (right plot).} 
\label{fig:simplest} } 

 We see from these plots that the fine-tuning is sizeable and 
increases with $f_2/f_1$. From the bound $M_T\geq 2\lambda_t f_1 f_2/f$ 
and the fact that $\delta m^2$ and $\delta_1\lambda$ cannot be arbitrarily 
large, it follows that $m_h^2$ is limited to a certain range. This range 
depends on the value of $f_2/f_1$: for $f_2=f_1$ one gets 163 GeV $\leq 
m_h \leq$ 606 GeV and a narrower range for larger $f_2/f_1$, as can 
be seen in 
fig.~\ref{fig:simplestf2f1}.

To access lower values of $m_h$ one can add a piece $\lambda_0$ to the 
Higgs quartic coupling in the potential (\ref{Vsch}). This new term can 
result from the unknown heavy physics at the cut-off $\Lambda$. For 
$\lambda_0<0$ one can get values of $m_h$ below the lower bounds discussed 
before. In the presence of such term we should also worry about the 
quadratically divergent contributions of scalars to the Higgs mass 
parameter. From 
\be 
\delta V_1^{\rm quad} = {\Lambda^2\over 32\pi^2} 
(m_h^2+3m_G^2+m^2_\eta²)\ ,
\ee 
where $m_h, m_G$ and $m_\eta$ are the 
tree-level masses of the Higgs, the electroweak Goldstones and $\eta$ 
respectively, one gets\footnote{Of 
course, this contribution is due to the fact that the Simplest model does 
not include 
additional fields to cancel the quadratic divergencies from loops of its 
scalar 
fields.} (after substituting $\Lambda=4\pi f_1$) 
\be 
\delta_q m^2=- {5f^2\over 8f_2^2}\mu_0^2+6\lambda_0 f_1^2\ . 
\ee 
The piece proportional to $\mu_0^2$ is not particularly dangerous and can 
even be interpreted as a redefinition of the original $\mu_0^2$ parameter, 
while the second term, proportional to the new coupling $\lambda_0$, can 
be sizeable, thus having a significant impact on the fine-tuning. 
In the presence of these quadratically divergent corrections we 
expect to have a contribution to the Higgs mass parameter of order 
$6\lambda_0 f_1^2$ already at the cut-off. Therefore we introduce such 
mass term in the potential, multiplied by some unknown coefficient $c$, 
from the beginning. As we did in previous models, we then split $c$ into 
an unknown `tree-level' contribution $c_0$ and a calculable radiative 
one-loop correction $c_1$, with $c=c_0+c_1=c_0+1$. Our potential is now 
\be V(h)={1\over 2} [\mu_0^2+\delta m^2+6(c_0+1)\lambda_0 f_1^2] h^2 
+{1\over 4}\left[\lambda_0+\delta_1 \lambda(h)-{f^2\over 3f_1^2 
f_2^2}\left(\delta m^2+{\mu_0^2\over 4}\right)\right]h^4+... \label{Vsch1} 
\ee and the set of input parameters is enlarged to \be 
\{\lambda_0,\lambda_1,\lambda_2,\mu^2,f_1,f_2, c_0\}\ . \label{parl0} \ee

Fig.~\ref{fig:simplest} shows the fine-tuning associated to this 
modified potential in the plane $\{c_0,\lambda_1\}$ for $m_h=115$ Gev 
(left plot) and $m_h=250$ GeV (right plot) for $f_2=f_1$. As expected, 
lower Higgs masses can now be reached, but there is a fine-tuning price to 
pay. As shown by the right plot, in the case of larger Higgs masses, 
already accessible for $\lambda_0=0$, the effect of the new parameters 
$c_0$ and $\lambda_0$ 
allows the fine-tuning to be reduced if such parameters are chosen 
appropriately, but the effect is never dramatic (for the sake 
of comparison, we show by a dashed line, the  
fine-tuning corresponding to $\lambda_0=0$). However, the 
fine-tuning gets worse in most of the parameter space.

From figs.~\ref{fig:simplestf2f1} and \ref{fig:simplest}, we can conclude 
that the fine-tuning in the Simplest LH model is similar to that of the 
models analyzed in previous sections: it is always significant and 
usually comparable to (or higher than) that
 of the Little Hierarchy problem [$\Delta \simgt {\cal O}(100)$]. Only for 
some small regions of parameter space is $\Delta$ comparable to the 
MSSM one ($\Delta \sim 20-40$ for $m_h\simlt 125$ GeV); usually it is much 
worse. The last point is 
illustrated by the scatter-plot of fig.~{\ref{fig:scatter}}, which shows 
the value of $\Delta$ vs. $m_h$ for random values of the parameters 
(\ref{parl0}) compatible with $v=246$ GeV. More precisely, we have set 
$f_1=f_2=1$ TeV and chosen at random $\lambda_0\in [-2,2]$, $\lambda_1\in 
[\lambda_t/\sqrt{2},15]$ and $c_0\in [-10,10]$. The solid line gives the 
minimal value of $\Delta$ as a function of $m_h$ and has been computed 
independently (rather than deduced from the scatter plot). Clearly, the 
density of points gets sparser near this lower bound.

\FIGURE[t]{\psfig{file=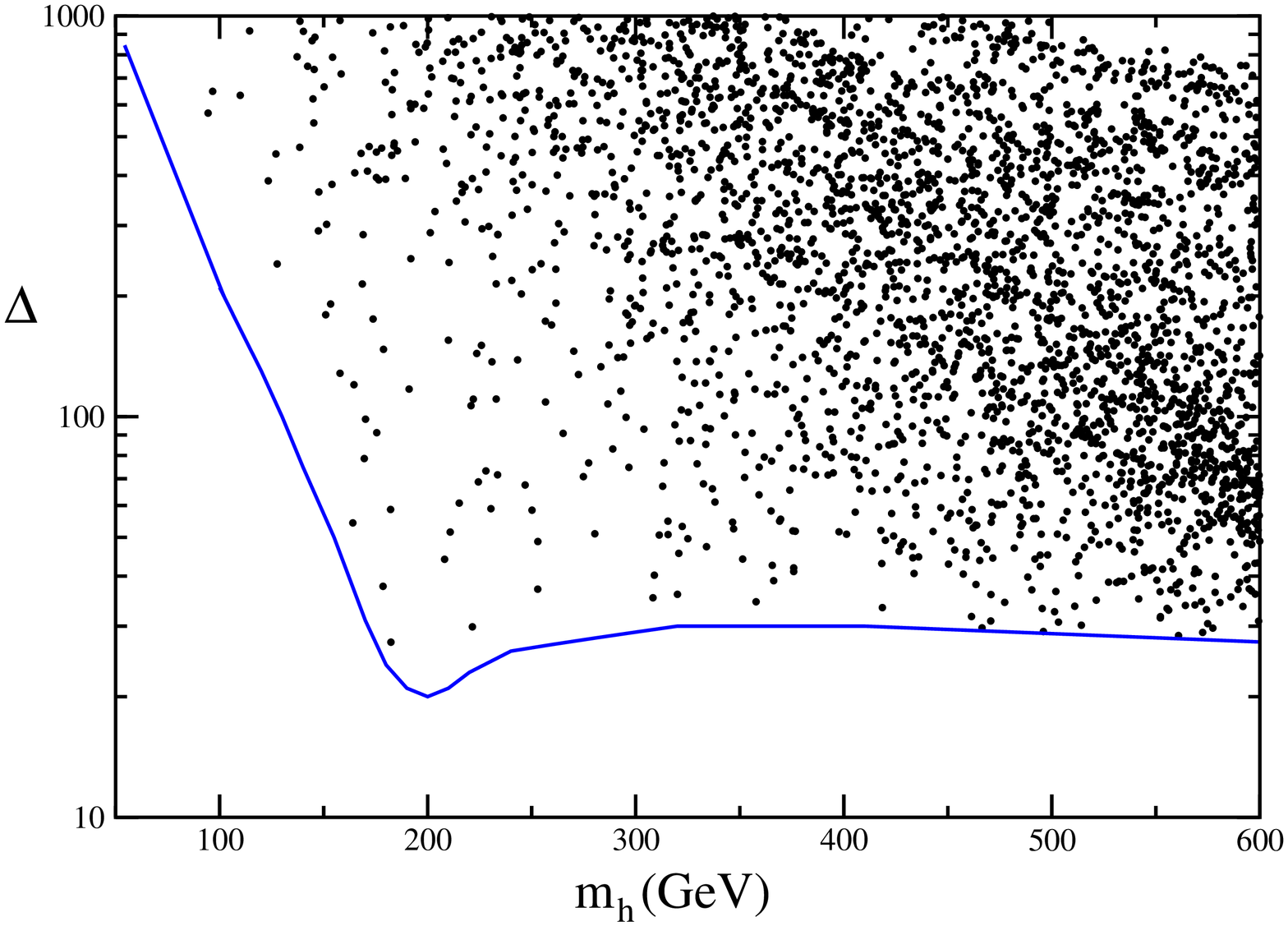,width=7cm,bbllx=1.6cm,bblly=0.cm,bburx=20.cm,bbury=23.cm,angle=-90} 
\caption{\footnotesize Scatter-plot of the fine-tuning in the Simplest 
Little Higgs model as a function of the Higgs mass.} \label{fig:scatter}} 

\section{Conclusions}

\FIGURE[t]{\psfig{file=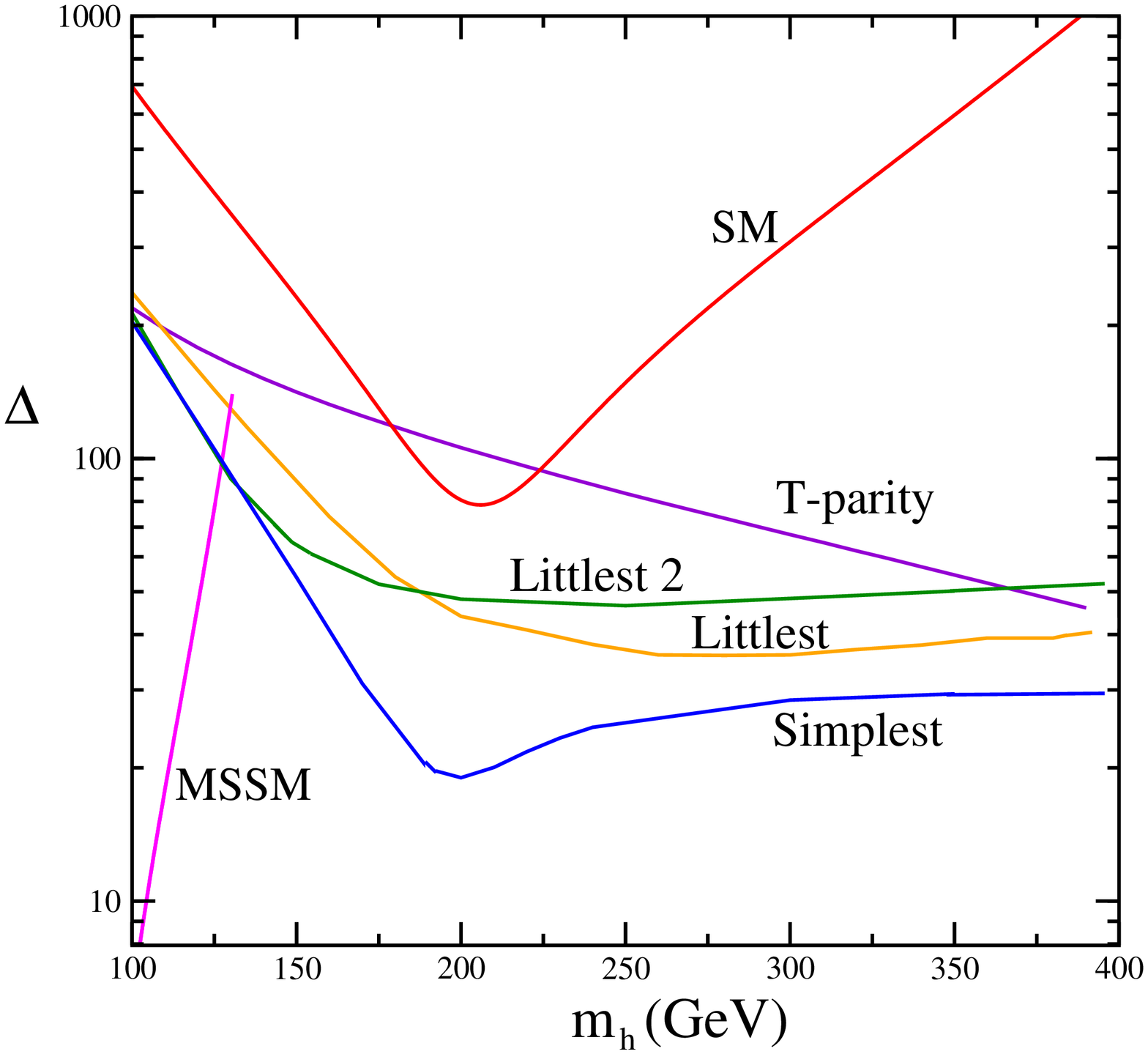,width=7cm,bbllx=2.cm,bblly=0.cm,bburx=20.cm,bbury=23.cm,angle=-90} 
\caption{\footnotesize 
Comparative summary of the fine-tuning vs. $m_h$ for different 
scenarios. The curves for Little Higgs models (lines labeled ``Littlest'', 
``Littlest 2'', ``$T$-parity'' and ``Simplest'') are lower bounds on the 
corresponding fine-tuning, see text for details.} 
\label{fig:comp}} 

We have rigorously analyzed the fine-tuning associated to 
the electroweak breaking process in Little Higgs (LH) scenarios, focusing
on four popular and representative models, corresponding to 
refs.~\cite{Littlest,Peskin,ChengLow,Sch}.

Although LH models solve parametrically the Little Hierarchy problem
[generating a Higgs mass parameter of order $f/(4\pi)$], our first 
conclusion is that these models generically have a substantial 
fine-tuning built-in, usually much higher than suggested by 
the rough considerations commonly made. 
 This is due to implicit tunings between parameters that can be overlooked 
at first glance but show up in a more systematic analysis. This does not 
demonstrate, of course, that all LH models are necessarily fine-tuned, but 
it stresses the need of a rigorous analysis in order to claim that a 
particular model is not fine-tuned, especially if a quantitative statement 
is attempted (e.g.~to compare its degree of fine-tuning with that of the 
MSSM). In this respect, the analysis presented here can also be helpful as 
a guide to the ingredients that typically increase the fine-tuning in LH 
models, in order to correct them in improved constructions.

We have quantified the degree of fine-tuning following the 'standard' 
criterion of Barbieri and Giudice \cite{BG}, through a fine-tuning 
parameter $\Delta$, that can be computed in each model ($\Delta \simeq 
100$ means a fine-tuning at the one percent level, etc.), finding that the 
four LH scenarios analyzed here present fine-tuning ($\Delta > 10$) in all 
 cases. The results are summarized in the plots of 
figs.~\ref{fig:littlest-a2} (for the Littlest Higgs), \ref{fig:peskin2}
(for the modified Littlest Higgs), \ref{fig:littlesTa} and
\ref{fig:littlesTmu} (for the Littlest Higgs with $T$-parity), and
\ref{fig:simplestf2f1} and \ref{fig:simplest} (for the Simplest Little 
Higgs). Actually, the fine-tuning is comparable to or higher than 
--sometimes much higher-- than the 
one associated to the Little Hierarchy problem of the SM (given by the 
blue line of fig.~\ref{fig:velt-a}) in most of the parameter space of 
these models. Since LH models have been designed to solve the 
Little Hierarchy problem, we believe this is a serious drawback. Likewise, 
the fine-tuning is usually worse than that of supersymmetric models 
($\Delta = 20-40$ for the MSSM and lower for other supersymmetric 
scenarios), which succeed at stabilizing a much larger hierarchy 
($\Lambda\simeq M_{GUT}$ or $M_{Planck}$ rather than $\Lambda\simeq 10$ 
TeV).

We can make the previous statements more precise. Fig.~\ref{fig:comp} 
shows the fine-tuning $\Delta$ as a function of $m_h$ for different 
scenarios. The curve labelled ``SM'' represents the fine-tuning 
of the Little Hierarchy problem in the SM, as discussed in the 
introduction. The ``MSSM'' line shows the fine-tuning of the 
MSSM\footnote{This curve has been obtained for large $\tan\beta$ (which 
minimizes the fine-tuning) but disregarding stop-mixing effects (which 
can help in reducing the fine-tuning). It also takes into account the 
most recent experimental value for the top mass \cite{top}, which 
makes the fine-tuning lower than in previous analyses.}. Then, for each 
LH model analyzed in sects.~2--4 we have plotted (lines 
labeled ``Littlest'', ``Littlest 2'', ``$T$-parity''
and ``Simplest'') the minimum value of $\Delta$ accessible by varying the
parameters of the model.
Usually, only in a quite small area of parameter space of each model
is the fine-tuning close to the lower bound shown,
so the LH curves in fig.~\ref{fig:comp} are a 
very conservative estimate of the fine-tuning in the 
corresponding LH models. This 
point is illustrated by fig.~{\ref{fig:scatter}} for the Simplest LH model 
(the best behaved): the lower line in that plot corresponds to the 
``Simplest'' line in fig.~\ref{fig:comp}. Now we see that the value of 
$\Delta$ for all these models is $\geq {\cal O} (100)$ in most of parameter 
space, and larger that $20-30$ in all cases. This fine-tuning is larger 
than the MSSM one, at least for the especially interesting range 
$m_h\simlt 130$ GeV. Notice here that 
$m_h\simgt 135$ GeV is not available in the MSSM if the supersymmetric 
masses are not larger than $\sim 1$ TeV. This limitation does not hold 
for other supersymmetric models, e.g. those with low-scale SUSY 
breaking, as discussed in ref.~\cite{CEH0}, which are 
definitely in better shape than LH models concerning fine-tuning issues.

Regarding the specific ingredients that potentially increase the 
fine-tuning in LH models, we stress two of them. First, the LH Lagrangian 
is generically enlarged with operators that have the same structure as 
those generated through the quadratically divergent radiative corrections 
to the potential (and  are necessary for the viability of the
models). Such operators have two contributions: the radiative one 
(calculable) and the 
'tree-level' one (arising from physics beyond the cut-off and unknown).  
Very often the required value of the coefficient in front of a given 
operator is much smaller than the calculable contribution, which implies a 
tuning (usually unnoticed) between the tree-level and the one-loop pieces 
(similar to the hierarchy problem in the SM). Second, the value of the 
Higgs quartic coupling, $\lambda$, receives several contributions which 
have a non-trivial dependence on the various parameters of the model.
Sometimes 
it is difficult, without an extra fine-tuning, to keep $\lambda$ small, 
as required to have $m_h$ in the region that is more interesting 
phenomenologically. 

\newpage

\appendix

\section{Fine-tuning estimates with constraints} 
\setcounter{equation}{0} 
\renewcommand{\theequation}{A.\arabic{equation}}

Let $F(x_i)$ be a quantity that depends on some input parameters $x_i$ 
$(i=1,...,N)$, considered as independent. The fine-tuning in $F$ 
associated to $x_i$ is $\Delta_i$, defined by \be {\delta F\over F} = 
\Delta_i {\delta x_i\over x_i}\ . \ee It is convenient for the following 
discussion to switch to vectorial notation and define \be \vec{\Delta} F 
\equiv \left\{ {\partial \log F\over \partial \log x_i}\right\} \ , \ee 
which is a vector of dimension $N$ with components $\Delta_i$, and is 
simply the gradient of $\log F$ in the $\{\log x_i\}$ space. Based on the 
statistical meaning of $\Delta_{i}$, we define the total fine-tuning 
associated to the quantity $F$ as \be \label{DefDelta} \Delta 
F\equiv\left[\sum_i\Delta_i^2\right]^{1/2}=||\vec{\Delta}F||\ . \ee

Next suppose that the $x_i$ are not independent but are instead related by 
a number of (experimental or theoretical) constraints 
$G^{(0)}_\alpha(x_i)=0$ ($\alpha=1,...,m$ with $m<N$) so that, when one 
computes the fine-tuning in $F$, one is only free to vary the input 
$x_i$'s in such a way that the constraints are respected. In order to 
compute the ``constrained fine-tuning" in $F$ we first define, for each 
constraint, the vector $\vec{\Delta}G_\alpha^{(0)}=\{ \partial 
G_\alpha^{(0)}/\partial\log x_i\}$ which is normal to the 
$G_\alpha^{(0)}=0$ hypersurface in the $\{\log x_i\}$ space. We then use 
the Gramm-Schmidt procedure to get from the vectors 
$\vec{\Delta}G_\alpha^{(0)}$ an orthonormal set, $\vec{\Delta}G_\alpha$, 
that satisfies 
\be 
\vec{\Delta}G_\alpha\cdot\vec{\Delta}G_\beta=\delta_{\alpha\beta}\ . \ee 
Then we can find the constrained fine-tuning simply projecting the 
unconstrained $\vec{\Delta} F$ on the $G_\alpha=0$ manifold [which 
coincides with the $G_\alpha^{(0)}=0$ manifold]: \be \left.\vec{\Delta} 
F\right|_G=\vec{\Delta} F 
-\sum_{\alpha}(\vec{\Delta}F\cdot\vec{\Delta}G_\alpha)\vec{\Delta}G_\alpha\ 
. \ee Finally, \be \label{constrDelta} \left.\Delta 
F\right|_G=\left|\left|\left.\vec{\Delta}F\right|_G\right|\right|= 
\left[(\Delta F)^2 - \sum_\alpha(\vec{\Delta} 
F\cdot\vec{\Delta}G_\alpha)^2\right]^{1/2}\ . \ee 
As was to be expected, the 
constrained fine-tuning, $\left.\Delta F\right|_G$, is always smaller that 
the unconstrained fine-tuning $\Delta F$.

The previous procedure can also be seen as a change of coordinates in the 
``euclidean'' $\{\log x_i\}$ space [which leaves eq.~(\ref{DefDelta}) 
invariant], such that the first $m$ new coordinates $\{\log y_\alpha\}$ 
span the same subspace as the $\vec{\Delta}G_\alpha^{(0)}$ vectors. These 
$m$ coordinates have to be simply eliminated from eq.~(\ref{DefDelta}), as 
they are fixed by the constraints, while the remaining ones are totally 
unconstrained. In this way the final expression (\ref{constrDelta}) is 
recovered.

Note that if $F$ does not depend on some of the parameters, say $\{x_a\}$, 
but some of the constraints do, the constrained fine-tuning will 
generically depend on the value of $\{x_a\}$, even if the other parameters 
remain the same. This is in fact a perfectly logical result. Notice that 
the fine-tuning quantity, $\Delta F$, measures the relative change of $F$ 
against the relative changes in the $x_i$ parameters. Imagine a function 
$F=F(x_1)$ and a constraint $G^{(0)}= x_1 + x_2 +x_3 -C = 0$. If 
$x_2,x_3\ll x_1$ the value of $x_1$ is essentially fixed and thus $\Delta 
F|_G$  should be small (if $x_2, x_3$ are allowed to change a 100\%, $x_1$ 
is 
only allowed to change in a very small relative range). In the opposite 
case, if $x_2,x_3\gg x_1$ (for the same value of $x_1$) the $x_1$ 
parameter can be freely varied and thus $\Delta F|_G\simeq \partial \log 
F/ 
\partial \log x_1$. Therefore, $\Delta F|_G$ does depend on $x_2$ and 
$x_3$ 
even if $F=F(x_1)$. We have found this effect in some of the 
scenarios studied (although it always had a mild impact on the final 
fine-tuning); see sect.~4, footnote 10.

\section{Formulas for Little Higgs models} \setcounter{equation}{0} 
\renewcommand{\theequation}{B.\arabic{equation}}

\subsection{The Littlest Higgs Model}

This model \cite{Littlest} is based on an $SU(5)/SO(5)$ nonlinear sigma 
model. The spontaneous breaking of $SU(5)$ down to $SO(5)$ is produced by 
the vacuum expectation value of a $5\times 5$ symmetric matrix field 
$\Phi$.  We follow \cite{Littlest} and choose \beq 
\langle\Phi\rangle=\Sigma_0 =\left(\begin{array}{ccc} 0 &0&I_2\\ 0&1&0\\ 
I_2&0&0 \end{array}\right)\ . \label{vev} \eeq This breaking of the global 
$SU(5)$ symmetry produces 14 Goldstone bosons which include the Higgs 
doublet field. These Goldstone bosons can be parametrized through the 
nonlinear sigma model field \beq \Sigma = e^{i \Pi/f} \Sigma_0 e^{i 
\Pi^T/f} = e ^{2i\Pi/f}\Sigma_0, \label{Sigma} \eeq with $\Pi = \sum_a 
\Pi^a X^a$, where $\Pi_a$ are the Goldstone boson fields and $X^a$ the 
broken $SU(5)$ generators. The model assumes a gauged $SU(2)_1\times 
U(1)_1\times SU(2)_2\times U(1)_2$ subgroup of $SU(5)$ with generators 
($\sigma^a$ are the Pauli matrices) \beq Q_1^a = \left(\begin{array}{cc} 
\sigma^a/2 &0_{2\times 3}\\ 0_{3\times 2}&0_{3\times 3} 
\end{array}\right)\ , \hspace{0.5cm} Q_2^a = \left(\begin{array}{cc} 
0_{3\times 3} &0_{3\times 2}\\ 0_{2\times 3}&-\sigma^{a^{*}}/2 
\end{array}\right)\ , \label{generators} \eeq and \be Y_1={1\over 10}{\rm 
diag}(-3,-3,2,2,2)\ , \;\;\;\; Y_2={1\over 10}{\rm diag}(-2,-2,-2,3,3)\ . 
\ee The vacuum expectation value in eq.~(\ref{vev}) breaks $SU(2)_1\times 
U(1)_1\times SU(2)_2\times U(1)_2$ down to the diagonal $SU(2)\times 
U(1)$, identified with the SM group.

The Goldstone and (pseudo)-Goldstone bosons in the hermitian matrix $\Pi$ 
in $\Sigma$ fall in representations of the SM group as \beq \Pi = 
\left(\begin{array}{ccc} \xi &\frac{H^{\dagger}}{\sqrt 2}&\phi^{\dagger}\\ 
\frac{H}{\sqrt 2}&0&\frac{H^{*}}{\sqrt 2}\\ \phi&\frac{H^{T}}{\sqrt 
2}&\xi^T \end{array}\right)+{1\over \sqrt{20}}\zeta^0{\rm 
diag}(1,1,-4,1,1)\ , \label{pi} \eeq where $H=(h^0,h^+)$ is the Higgs 
doublet; $\phi$ is a complex $SU(2)$ triplet given by the symmetric 
$2\times 2$ matrix: \be \phi=\left[ \begin{array}{cc} \phi^0 & {1\over 
\sqrt{2}} \phi^+\\ {1\over \sqrt{2}} \phi^+ & \phi^{++} \end{array} 
\right]\ , \ee the field $\zeta^0$ is a singlet which is the Goldstone 
associated to the $U(1)_1\times U(1)_2\rightarrow U(1)_Y$ breaking and 
finally, $\xi$ is the real triplet of Goldstone bosons associated to 
$SU(2)_1\times SU(2)_2\rightarrow SU(2)$ breaking: \be \xi={1\over 2} 
\sigma^a\xi^a=\left[ \begin{array}{cc} {1\over 2}\xi^0 & {1\over \sqrt{2}} 
\xi^+\\ {1\over \sqrt{2}} \xi^- & -{1\over 2}\xi^{0} \end{array} \right]\ 
. \ee All the fields in $\Pi$ as written above are canonically normalized.

The kinetic part of the Lagrangian is 
\beq 
{\cal L}_{kin} = 
\frac{f^2}{8} {\rm Tr}[(D_{\mu}\Sigma)(D^{\mu}\Sigma)^{\dagger}]\ , 
\label{Lkin} 
\eeq 
where \beq D_{\mu}\Sigma = \partial_{\mu} \Sigma - 
i\sum_{j=1}^2 g_j W_j^a(Q_j^a \Sigma + \Sigma Q_j^{a T})  - i \sum_{j=1}^2 
g_j' B_y(Y_j \Sigma + \Sigma Y_j^T). \eeq

In this model, additional fermions are introduced in a vector-like 
coloured pair $t',{t'}^c$ to cancel the Higgs mass quadratic divergence 
from top loops (other Yukawa couplings are neglected). The relevant part 
of the Lagrangian containing the top Yukawa coupling is given by \beq 
{\cal L}_{f} = {1\over 2}\lambda_{1} f \epsilon_{ijk} \epsilon_{xy} 
\chi_i \Sigma_{jx} \Sigma_{ky} u'^{c} + \lambda_{2} f t' {t'}^c + h.c., 
\label{Lf} \eeq where $\chi_i = (t, b, t')$, indices $i,j,k$ run from 1 to 
3 and $x,y$ from 4 to 5, and $\epsilon_{ijk}$ and $\epsilon_{xy}$ are the 
completely antisymmetric tensors of dimension 3 and 2, respectively.

As mentioned in the text, by considering gauge and fermion loops one sees 
that the Lagrangian should also include gauge invariant terms of the form, 
\bea -\Delta {\cal L}\; =\;  V & \equiv & c\ {\cal O}_V(\Sigma) + c'\ 
{\cal O}_F(\Sigma) \phantom{\left[{1\over 1}\right]_0^1}\nonumber\\ &=& c 
f^4 \sum_{i=1,2}g_i^2 \sum_a {\rm Tr}[(Q_i^{a}\Sigma)(Q_i^{a}\Sigma)^*] + 
c f^4 \sum_{i=1,2}g_i'^2 {\rm Tr}[(Y_i\Sigma)(Y_i\Sigma)^*]\nonumber\\ & 
-& {1\over 8} c' f^4 \lambda_1^2 
\epsilon^{wx}\epsilon_{yz}\Sigma_{iw}\Sigma_{jx} \Sigma^{iy *}\Sigma^{jz 
*}\ , \label{potential} \eea with $c$ and $c'$ assumed to be constants of 
${\cal O}(1)$. The analysis of the spectrum and Higgs potential for this 
model is presented in section~2, after eq.~(\ref{DeltaL1}).

\subsection{A Modified Version of the Littlest Higgs Model}

This model is also based on the $SU(5)/SO(5)$ Littlest Higgs 
\cite{Littlest}, but modified \cite{Peskin} in such a way that only one 
abelian $U(1)$ factor (identified with hypercharge) is gauged. The 
$SU(2)_1\times SU(2)_2$ generators are as in the Littlest model 
[eq.~(\ref{generators})] and the hypercharge generator is $Y={\rm 
diag}(1,1,0,-1,-1)/2$. The field content of the hermitian matrix $\Pi$ in 
$\Sigma$ is the same as in the Littlest Higgs model but now the field 
$\zeta^0$ [Goldstone associated to the breaking of the $U(1)$ symmetry 
left ungauged] is not absorbed by the Higgs mechanism (there is no $B'$ 
now) and remains in the physical spectrum. In any case, this field plays 
no significant role in the discussion (it can be given a small mass to 
avoid phenomenological problems by adding explicit breaking terms 
\cite{UV}).

The kinetic part of the Lagrangian is as in the Littlest Higgs, 
eq.~(\ref{Lkin}) model but 
now with 
\beq 
D_{\mu}\Sigma = \partial_{\mu} \Sigma - i\sum_{j=1}^2 g_j 
W_j^a(Q_j^a \Sigma + \Sigma Q_j^{a T})  - i g' B_Y(Y \Sigma + \Sigma Y^T). 
\eeq 
The fermionic couplings in the Lagrangian can be kept as in the 
Littlest Higgs model also. Then the scalar operators ${\cal O}_F(\Sigma)$ 
and ${\cal O}_V(\Sigma)$, induced by fermion 
and gauge boson loops have the same form of eq.~(\ref{potential}) but with 
the $U(1)$ part limited to $U(1)_Y$ only. The main difference with respect 
to the Littlest Higgs case is that now the Higgs boson gets a small tree 
level mass of order $g'^2f^2$ through the ${\cal O}_V(\Sigma)$ operator.

The $h$-dependent field masses, needed for the calculation of the one-loop 
Higgs potential, are the following. In the gauge boson 
sector we have \be m^2_{W'}(h)={1\over 4}(g_1^2+g_2^2)f^2-{1\over 
4}g^2h^2+{\cal O}(h^4/f^2)\ , \ee with no $B'$ gauge boson. In the fermion 
sector, the heavy Top has mass \be m_T^2(h)=M_T^2+{\cal 
O}(h^2)=(\lambda^2_1+\lambda^2_2)f^2-{1\over 2}\lambda_t^2h^2+{\cal 
O}(h^4/f^2)\ . \ee In the scalar sector, decomposing $h^0\equiv 
(h^{0r}+ih^{0i})/\sqrt{2}$ and $\phi^{0}\equiv 
i(\phi^{0r}+i\phi^{0i})/\sqrt{2}$ and using $\lambda'_a\equiv 
cg_2^2-c'\lambda_1^2$ and $\lambda'_b\equiv cg_1^2$, combined in 
$\lambda'_+\equiv \lambda'_a+\lambda'_b$ and $\lambda'_-\equiv 
\lambda'_a-\lambda'_b$, the masses are as follows. Writing simultaneously 
the relevant part of the mass matrices in the $CP$-even sector (using the 
basis $\{h^{0r},\phi^{0r}\}$), the $CP$-odd sector (in the basis 
$\{h^{0i},\phi^{0i}\}$) and the charged sector (in the basis 
$\{h^+,\phi^+\}$), we get \bea M^2_\kappa(h)&=&\left[ \begin{array}{cc} 
{1\over 4}a_\kappa \lambda'_+h^2+{1\over\sqrt{2}}s_\kappa\lambda'_- f 
t+{\cal O}(h^4/f^2) & b_\kappa \lambda'_-fh+{\cal O}(h^2)\\ &\\ b^*_\kappa 
\lambda'_-fh+{\cal O}(h^2)& \lambda'_+\left(f^2-c_\kappa h^2\right)+{\cal 
O}(h^4/f^2) \end{array} \right]\nonumber\\ &&\nonumber\\ &+&c 
g'^2\left[\begin{array}{cc} f^2-d_\kappa h^2+{\cal O}(h^4/f^2) & {\cal 
O}(h^2)\\ &\\ {\cal O}(h^2)& 4f^2-e_\kappa h^2+{\cal O}(h^4/f^2) 
\end{array} \right]\ , \label{massmatp} \eea where the index 
$\kappa=\{0r,0i,+\}$ labels the different sectors. The numbers $a_\kappa$, 
$b_\kappa$, $c_\kappa$ and $s_\kappa$ are as in (\ref{massmat}) while 
$d_\kappa=\{1,1/6,1/6\}$ and $e_\kappa=13|b_\kappa|^2/3$. We have also 
included in these mass matrices the contribution of the triplet VEV, 
$t\equiv\langle\phi^{0r}\rangle$, with \be t\simeq -{1\over 
2\sqrt{2}}{\lambda'_-h^2\over (\lambda'_++4cg'^2)f}\ . \ee As in the 
Littlest Higgs model, the off-diagonal entries in (\ref{massmatp}) are due 
to the $h^2\phi$ coupling which causes mixing between $h$ and $\phi$ after 
electroweak symmetry breaking. This effect is negligible for the heavy 
triplet [at order $h^2$ in the masses, the components $\phi^{0r}$ and 
$\phi^{0i}$ can still be combined in a complex field $\phi^{0}$]. We call 
${h'}^{0r}$, ${h'}^{0i}$ and ${h'}^+$ the light mass eigenvalues of 
(\ref{massmat}) in the different sectors. The explicit masses for the 
different components of the triplet field are then\footnote{In writing the 
expansions for these masses we are assuming $cg'^2f^2\sim \lambda h^2 \ll 
\lambda'_+f^2$.} 
\be \left[ \begin{array}{c} m_{\phi^0}^2(h)\\ 
m_{\phi^+}^2(h)\\ m_{\phi^{++}}^2(h) \end{array}\right] =M_{\phi}^2+{\cal 
O}(h^2)=(\lambda'_++4cg'^2)f^2 -\left[ \begin{array}{c} 2\\ 1\\ 0 
\end{array}\right] \left(\lambda+{17\over 12}c g'^2\right) h^2+ {\cal 
O}(h^4/f^2)\ . \label{masatripa} \ee For ${h'}^{0r}$, ${h'}^{0i}$ and 
${h'}^+$ we get 
\bea \left[ \begin{array}{c} m^2_{{h'}^{0r}}(h)\\ 
m^2_{{h'}^{0i}}(h)\\ m^2_{h^+}(h)\end{array}\right] &=&M_s^2+{\cal O}(h^2) 
=cg'^2f^2 + \left[ \begin{array}{c} 3\\ 1\\ 1 \end{array}\right]\lambda 
h^2+ \left[ \begin{array}{c} 0\\ 1\\ 1 \end{array}\right]{1\over 
6}cg'^2h^2+{\cal O}(h^4/f^2)\ .\nonumber\\ && \eea From the previous 
expressions for the masses one can check that the cancellation of $h^2$ 
terms in ${\rm Str} M^2$ works except for the $g'$-dependent terms, as 
expected. The presence of the coupling $g'$, which does not respect the 
$SU(3)_{1,2}$ symmetries, complicates the structure of couplings in the 
Higgs sector. For instance, the Higgs quartic coupling after integrating 
out the heavy triplet is given by \be \lambda={1\over 
4}\left[\lambda'_a+\lambda'_b-{4\over 3}c g'^2- 
{(\lambda'_a-\lambda'_b)^2\over(\lambda'_a+\lambda'_b+4cg'^2)} \right]\ , 
\ee to be compared with the theoretically cleaner formula (\ref{lambda}) 
that holds in the Littlest Higgs case. All mass formulas and couplings 
written above reproduce those of the Littlest Higgs model in the limit 
$\lambda'_{a,b}\rightarrow \lambda_{a,b}$ and $g'\rightarrow 0$. After 
electroweak symmetry breaking some kinetic terms are non-canonical due to 
${\cal O}(h^2/f^2)$ corrections from non-renormalizable operators. The 
masses above include effects from field redefinitions necessary to render 
canonical all fields.\footnote{An automatic way of taking care of this 
complication is presented in ref.~\cite{ELR}.}

\subsection{A Little Higgs Model with $T$-parity}

This model, proposed in \cite{ChengLow}, is also based on the 
$SU(5)/SO(5)$ structure of the Littlest Higgs model, with the same gauge 
and scalar field content (see Appendix~B.1). The gauge kinetic part of the 
Lagrangian is as in eq.~(\ref{Lkin}) with $T$-parity requiring 
$g_1=g_2=\sqrt{2}g$ and $g'_1=g'_2=\sqrt{2}g'$. Imposing $T$-invariance on 
the fermionic sector requires the introduction of several new degrees of 
freedom. Those relevant for making the fermionic Lagrangian of 
eq.~(\ref{Lf})  $T$-symmetric are a new vector-like pair of coloured 
doublets $\tilde{q}_3,\tilde{q}_3^c$ ($T$-even) plus two new coloured 
singlets ${u'}^c_T$ (the $T$-image of ${u'}^c$) and $U$ (which is 
$T$-odd). The fermionic Lagrangian reads \cite{ChengLow} 
\be 
{\cal L}_{f}= 
{1\over 4}\lambda_{1} f \epsilon_{ijk} \epsilon_{xy} \left[(\xi Q)_i 
\Sigma_{jx} \Sigma_{ky} u'^{c} + (\tilde{\xi}Q)_i \tilde{\Sigma}_{jx} 
\tilde{\Sigma}_{ky} u'^{c}_{T} \right]+ \lambda_{2} f t' {t'}^c +{1\over 
\sqrt{2}}\lambda_3 f U(u'^{c}-u'^{c}_{T}) + h.c., 
\label{LfT} 
\ee 
plus (heavy) mass terms for $\tilde{q}_3$. Here we have used 
$Q\equiv(q_3,t',\tilde{q}_3)^T$, $\xi\equiv \exp[i\Pi/f]$, 
$\tilde\xi\equiv \Omega\exp[i\Pi/f]\Omega$ [with $\Omega\equiv {\rm 
diag}(1,1,-1,1,1)$] and $\tilde\Sigma\equiv \tilde\xi^2\Sigma_0$. The 
index convention is as in (\ref{Lf}). Finally, the scalar operators of 
(\ref{potential}) turn out to be given by \bea -\Delta {\cal L}\; =\;  V & 
= & 2 c g^2 f^4 \sum_{i=1,2} \sum_a {\rm 
Tr}[(Q_i^{a}\Sigma)(Q_i^{a}\Sigma)^*] + 2 c g'^2 f^4 \sum_{i=1,2} {\rm 
Tr}[(Y_i\Sigma)(Y_i\Sigma)^*]\nonumber\\ & -& {1\over 16} c' f^4 
\lambda_1^2 \epsilon^{wx}\epsilon_{yz}\left( 
\Sigma_{iw}\Sigma_{jx}\Sigma^{iy *}\Sigma^{jz *}+ 
\tilde\Sigma_{iw}\tilde\Sigma_{jx}\tilde\Sigma^{iy *}\tilde\Sigma^{jz *} 
\right)\ , \label{potentialT} \eea which is simply a $T$-invariant version 
of (\ref{potential}).

In this model, the squared masses to ${\cal O}(h^2)$, needed for the 
calculation of the one-loop Higgs potential, are very similar to 
those in the Littlest Higgs model. In the gauge boson sector they are 
exactly the same as in (\ref{masas1}), with gauge couplings related by 
eq.~(\ref{gT}): \bea \label{masas1T} m_{W'}^2(h)&=&M_{W'}^2+{\cal 
O}(h^2)=g^2 f^2 - {1\over 4}g^2h^2 + {\cal O}(h^4/f^2)\ ,\nonumber\\ 
m_{B'}^2(h)&=&M_{B'}^2+{\cal O}(h^2)={1\over 5}g'^2 f^2-{1\over 4}g'^2h^2 
+ {\cal O}(h^4/f^2)\ . \eea 
In the fermion sector, the only mass 
relevant for our purposes is that of the heavy Top which, to order $h^2$, 
remains the same as in the Littlest Higgs model: \be 
m_{T}^2(h)=M_{T}^2+{\cal O}(h^2)=(\lambda_1^2+\lambda_2^2)f^2- {1\over 2} 
\lambda_t^2 h^2 + {\cal O}(h^4/f^2)\ . \ee The squared masses of the other 
heavy fermions do not have an $h^2$-dependence.

In the scalar sector, an important difference with respect to the Littlest 
Higgs model is that now there is no $\phi h^2$ coupling. As a result, the 
Higgs quartic coupling does not get modified after decoupling the triplet 
field and is simply given by \be \lambda={1\over 4}(\lambda_a+\lambda_b)\ 
, \label{lambdaT2} \ee to be compared with eq.~(\ref{lambda}). Another 
direct consequence of not having a $\phi h^2$ coupling is the absence of 
the off-diagonal entries in the scalar mass matrices in the $CP$-even, 
$CP$-odd and charged sectors. Using the same conventions of 
eq.~(\ref{massmat}), these mass matrices are given by \be 
M^2_\kappa(h)=\left[ \begin{array}{cc} a_\kappa \lambda h^2+{\cal 
O}(h^4/f^2) & 0\\ 0& 4\lambda(f^2-c_\kappa h^2)+{\cal O}(h^4/f^2) 
\end{array} \right]\ , \label{massmatT} \ee with the constants $a_\kappa$ 
and $c_\kappa$ exactly as in the Littlest Higgs model, 
eq.~(\ref{massmat}). The explicit masses for the different components of 
the heavy triplet field are still given by (\ref{masatripa}), and making 
use of (\ref{lambdaT2}) they simply read \be \left[ \begin{array}{c} 
m_{\phi^0}^2(h)\\ m_{\phi^+}^2(h)\\ m_{\phi^{++}}^2(h) \end{array}\right] 
=M_{\phi}^2+{\cal O}(h^2)=4\lambda f^2 -\left[ \begin{array}{c} 2\\ 1\\ 0 
\end{array}\right] \lambda h^2+{\cal O}(h^4/f^2)\ . \label{masatripT} \ee 
For the light eigenvalues of (\ref{massmatT}), which now do not mix with 
the triplet components, we simply get $m^2_{h^{0r}}(h)=3\lambda h^2$, 
$m^2_{h^{0i}}(h)=m^2_{h^+}(h)=\lambda h^2$, as in the Standard Model.

\subsection{The Simplest Little Higgs Model}

This is a model proposed in \cite{Sch} which is based on $[SU(3)\times 
U(1)]^2/[SU(2)\times U(1)]^2$, with a gauged $[SU(3)\times U(1)]$ subgroup 
broken down to the electroweak $SU(2)\times U(1)$. This spontaneous 
symmetry breaking produces 10 Goldstone bosons, 5 of which are eaten by 
the Higgs mechanism to make massive a complex $SU(2)$ doublet of extra 
$W'$s, $(W'^{\pm}, W'^0)$, and an extra $Z'$. The remaining 5 degrees of 
freedom are: $H$ [an $SU(2)$ doublet to be identified with the SM Higgs] 
and $\eta$ (a singlet).
 
Explicitly, the spontaneous breaking is produced by the VEVs of two scalar 
triplet fields, $\Phi_1$ and $\Phi_2$: \be \langle\Phi_1\rangle=\left( 
\begin{array}{c} 0\\ 0\\ f_1 \end{array} \right)\ , \;\;\;\;\; 
\langle\Phi_2\rangle=\left( \begin{array}{c} 0\\ 0\\ f_2 \end{array} 
\right)\ . \label{vac} \ee These triplets transform under the global 
symmetry as \be \Phi_1\rightarrow e^{-i\alpha_1/3}U_1 \Phi_1\ , \; \; \; 
\; \; \Phi_2\rightarrow e^{-i\alpha_2/3}U_2 \Phi_2\ , \ee where $U_i$ is 
an $SU(3)_i$ matrix and $e^{-i\alpha_i/3}$ are $U(1)_i$ rotations, with 
gauge transformations corresponding to the diagonal $U_1=U_2$, 
$\alpha_1=\alpha_2$. Using the broken generators, the Goldstone 
fluctuations around the vacuum (\ref{vac}) can be written as \be 
\Phi_i=\exp \left[{i\over f}\left( \begin{array}{ccc} 0 & 0 & h_i^+\\ 0 & 
0 & h_i^0\\ h_i^- & h_i^{0*} & \eta_i/\sqrt{2} \end{array} \right) 
\right]\left( \begin{array}{c} 0\\ 0\\ f_i \end{array} \right)\ , \ee for 
$i=1,2$, with $f^2=f_1^2+f_2^2$. Identifying explicitly the linear 
combinations of $h_i$ and $\eta_i$ that correspond to the eaten Goldstones 
$(G^\pm,G^0,G_S)$ and the physical fields $(H,\eta)$ one gets 
\bea 
\Phi_1&=&\exp \left[{i\over f}\left( 
\begin{array}{ccc} 
0 & 0 & G^+\\ 
0 & 0 & G^0\\ 
G^- & G^{0*} & G_S/\sqrt{2} 
\end{array} 
\right)
+{i f_2\over f f_1} \left( 
\begin{array}{ccc} 
0 & 0 & h^+\\ 
0 & 0 & h^0\\ 
h^- & h^{0*} & \eta/\sqrt{2} 
\end{array} \right) \right] 
\left( \begin{array}{c} 
0\\ 0\\ f_1 \end{array} \right)\ ,\nonumber\\ 
&&\\ 
\Phi_2&=&\exp \left[{i\over f}\left( 
\begin{array}{ccc} 0 & 0 & G^+\\ 0 & 0 & G^0\\ G^- & G^{0*} & 
G_S/\sqrt{2} \end{array} \right)-{i f_1\over f f_2} \left( 
\begin{array}{ccc} 0 & 0 & h^+\\ 0 & 0 & h^0\\ h^- & h^{0*} & 
\eta/\sqrt{2} \end{array} \right) \right] \left( \begin{array}{c} 0\\ 0\\ 
f_2 \end{array} \right)\ .\nonumber 
\eea

The scalar kinetic part of the Lagrangian is \be {\cal L}_k=|D_\mu 
\Phi_1|^2+|D_\mu \Phi_2|^2\ , \label{Lk} \ee with \be D_\mu 
\Phi_i=\partial_\mu \Phi_i - i g W_\mu^a T^a \Phi_i + {i\over 3}g_x 
B_\mu^x \Phi_i\ , \label{Lk1} \ee corresponding to the $SU(3)\times 
U(1)_x$ gauged group. Obviously, $g$ corresponds to the $SU(2)$ gauge 
coupling while the relation between $g, g_x$ and the $U(1)_Y$ gauge 
coupling $g'$ is given by (\ref{gpgx}), which simply fixes $g_x$ in terms 
of $g$ and $g'$.

In order to write the one-loop Higgs potential, one can compute from 
(\ref{Lk}) the masses of the gauge bosons in terms of 
$\Phi_{1,2}$. For this we find convenient to define the operator 
\be 
{\cal O}_{12}\equiv {1\over 
f^2}\left(f_1^2f_2^2-|\Phi_1^\dagger\Phi_2|^2\right)\ . 
\ee 
In a 
background of $\langle h^0\rangle=h/\sqrt{2}$ and $\eta$, this operator 
can be 
expanded as 
\be 
{\cal O}_{12}={1\over 2} h^2 -{1\over 48} {f^2\over f_1^2 
f_2^2} h^2(4 h^2+\eta^2)+... 
\ee 
Generically, one gets masses of the form 
\be m^2_{H,L}={M^2\over 2}\left[1\pm\sqrt{1-4\kappa_{12}^2{\cal 
O}_{12}/M^2}\right]\ , \label{genmass} \ee where the subindices $H,L$ 
stand for heavy and light masses, $M$ is a generic mass of order $f$ and 
$\kappa_{12}$ is some combination of couplings. An expansion in powers of 
${\cal O}_{12}$ gives 
\bea 
m_H^2&=&M^2-\kappa_{12}^2{\cal O}_{12}+{\cal O}({\cal O}_{12}^2)\ , 
\nonumber\\ 
m_L^2&=&\kappa_{12}^2{\cal O}_{12}+{\cal O}({\cal O}_{12}^2)\ . 
\eea

Besides the massless photon, the rest of gauge bosons have the following 
masses. For the charged $(W^\pm,W'^\pm)$, formula (\ref{genmass}) holds 
with \be M^2=M_{W'}^2\equiv {1\over 2} g^2 f^2\ , \; \; \; \; 
\kappa_{12}^2={1\over 2} g^2\ . \ee Expanding in powers of $h$, one 
reproduces (\ref{mWWp}). For the $(Z'^0,Z^0)$ pair, again the masses are 
given by formula (\ref{genmass}), now with \be M^2=M_{Z'}^2\equiv {2 
g^2\over 3-t_w^2} f^2\ , \; \; \; \; \kappa_{12}^2={1\over 2} (g^2+g'^2)\ 
, 
\ee where $t_w\equiv g'/g$. An expansion in powers of $h$ reproduces 
(\ref{mZZp}). Finally, for the complex $W'^0$ 
\be 
m_{W'^0}^2=M_{W'}^2\ ,\;\;\; \kappa_{12}^2=0\ .
\ee

In the fermion sector, the Yukawa part of the Lagrangian, reads \be {\cal 
L}_Y=\lambda_1 u_1^c \Phi_1^\dagger \Psi_Q+\lambda_2 u_2^c \Phi_2^\dagger 
\Psi_Q +{\rm h.c.}\ , \ee with generation indices suppressed (we only care 
about the third family). Here $\Psi_Q$ is an $SU(3)$ triplet (with 
$x$-charge $1/3$) that contains the usual quark doublet while $u_{1,2}^c$ 
are $SU(3)$ singlets (with $x$-charge $-2/3$). A combination of $u_1^c$ 
and $u_2^c$ corresponds to the SM top quark field while the orthogonal 
combination gets a heavy mass with the third component of $\Psi_Q$. 
The explicit masses of these fields follow the pattern of (\ref{genmass}) 
with \be M^2=M_T^2\equiv \lambda_1^2f_1^2+\lambda_2^2f_2^2\ , \; \; \; \; 
\kappa_{12}^2=\lambda_t^2\ , \ee where $\lambda_t$ is the SM top Yukawa 
coupling, given by \be {f^2\over \lambda_t^2} = {f_1^2\over \lambda_2^2}+ 
{f_2^2\over \lambda_1^2}\ . \ee An expansion in powers of $h$ gives 
(\ref{mtT}).

From the generic formula for the masses in eq.~(\ref{genmass}) one sees 
that ${\rm Str} M^2$ is field independent. Therefore, and in contrast with 
previous models, one-loop quadratically divergent corrections from gauge 
or fermion loops do not induce scalar operators to be added to the 
Lagrangian. (This is not the case for scalars, see section~5).

Less divergent one-loop corrections do induce both a mass term and a 
quartic coupling for the Higgs, as explicitly shown in the main text. Here 
we present the one-loop potential in terms of the fields $\Phi_1$ and 
$\Phi_2$. In the $\overline{\rm MS}$ scheme with the renormalization scale 
set to $Q=\Lambda$, it is straightforward to compute the one-loop 
potential including fermion and gauge 
boson loops, once the masses are known as 
functions of ${\cal O}_{12}$. Performing an expansion in powers of ${\cal 
O}_{12}$, this 
potential reads 
\be 
V=\delta m^2 {\cal O}_{12} + \delta_1 \lambda( {\cal 
O}_{12})  {\cal O}_{12}^2 + ... 
\ee 
with $\delta m^2$ as given in 
(\ref{dm2}) and $\delta_1 \lambda( {\cal O}_{12})$ as given by 
(\ref{deltal}) with the $h$-dependence coming through the dependence of 
the masses on  ${\cal O}_{12}$, see eq.~(\ref{genmass}).
Expanding further in powers of $h$ and 
$\eta$, we get \be \label{Vrada} V(h)={1\over 2} \delta m^2 h^2 +{1\over 
4}\left[\delta_1 \lambda(h)-{\delta m^2\over 3} {f^2\over f_1^2 
f_2^2}\right]h^4 -{\delta m^2\over 48} {f^2\over f_1^2 f_2^2}h^2\eta^2 
+... \ee which reproduces (\ref{Vrad}) and gives also the $\eta$ terms.

Finally, a mass operator is introduced in the tree level potential to get 
a correct electroweak symmetry breaking \cite{Sch} 
\be 
\delta_0 V=\mu^2 {\cal O}_X\equiv 
\mu^2 (2f_1f_2-\Phi_1^\dagger\Phi_2 - \Phi_2^\dagger\Phi_1 )\ , \ee which, 
in terms of $h$ and $\eta$, gives \be \delta_0 V= {1\over 
2}\mu_0^2(h^2+\eta^2)-{1\over 48} {\mu_0^2f^2\over f_1^2f_2^2}(h^4+ 
3h^2\eta^2+\eta^4)+... \label{delta0Va} \ee with $\mu_0^2\equiv \mu^2 f^2/ 
(f_1 f_2)$.

As explained in the main text, by choosing $\mu^2_0>0$ we get a positive 
contribution to the Higgs mass parameter that can compensate the heavy Top 
contribution in $\delta m^2$. The tree-level value of the Higgs quartic 
coupling $\lambda$ from (\ref{delta0Va}) is then negative but the large 
(and positive) radiative corrections to $\lambda$ can easily overcome this 
effect. We also see that (\ref{delta0Va}) gives a mass of order $\mu_0$ to 
the $\eta$ field [this field had no mass term in (\ref{Vrada})].



\begin{thebibliography}{99} 
%
\bibitem{CEHI} 
J.~A.~Casas, J.~R.~Espinosa and I.~Hidalgo, ``Implications for New Physics 
from Fine-Tuning Arguments: I. Application to SUSY and Seesaw Cases,'' 
JHEP {\bf 0411} (2004) 057 [hep-ph/0410298]. 
%
\bibitem{veltman} 
R.~Decker and J.~Pestieau, ``Lepton Selfmass, Higgs Scalar And Heavy 
Quark Masses,'' Lett.\ Nuovo Cim.\ {\bf 29} (1980) 
560;\\ 
M.~J.~G.~Veltman, ``The Infrared - Ultraviolet 
Connection,'' Acta Phys.\ Polon.\ B {\bf 12} (1981) 437. 
%
\bibitem{LEParadox} R.~Barbieri 
and A.~Strumia, ``The 'LEP paradox','' [hep-ph/0007265];\\ 
R.~Barbieri, A.~Pomarol, R.~Rattazzi and A.~Strumia, ``Electroweak 
symmetry breaking after LEP1 and LEP2,'' Nucl.\ Phys.\ B {\bf 703}, 127 
(2004) [hep-ph/0405040]. 
%
\bibitem{UV} 
E.~Katz, J.~y.~Lee, A.~E.~Nelson and D.~G.~E.~Walker, 
``A composite little Higgs model,'' [hep-ph/0312287]. 
%
\bibitem{UVmore}
D.~E.~Kaplan, M.~Schmaltz and W.~Skiba,
``Little Higgses and turtles,''
Phys.\ Rev.\ D {\bf 70} (2004) 075009
[hep-ph/0405257];\\
P.~Batra and D.~E.~Kaplan,
``Perturbative, non-supersymmetric completions of the little Higgs,''
[hep-ph/0412267].
%
\bibitem{BG} R.~Barbieri and G.~F.~Giudice, 
``Upper Bounds On Supersymmetric Particle Masses,'' Nucl.\ Phys.\ B {\bf 
306} (1988) 63. 
\bibitem{BC} For discussions on the validity of this approach, see
B.~de Carlos and J.~A.~Casas,
``One Loop Analysis Of The Electroweak Breaking In Supersymmetric Models
And The Fine Tuning Problem,''
Phys.\ Lett.\ B {\bf 309} (1993) 320
[hep-ph/9303291];\\
%
G.~W.~Anderson and D.~J.~Casta\~no,
``Measures of fine tuning,''
Phys.\ Lett.\ B {\bf 347} (1995) 300
[hep-ph/9409419];\\
P.~Ciafaloni and A.~Strumia,
``Naturalness upper bounds on gauge mediated soft terms,''
Nucl.\ Phys.\ B {\bf 494} (1997) 41
[hep-ph/9611204].
%
%
\bibitem{KM} 
C.~F.~Kolda and H.~Murayama, 
``The Higgs mass and new physics scales in the minimal standard model,'' 
JHEP {\bf 0007} (2000) 035 [hep-ph/0003170]. 
%
\bibitem{EJ} M.~B.~Einhorn and 
D.~R.~T.~Jones, ``The Effective Potential And Quadratic Divergences,'' 
Phys.\ Rev.\ D {\bf 46} (1992) 5206. 
%
\bibitem{CEH0} A.~Brignole, J.~A.~Casas, 
J.~R.~Espinosa and I.~Navarro, ``Low-scale supersymmetry breaking: 
Effective description, electroweak breaking and phenomenology,'' Nucl.\ 
Phys.\ B {\bf 666} (2003) 105 [hep-ph/0301121];\\ 
J.~A.~Casas, J.~R.~Espinosa and 
I.~Hidalgo, ``The MSSM fine tuning problem: A way out,'' JHEP {\bf 0401}, 
008 (2004) [hep-ph/0310137]. 
%
\bibitem{RETAHILA} 
D.~Comelli and C.~Verzegnassi, 
``One loop corrections to the lightest Higgs mass in the minimal eta model 
with a heavy Z-prime,'' Phys.\ Lett.\ B {\bf 303} (1993) 277;\\ 
J.~R.~Espinosa and M.~Quir\'os, ``Upper bounds on the lightest Higgs 
boson mass in general  supersymmetric Standard Models,'' Phys.\ Lett.\ B 
{\bf 302} (1993) 51 [hep-ph/9212305];\\ 
M.~Cveti\v c, D.~A.~Demir, J.~R.~Espinosa, L.~L.~Everett and P.~Langacker, 
``Electroweak breaking and the mu problem in supergravity models with an 
additional U(1),'' Phys.\ Rev.\ D {\bf 56} (1997) 2861 [Erratum-ibid.\ D 
{\bf 58} (1998) 119905] [hep-ph/9703317];\\ 
P.~Batra, A.~Delgado, D.~E.~Kaplan and T.~M.~Tait, 
``The Higgs mass bound in gauge extensions of the minimal supersymmetric 
standard model,'' [hep-ph/0309149];\\ 
%
M.~Drees, ``Supersymmetric Models With Extended Higgs 
Sector,'' Int.\ J.\ Mod.\ Phys.\ A {\bf 4} (1989) 3635;\\ 
J.~R.~Ellis, J.~F.~Gunion, 
H.~E.~Haber, L.~Roszkowski and F.~Zwirner, ``Higgs Bosons In A Nonminimal 
Supersymmetric Model,'' Phys.\ Rev.\ D {\bf 39} (1989) 844;\\ 
P.~Binetruy and C.~A.~Savoy, ``Higgs 
And Top Masses In A Nonminimal Supersymmetric Theory,'' Phys.\ Lett.\ B 
{\bf 277} (1992) 453;\\ 
J.~R.~Espinosa and M.~Quir\'os, ``On Higgs boson 
masses in nonminimal supersymmetric standard models,'' Phys.\ Lett.\ B 
{\bf 279} (1992) 92;\\ 
``Gauge unification and the supersymmetric light Higgs mass,'' Phys.\ 
Rev.\ Lett.\ {\bf 81} (1998) 516 [hep-ph/9804235];\\ 
G.~L.~Kane, C.~F.~Kolda and J.~D.~Wells, ``Calculable upper limit on the 
mass of the lightest Higgs boson in any perturbatively valid 
supersymmetric theory,'' Phys.\ 
Rev.\ Lett.\ {\bf 70} (1993) 2686 [hep-ph/9210242];\\ 
%
M.~Bastero-Gil, C.~Hugonie, 
S.~F.~King, D.~P.~Roy and S.~Vempati, ``Does LEP prefer the NMSSM?,'' 
Phys.\ Lett.\ B {\bf 489} (2000) 359 [hep-ph/0006198]. 
%
\bibitem{Littlest} 
N.~Arkani-Hamed, A.~G.~Cohen, E.~Katz and A.~E.~Nelson, ``The littlest 
Higgs,'' JHEP {\bf 0207} (2002) 034 [hep-ph/0206021]. 
%
\bibitem{Peskin} 
M.~Perelstein, M.~E.~Peskin and A.~Pierce, ``Top quarks and electroweak 
symmetry breaking in little Higgs models,'' Phys.\ Rev.\ D {\bf 69} (2004) 
075002 [hep-ph/0310039]. 
\bibitem{ChengLow} 
H.~C.~Cheng and I.~Low, ``Little 
hierarchy, little Higgses, and a little symmetry,'' JHEP {\bf 0408} (2004) 
061 [hep-ph/0405243]. 
\bibitem{Sch} 
M.~Schmaltz, ``The simplest little 
Higgs,'' JHEP {\bf 0408}, 056 (2004) [hep-ph/0407143]. 
%
\bibitem{pew}
C.~Csaki, J.~Hubisz, G.~D.~Kribs, P.~Meade and J.~Terning,
``Big corrections from a little Higgs,''
Phys.\ Rev.\ D {\bf 67}, 115002 (2003)
[hep-ph/0211124];\\
``Variations of little Higgs models and their electroweak constraints,''
Phys.\ Rev.\ D {\bf 68}, 035009 (2003)
[hep-ph/0303236];\\
J.~L.~Hewett, F.~J.~Petriello and T.~G.~Rizzo,
``Constraining the littlest Higgs,''
JHEP {\bf 0310} (2003) 062
[hep-ph/0211218];\\
T.~Han, H.~E.~Logan, B.~McElrath and L.~T.~Wang,
``Phenomenology of the little Higgs model,''
Phys.\ Rev.\ D {\bf 67}, 095004 (2003)
[hep-ph/0301040];\\
M.~C.~Chen and S.~Dawson,
``One-loop radiative corrections to the rho parameter in the littlest 
Higgs model,''
Phys.\ Rev.\ D {\bf 70} (2004) 015003
[hep-ph/0311032].
%
\bibitem{NDA} A.~Manohar and 
H.~Georgi, ``Chiral Quarks And The Nonrelativistic Quark Model,'' Nucl.\ 
Phys.\ B {\bf 234} (1984) 189;\\ 
H.~Georgi, 
{\it Weak Interactions and Modern Particle Theory}, Benjamin/Cummings, 
(Menlo Park, 1984);\\ 
H.~Georgi and L.~Randall, 
Phys.\ B {\bf 276} (1986) 241. 
%
\bibitem{langacker} 
J.~Erler and P.~Langacker, ``Electroweak model and constraints on new 
physics,'' [hep-ph/0407097];\\ 
in 
S.~Eidelman {\it et al.} [Particle Data Group], 
``Review of particle physics,'' Phys.\ Lett.\ B {\bf 592} (2004) 1. 
%
\bibitem{SS} M.~Soldate and R.~Sundrum, 
``Z Couplings To Pseudogoldstone Bosons Within 
Extended Technicolor,'' Nucl.\ Phys.\ B {\bf 340} (1990) 1. 
%
\bibitem{ELR} J.~R.~Espinosa, 
M.~Losada and A.~Riotto, ``Symmetry nonrestoration at high temperature in 
little Higgs models,'' [hep-ph/0409070]. 
%
\bibitem{top}
P.~Azzi {\it et al.}  [CDF Collaboration],
``Combination of CDF and D0 results on the top-quark mass,''
[hep-ex/0404010].
%
\end{thebibliography}
\end{document}